\documentclass[aps,prl,twocolumn,superscriptaddress,longbibliography]{revtex4-1}

\usepackage{mathrsfs}
\usepackage{tabularx}
\usepackage{slashed}
\usepackage{amsmath}
\usepackage{ulem}
\usepackage{amsfonts}
\usepackage{graphicx,color}
\usepackage{times}
\usepackage{braket}
\usepackage[caption=false]{subfig}
\usepackage[colorlinks,linkcolor=red,citecolor=blue]{hyperref}
\newcommand{\comment}[1]{}

\AtBeginDocument{%
	\newwrite\bibnotes
	\def\bibnotesext{Notes.bib}
	\immediate\openout\bibnotes=\jobname\bibnotesext
	\immediate\write\bibnotes{@CONTROL{REVTEX41Control}}
	\immediate\write\bibnotes{@CONTROL{%
			apsrev41Control,author="08",editor="1",pages="1",title="0",year="1"}}
	\if@filesw
	\immediate\write\@auxout{\string\citation{apsrev41Control}}%
	\fi
}%

\begin{document}
	
	\title{Multiferroicity and Topology in Twisted Transition Metal Dichalcogenides}
	
	\author{Ahmed Abouelkomsan}
   	\thanks{ahmed.abouelkomsan@fysik.su.se}
 	\affiliation{Department of Physics, Stockholm University, AlbaNova University Center, 106 91 Stockholm, Sweden}
	\affiliation{Department of Physics, Massachusetts Institute of Technology, Cambridge, Massachusetts 02139, USA}
	\author{Emil J. Bergholtz}
	\affiliation{Department of Physics, Stockholm University, AlbaNova University Center, 106 91 Stockholm, Sweden}
	\author{Shubhayu Chatterjee}
	\affiliation{Department of Physics, University of California, Berkeley, California 94720, USA}
 \affiliation{Department of Physics, Carnegie Mellon University, Pittsburgh, PA 15213, USA}
	
	\date{\today}
	\begin{abstract}
		Van der Waals heterostructures have recently emerged as an exciting platform for investigating the effects of strong electronic correlations, including various forms of magnetic or electrical orders. Here, we perform an unbiased exact diagonalization study of the effects of interactions on topological flat bands of twisted transition metal dichalcogenides (TMDs) at odd integer fillings. For hole-filling $\nu_h = 1$, we find that the Chern insulator phase, expected from interaction-induced spin-valley polarization of the bare bands, is quite fragile, and gives way to spontaneous multiferroic order --- coexisting ferroelectricity and ferromagnetism, in presence of long-range Coulomb repulsion. 
		We provide a simple real-space picture to understand the phase diagram as a function of interaction range and strength. 
		Our findings establish twisted TMDs as a novel and highly tunable platform for multiferroicity, and we outline a potential route towards electrical control of magnetism in the multiferroic phase.
	\end{abstract}
	
	\maketitle
	\textit{Introduction}. 
	In recent years, moir\'e materials have emerged as highly versatile Van der Waals heterostructures \cite{geimVanWaalsHeterostructures2013,ladoDesignerQuantumMatter2021}. 
	Through a wide range of experimental tuning knobs, the interaction strength can be enhanced relative to the bandwidth. 
	Consequently, a variety of correlated phases have been observed in two separate classes of moir\'e materials, graphene-based heterostructures such as twisted bilayer graphene \cite{balentsSuperconductivityStrongCorrelations2020,caoCorrelatedInsulatorBehaviour2018,caoUnconventionalSuperconductivityMagicangle2018,chenSignaturesTunableSuperconductivity2019,luSuperconductorsOrbitalMagnets2019,yankowitzTuningSuperconductivityTwisted2019} and heterostructures made from semiconductor transition metal dichalcogenides (TMDs) \cite{tangSimulationHubbardModel2020a,xuTunableBilayerHubbard2022,wangCorrelatedElectronicPhases2020,xuCorrelatedInsulatingStates2020a,huangCorrelatedInsulatingStates2021,reganMottGeneralizedWigner2020a,liContinuousMottTransition2021,ghiottoQuantumCriticalityTwisted2021,liQuantumAnomalousHall2021,zhaoRealizationHaldaneChern2022,taoValleycoherentQuantumAnomalous2022} either by using the same TMDs (homobilayers) or different TMDs (heterobilayers).	
    These correlated phases encompass unconventional forms of magnetism driven by both orbital and spin degrees of freedom \cite{polshynElectricalSwitchingMagnetic2020,chenTunableCorrelatedChern2020,chenTunableOrbitalFerromagnetism2022,wangLightinducedFerromagnetismMoire2022,linSpinorbitDrivenFerromagnetism2022a,sharpeEmergentFerromagnetismThreequarters2019,liuOrbitalMagneticStates2021,serlinIntrinsicQuantizedAnomalous2020,chenTunableOrbitalFerromagnetism2022} or different forms of electric order such as ferroelectricity, charge density waves and Wigner crystals \cite{xuCorrelatedInsulatingStates2020a,huangCorrelatedInsulatingStates2021,reganMottGeneralizedWigner2020a,polshynTopologicalChargeDensity2022}. 
	However, multiferroicity --- the simultaneous presence of magnetic and electric order, is highly desirable, as it enables electrical manipulation of magnetism, and vice versa \cite{eerensteinMultiferroicMagnetoelectricMaterials2006}. 
	The high tunability of moir\'e materials combined with prospects of spontaneous multiferroicity would therefore not only sharpen our understanding of strong electronic correlations, but also open the door to next-generation spintronics devices \cite{parsonnetNonvolatileElectricField2022,manipatruniScalableEnergyefficientMagnetoelectric2019,chenDissipationlessMultiferroicMagnonics2015}.  
	
\begin{figure}[t!]
		\centering
		\includegraphics[width=\linewidth]{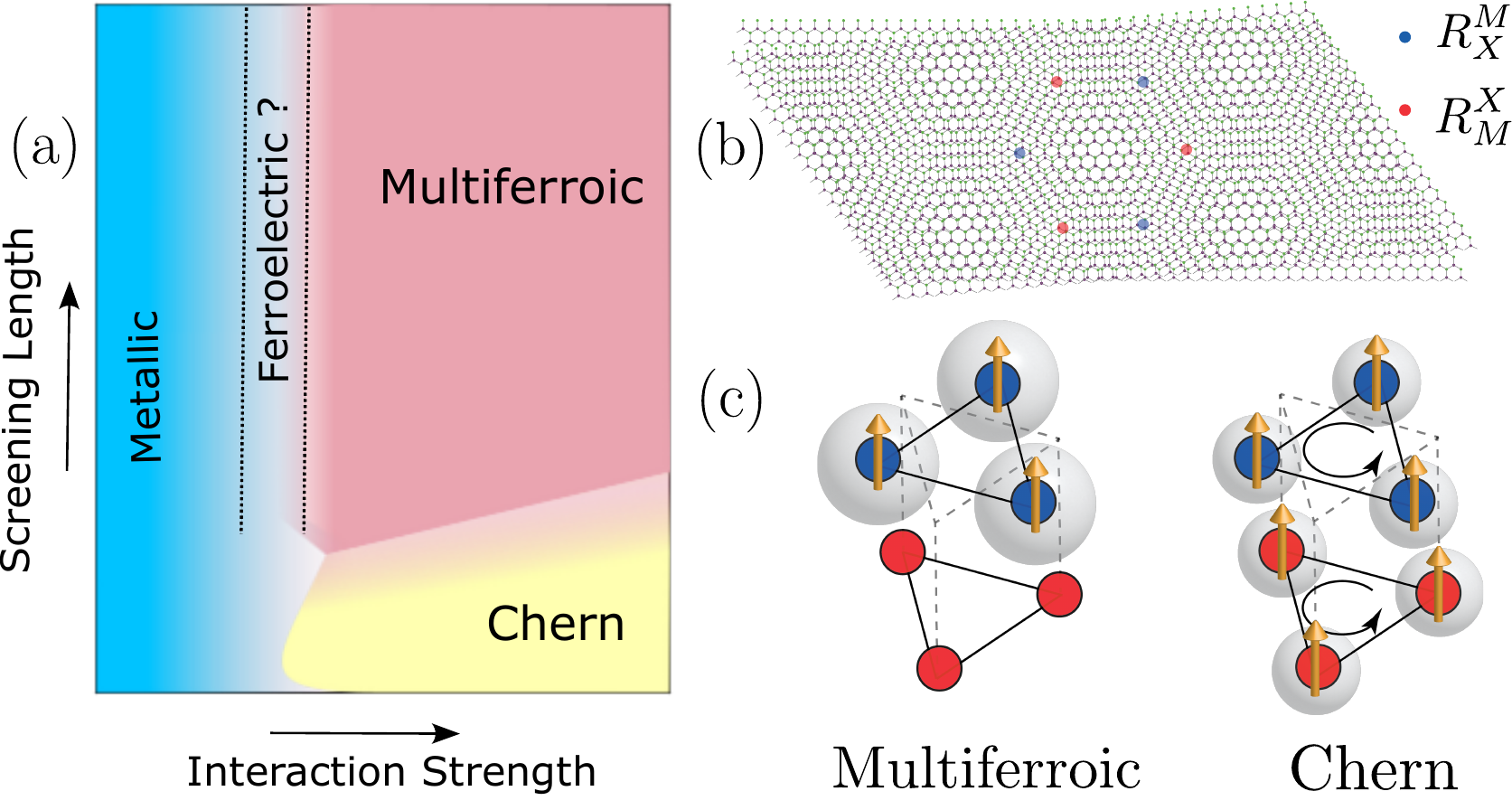}
		\caption{(a) Schematic phase diagram at hole integer filling $\nu_h = 1$ as a function of screening length and interaction strength of Coulomb interactions. 
        (b) Moir\'e TMD homobilayer in real space with two interpenetrating triangular lattices from the two layers forming a honeycomb lattice that carries most of the electronic spectral weight in the active bands. (c) Schematic real-space description of the multiferroic and Chern insulator phases.} 
		\label{fig_illus}
	\end{figure}

\begin{figure}[t]
	\centering
	\includegraphics[width=\linewidth]{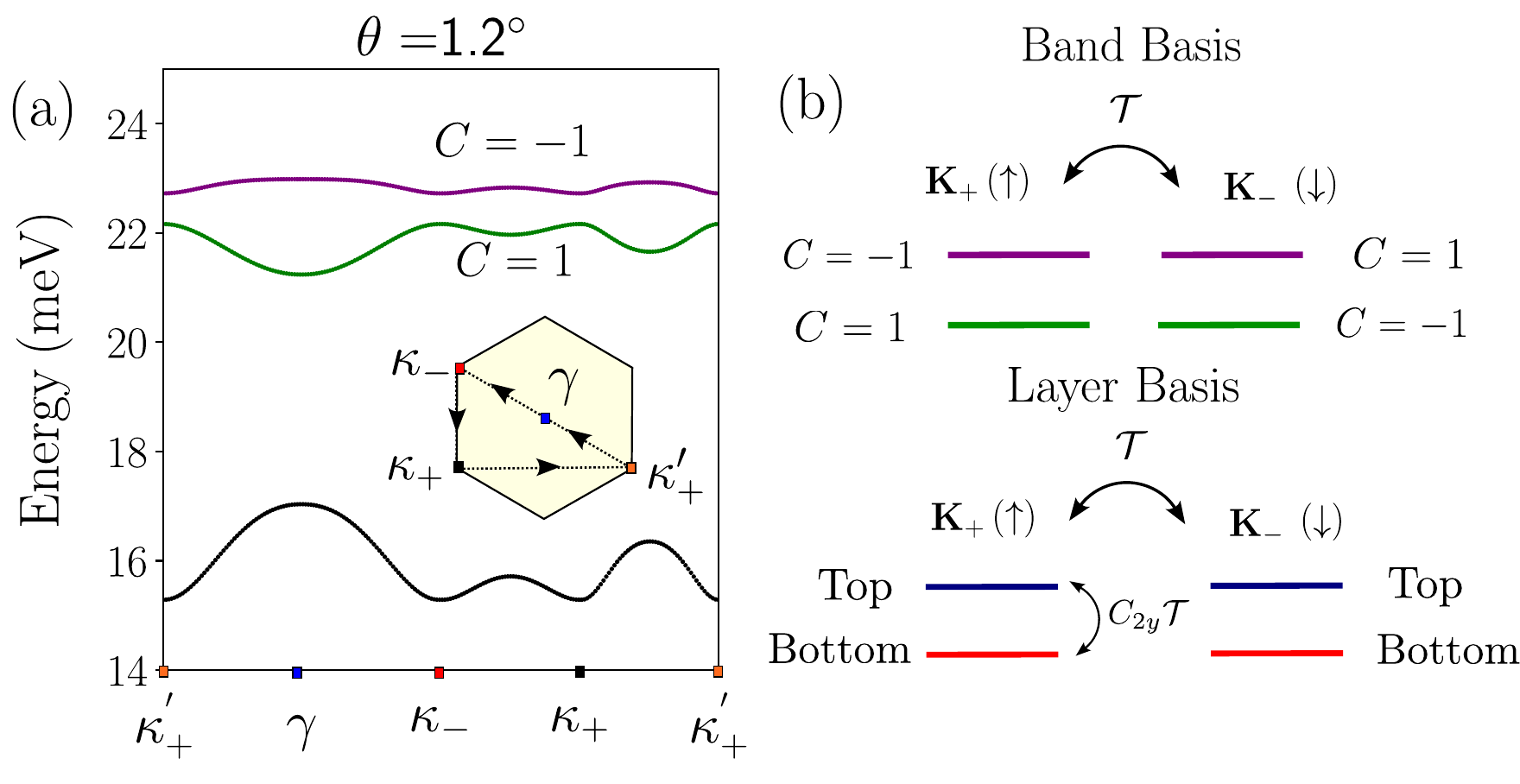}
	\caption{ (a) Band structure \cite{wuTopologicalInsulatorsTwisted2019a} of twisted $\mathrm{MoTe}_2$ at twist angle $\theta = 1.2^\circ$. (b) A schematic represnetation of the problem considered in both band and layer basis.}
	\label{fig_bandstructure}
\end{figure}

	In this Letter, we introduce twisted TMD bilayers as a candidate for interaction-driven spontaneous multiferroic order. 
	To this end, we consider the flat bands of TMD homobilayers at small twist angles, and present an unbiased exact diagonalization study of the interacting phase diagram at odd integer hole-fillings ($\nu_h = 1, 3$). 
	The key feature of the phase diagram at $\nu_h = 1$ (Fig. \ref{fig_illus}(a)) is the robust presence of multiferroic order --- ferromagnetism from spin-valley polarization and ferroelectricity from layer polarization, driven by long-range Coulomb repulsion. 
 Decreasing the range of the Coulomb interaction by tuning the gate-distance drives the system through a topological phase transition from the multiferroic phase to a ferromagnetic Chern insulator.
	We provide an intuitive understanding of the competition between the two phases via a simple real space description of the topological flat bands.
    Finally, leveraging multiferroicity, we propose a route towards electrical control of magnetism in twisted TMDs. 
\\ 
	\textit{Model}. Our starting point is continuum models \cite{wuTopologicalInsulatorsTwisted2019a,panBandTopologyHubbard2020a} developed for the valence bands of twisted TMD homobilayers.
	We are particularly interested in the regime where the \textit{active} bands are topologically non-trivial and carry opposite Chern numbers in a given valley, together with their time-reversed partners in the opposite valley. 
    To be specific, we focus on $\rm MoTe_2$ where this regime is accessible at twist angles $\theta = 1^\circ - 1.5^\circ$ \cite{wuTopologicalInsulatorsTwisted2019a} as shown in  Fig. \ref{fig_bandstructure}(a), although 
    our results apply more generally to other twisted TMD homobilayers such as twisted $\textrm{WSe}_2$ \cite{panBandTopologyHubbard2020a} or when the low-lying energy bands admit a specific real space description as we elaborate below.
	The strong spin-orbit coupling in TMD valence bands \cite{zhuGiantSpinorbitinducedSpin2011} leads to an effective spin-valley locking where the spin up (down) is tied to valley $\mathbf{K}_+$ ($\mathbf{K}_-$). 

The active pair of bands in each valley are very close in energy while there is a much bigger gap to the remaining valence bands (Fig. \ref{fig_bandstructure}(a)).
Therefore we project interactions \cite{supplement} to the active bands, leading to the following Hamiltonian:
	\begin{equation}
		\label{eq:intHam}
			\! H \! = \!  \sum_{\mathbf{k}\alpha\tau} -\epsilon_{\alpha\tau}(\mathbf{k}) c^{\dagger}_{\alpha\tau}(\mathbf{k}) c_{\alpha\tau}(\mathbf{k}) + \frac{1}{N_c} \! \sum_{\mathbf{q}\tau_1 \! \tau_2 } \!  U(\mathbf{q}) \! \!  : \! \rho_{\tau_1}(\mathbf{q}) \rho_{\tau_2}(-\mathbf{q})\! \! :\! 
	\end{equation}
	where $\alpha = 1 \> (2) $ denotes the top (bottom) band and $\tau = \pm$ denotes the valley (equivalently the spin $S_z$ because of the effective spin-valley locking). 
	$c^\dagger_{\alpha \tau}$ are hole band creation operators, and $\epsilon_{\alpha \tau}(\mathbf{k})$ is the non-interacting band dispersion. 
	The second term represents density-density interactions between the holes with the projected density operator in one valley given by $\rho_{\tau}(\mathbf{q}) = \sum_{\mathbf{k} \alpha\beta} \lambda^{\alpha\beta}_{\tau}(\mathbf{k}+\mathbf{q},\mathbf{k}) c^{\dagger}_{\alpha\tau}(\mathbf{k} + \mathbf{q}) c_{\beta\tau}(\mathbf{k}) $ where $\lambda^{\alpha\beta}_{\tau}(\mathbf{k}+\mathbf{q},\mathbf{k}) \equiv \langle u_{\alpha \tau}(\mathbf{k}+\mathbf{q})|u_{\beta \tau}(\mathbf{k}) \rangle$ are the form factors calculated from the Bloch eigenstates $u_{\alpha\tau}(\mathbf{k})$ of the bands. $N_c$ denotes the number of moir\'e unit cells.	We model the interaction $U(\mathbf{q})$ as the dual-gated Coulomb repulsion: $U(\mathbf{q}) = 2 \pi U_0 \tanh(d_{g} |\mathbf{q}|)/(\sqrt{3}|\mathbf{q}|a_M)$ where $a_M$ is the moir\'e lattice constant. 
    The screening length $d_g$ represents the distance between the gates and the moir\'e superlattice and $U_0 = 1/(4\pi\epsilon \epsilon_0 a_M)$ is the strength of interaction, $\epsilon$ being the dielectric constant.
	The Hamiltonian \eqref{eq:intHam} is normal-ordered with respect to charge neutrality. In addition, it has the following key symmetries: a $\pi$-rotation about an in-plane axis --- $C_{2y}$ that swaps the two layers, time reversal symmetry $\mathcal{T}$ that flips the spin-valley degree of freedom, and $U(1) \times U(1)$ symmetry corresponding to independent charge conservation within each valley.

    To diagnose the possible competing phases, it is useful construct a maximally layer-polarized bases for the flat bands.
    To this end, we 
    define $\tilde{c}^\dagger_{\beta \tau}(\mathbf{k}) = \sum_{\alpha} U_{\beta \alpha} (\mathbf{k}) c^{\dagger}_{\alpha \tau}(\mathbf{k}) $, where $U_{\beta \alpha}(\mathbf{k})$ is a $2 \times 2$ unitary matrix that is optimized to maximize the layer polarization in each valley \cite{devakulMagicTwistedTransition2021}. 
    The layer basis vectors within a single valley are related to each other by a combined $C_{2y} \mathcal{T}$ symmetry (Fig.\ref{fig_bandstructure}(b)), thus spontaneous ferroelectric order can be conveniently detected numerically through $C_{2y}$ breaking in the ground state. 
    \begin{figure}[t]
		\centering
		\includegraphics[width=\linewidth]{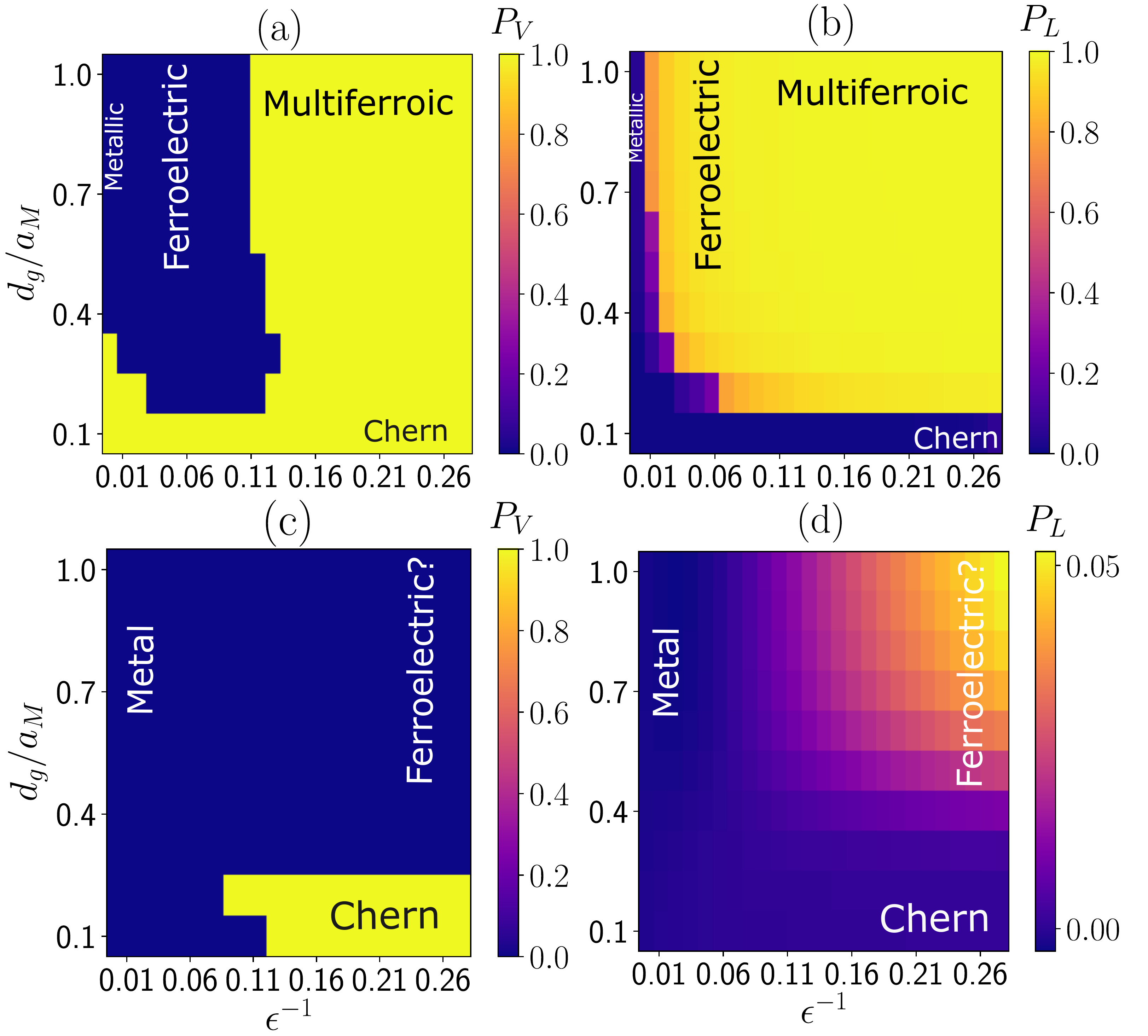}
		\caption{ED phase diagram on cluster C6 \cite{supplement,Vznote} for (a-b) hole-filling $\nu_h = 1$ and (c-d) hole-filling $\nu_h = 3$, as a function of (i) the ratio of the gate-induced screening length  to the moir\'e lattice constant $d_g/a_M$, and (ii) the inverse of the dielectric constant $\epsilon^{-1}$.
        $P_V$ and $P_L$ are the valley polarization and layer polarization densities respectively, as defined by Eq.~ \eqref{eq:observables}. 
        A multiferroic phase, with both $P_v, P_L \neq 0$, dominates the phase diagram for $\nu_h = 1$.} 
		\label{fig_phasediagram}
	\end{figure}
	
	\textit{Results.}
	We perform an unbiased momentum space exact diagonalization (ED) study \cite{supplement} of the Hamiltonian \eqref{eq:intHam} focusing on odd integer filling $\nu_h = 1$ (one hole per moir\'e unit cell) and $\nu_h = 3$ (three holes per moir\'e unit cell) \cite{supplement}.
	To characterize the obtained many-body ground states, we define the following two observables: 
	\begin{equation}
	\label{eq:observables}
	\begin{aligned}
		P_{V} =  \dfrac{1}{N_e} \sum_{\mathbf{k} \alpha }  \braket{n_{\alpha +}(\mathbf{k})} - \braket{n_{\alpha -}(\mathbf{k})} \\
		P_{L} =  \dfrac{1}{N_e} \sum_{\mathbf{k} \tau }  \braket{\tilde{n}_{2 \tau}(\mathbf{k})} - \braket{\tilde{n}_{1 \tau}(\mathbf{k})}
	\end{aligned}
	\end{equation} where $n_{\alpha\tau}(\mathbf{k}) = c^\dagger_{\alpha \tau}(\mathbf{k})c_{\alpha \tau}(\mathbf{k})$ is the $\alpha$ band occupation, $\tilde{n}_{\beta\tau}(\mathbf{k}) = \tilde{c}^\dagger_{\beta \tau}(\mathbf{k})\tilde{c}_{\beta \tau}(\mathbf{k})$ is the $\beta$ layer occupation and $N_e$ is the number of electrons. 
    The quantities in Eq.~\eqref{eq:observables} probe distinct properties of the many-body ground state. 
    $P_V$ measures spin-valley polarization and 
    detects ferromagnetism, while $P_L$ measures layer polarization and therefore probes ferroelectricity.
    Simultaneous non-zero values for both $P_V$ and $P_L$ thus indicate multiferroic order.
    
In Figs.~\ref{fig_phasediagram}(a) and ~\ref{fig_phasediagram}(b), we present the phase diagram at filling $\nu_h = 1$ as a function of two experimentally tunable parameters: (i) the ratio of the screening length of the dual-gated Coulomb interaction to the moir\'e lattice constant $d_g/a_M$ and (ii) the inverse of the dielectric constant $\epsilon^{-1}$. When $\epsilon^{-1}$ is extremely small, we find the expected metallic phase. 
	The metallic phase has no spin-valley polarization (c.f Fig. \ref{fig_phasediagram} (a)) and the many-body ground state occupation is found to be concentrated in the lower band in each valley, indicating that the system simply minimizes the non-interacting kinetic energy in this regime. 
	As $\epsilon^{-1}$ is increased, we find ubiquitous ferromagnetic order due to spin-valley polarization, as evident in figure \ref{fig_phasediagram}(a).
	The spin-valley ferromagnetism is of the Ising type and corresponds to spontaneous symmetry breaking of time-reversal symmetry.
	In addition to ferromagnetism, we find the ground state in the majority   of the phase diagram to show strong layer polarization, i.e, ferroelectricity, by spontaneously breaking the $C_{2y}$ symmetry. 
	To characterize this, we compute the layer polarization $P_L$ defined in Eq.~\eqref{eq:observables} and find that it is non-zero, as shown in figure \ref{fig_phasediagram}(b).  The simultaneous occurence of spin-valley ferromagnetism and ferroelectricity corresponds to the multiferroic phase which is observed for $d_g/a_M \geq  0.2$ and $\epsilon^{-1} \geq 0.11$ \cite{eps_note}.


			\begin{figure}[t!]
		\centering
		\includegraphics[width=\linewidth]{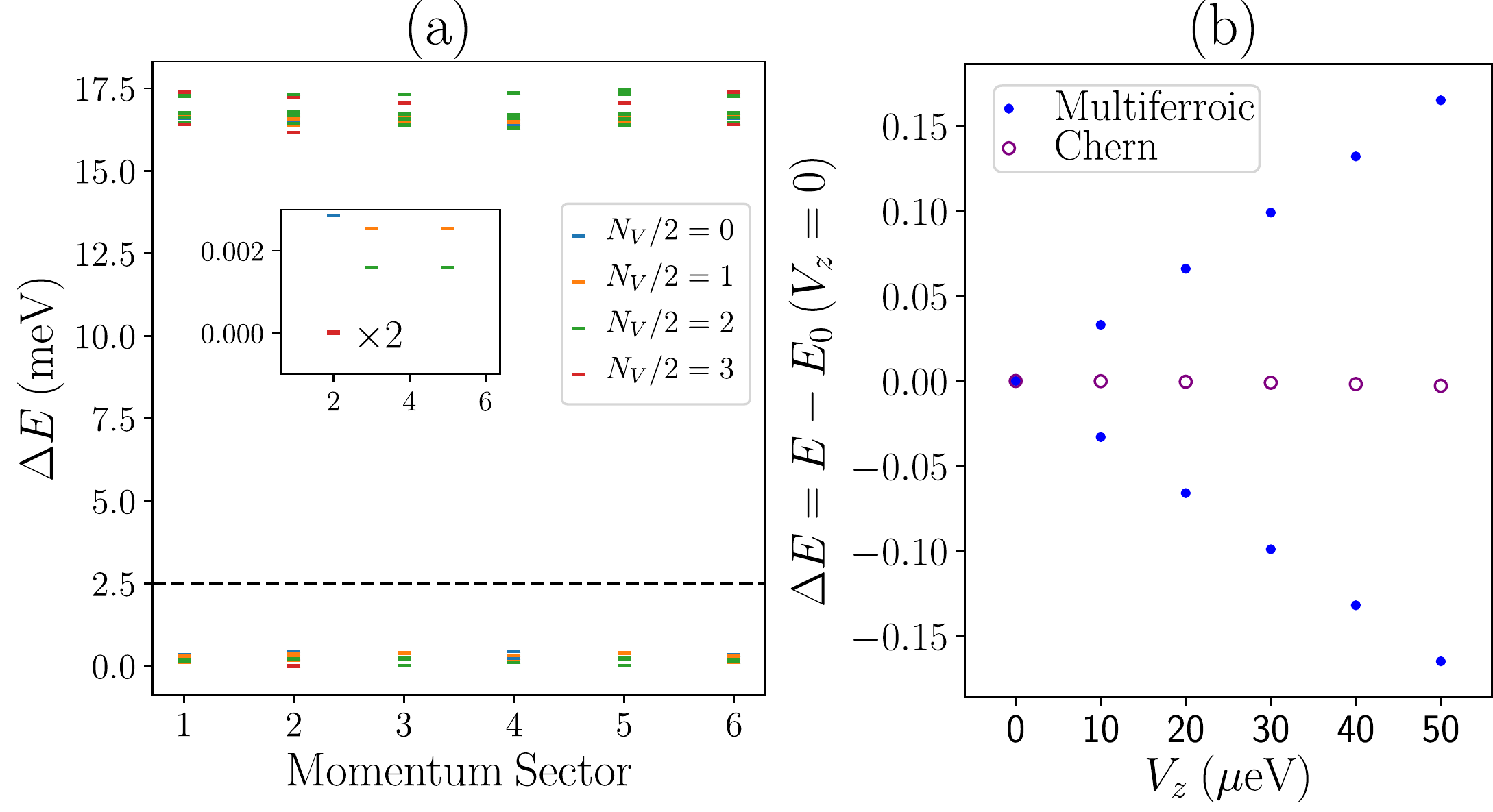}
		\caption{(a) ED spectrum from cluster C6 \cite{supplement} in the multiferroic phase for $\epsilon^{-1}= 0.287$, $d_g/a_M = 1$ at filling $\nu_h = 1$. $N_{V}$ labels the different spin-valley sectors, $N_{V} = (N_{+} - N_{-})$. The inset shows the low-lying manifold of states. There are 84 states below the dashed line  matching a Hilbert space dimension $\textrm{dim}_{\mathcal{H}} = 2 \times \sum_{N_V \geq 0} {N_c \choose N_V/2} = 84$ for $N_c = 6$ unit cells where the factor of $2$ accounts for the two different sublattices (b) Energy splitting of the ground states of the different phases upon the application of a small displacement field $V_z$ that explicitly breaks the $C_{2y}$ symmetry. }
		\label{fig_EDspectrum}
	\end{figure} 
	Further evidence that the layer-swapping $C_{2y}$ symmetry is broken in the multiferroic phase may be gleaned from the response of the ground states to a symmetry breaking perturbation in the form of a tiny out of plane displacement field $V_z$.
	When $V_z = 0$, the many-body multiferroic ground state manifold is two-fold degenerate (see inset of Fig. \ref{fig_EDspectrum}(a)) if we fix spin-valley polarization. 
	The two degenerate states are related by $C_{2y}$, and correspond to larger carrier density in the top or bottom layer. 
	The displacement field $V_z$ is odd under $C_{2y}$, and therefore splits this degeneracy. 
	Accordingly, as we increase $V_z$, we find the two-fold degenerate ground states of the multiferroic phase split linearly in $V_z$ and flow in opposite directions as shown in Fig. \ref{fig_EDspectrum}(b) --- providing comprehensive evidence of $C_{2y}$ breaking.
 This behavior of the multiferroic ground state manifold stands in sharp contrast to the ground states of the competing Chern insulator phase, that is insensitive to small displacement fields. 
	Lastly, we note that the parts of the phase diagram where we observe layer polarization also exhibit significantly large off-diagonal components in the band basis \cite{supplement}. 
	This is consistent with our intuition that the layer bases - an approximate eigenbasis for the multiferroic - strongly mix the top and bottom bands. 
	Indeed, such band-mixing is expected for large values of $\epsilon^{-1}$ where multiferroicity is observed --- interactions overcome the small band gap between the two active bands.
 
 Upon decreasing the screening length $d_g$, the range of Coulomb interaction is shortened in real space and we observe a transition to a Chern insulator phase obtained by hole filling one of the upper Chern band within one valley at $\nu_h = 1$.
 While the Chern insulator is still ferromagnetic in both orbital (valley) and spin sectors, as indicated in Fig. \ref{fig_phasediagram}(a), the ferroelectricity disappears as the Chern bands themselves have almost equal spectral weight in both layers.

In addition to the fully symmetric metal, spin-valley polarized Chern insulator and multiferroic phases, we also find signatures of an intermediate ferroelectric phase --- characterized by non-zero layer polarization and vanishing valley polarization, between the metallic and the multiferroic phases.
In contrast to the rest of the phases, we find the existence of this intermediate ferroelectric phase at $\nu_h = 1$ to be sensitive to the choice of the momentum space cluster used for ED \cite{supplement}.
Henceforth, we focus on understanding the rest of the phase diagram, which is robust to the choice of momentum cluster. 
 
The observed structure of the phase diagram in Fig \ref{fig_phasediagram}(a-b) for $\nu_h = 1$, in particular the competition between the Chern insulator and the multiferroic, can be understood from an approximate real space description of the flat bands.
The layer-projected wavefunctions of the two active bands in each spin-valley sector are found \cite{wuTopologicalInsulatorsTwisted2019a,panBandTopologyHubbard2020a} to be localized on atomic sites that form a triangular lattice in each layer (Fig. \ref{fig_illus}(b)).
In the top layer, these atomic sites are denoted by $R^{M}_X$, as metal ($M$) atoms of the top layer are locally aligned with chalcogen ($X$) atoms of the bottom layer within the moir\'e unit cell. 
In the bottom layer, these sites are denoted by $R^{X}_M$, as the chalcogen ($X$) atoms of the top layer are locally aligned with the metal ($M$) atoms of the bottom layer. 
Taken together, the two inter-penetrating triangular lattices from the two layers form a honeycomb lattice at the moir\'e lengthscale. 

This lattice structure combined with the opposite Chern $C = \pm 1$ bands allows the construction of a tight-binding model that provides a realization of Kane-Mele physics \cite{kaneQuantumSpinHall2005} at the non-interacting level. 
Since states in the two layers are localized on the two sublattices $R_X^M$ and $R_M^X$, layer polarization implies sublattice polarization. 
The reason for sublattice polarization for long-range repulsion is quite intuitive.
For large screening lengths $d_g$, the interaction is essentially long range --- there is significant nearest-neighbor repulsion between the electrons in addition to on-site repulsion. 
In this regime, the electrons can minimize both on-site and nearest neighbor repulsive interactions by localizing on a single triangular sublattice of the effective honeycomb lattice (Fig.~\ref{fig_illus}(c)), leading to ferroelectric order.

The aforementioned real space picture is strongly reflected in the many-body spectrum of the multiferroic phase (Fig. \ref{fig_EDspectrum}(a)) where we observe a low energy manifold of states separated by a large gap from the rest of the spectrum. 
The number of the low-lying states in this manifold is always consistent with the expected dimension of a Hilbert space that is the sum of two subspaces, one for each sublattice. Each subspace consist of states with single hole occupancy for $\nu_h = 1$.

On the other hand, decreasing the screening length $d_g$ so that the range of the Coulomb interaction is much smaller than the distance between the sublattices implies that only the on-site (Hubbard) repulsion is important.
In this regime, phases with uniform electron density on both sublattices, such as the Chern insulator, become energetically favorable, as observed in our numerics. 




  A few comments are in order. While we focused on twisted TMD homobilayers, our intuitive understanding of our numerical results that are based on the full continuum model, in terms of a simple real space picture consisting of layers as sublattices on a honeycomb lattice, suggests that multiferroicity could be looked for in models that admit such interpretation. 
This expands the scope of our study to other TMD moir\'e systems such as TMD heterobilayers where the moir\'e potential extrema could effectively give to a honeycomb lattice description \cite{devakulQuantumAnomalousHall2022a,zhangMoirQuantumChemistry2020,zhangElectronicStructuresCharge2021a,slagleChargeTransferExcitations2020}. 
This alleviates the need 
to make samples with tiny twist angles, a challenge so far 
for TMD heterostructures. 
	
	Our results highlight a screening-dependent competition between the quantum anomalous Hall effect and multiferroicity that could be tuned in experiments by changing electrostatic gate distances. 
    Since the Chern insulator phase requires small gate distances $d_g/a_M < 0.2$, we expect multiferroicity to be the dominant phase at filling $\nu_h = 1$ when the band gap between the active flat bands is minimal.
	Further, the multiferroic insulator breaks only discrete symmetries $C_{2y}$ and $\mathcal{T}$, and is therefore robust to thermal fluctuations. 
    We quantify this robustness by estimating the energy cost to transfer a hole to the opposite layer in the maximally polarized ferroelectric phase to be $\approx$ 17 meV for $\epsilon^{-1} \approx 0.23 $ \cite{supplement}, concomitant with the interaction scale $U_0$.

The moir\'e superlattice plays a crucial role in the energetic stabilization of ferroelectricity, as it endows the effective honeycomb lattice with a moir\'e lengthscale $a_M$ that is much larger than the microscopic inter-layer spacing. 
Without the moir\'e potential, two adjacent sites belonging to the same sublattice (layer) would typically be much closer in real space, relative to two sites in different sublattices (layers).
Accordingly, the Coulomb repulsion between electrons at adjacent sites in the same layer would dominate inter-layer repulsion, and sublattice/layer polarization would be energetically penalized, thereby precluding ferroelectric order.
	
	 We note that multiferroicity in twisted TMDs has been suggested in Ref. \onlinecite{haavistoTopologicalMultiferroicOrder2022a} based on a mean-field study of a simplified nearest neighbour hopping model on honeycomb lattice. 
	 However, such a model cannot accurately capture band topology, 
  or the detailed structure of the form-factors (Bloch wavefunction overlaps) that play
  crucial roles in the phase diagram of moir\'e systems \cite{bultinckGroundStateHidden2020a,bernevigTwistedBilayerGraphene2021a,lianTwistedBilayerGraphene2021}.
	 Our study uses a realistic continuum model that has the advantage of accurately capturing the band topology and form-factors. 
	 This enables us to study the energetic competition between the multiferroic and other phases with non-trivial topology, such as the Chern insulator.

Our results are robust to the effects of band mixing with the third valence band (Fig. \ref{fig_bandstructure}(a)). 
We find that the phase diagram, especially the extent of the multiferroic phase, to be qualitatively similar upon performing ED with three bands per valley,
thereby justifying projecting the interactions to two bands per valley \cite{supplement}.
Analogously, the phase diagram obtained via ED on a disinct cluster in momentum space (C8) is similar, and exhibits robust multiferroicity at larger interaction range \cite{supplement}. 
On the other hand, the multiferroic phase does not appear in the phase diagram (Figs.~\ref{fig_phasediagram}(c), ~\ref{fig_phasediagram}(d)) at filling $\nu_h = 3$, corresponding to hole-filling of three of the four bands, as the layer-polarized ferroelectric phase is valley-unpolarized in the parameter regime we studied, in contrast to $\nu_h = 1$. Since the interacting Hamiltonians at $\nu_h = 1$ and $\nu_h = 3$ are identical up to single-particle terms after a particle-hole transformation, we ascribe the lack of strong correlation effects at $\nu_h = 3$ to an interaction-induced enhancement of bandwidth \cite{supplement}.

Finally, we note that the bare band-structure --- a pair of flat bands with contrasting Chern numbers in each valley as shown in Fig. \ref{fig_bandstructure} --- is deceptively similar to the spinless chiral limit of magic angle twisted bilayer graphene (MATBG) \cite{tarnopolskyOriginMagicAngles2019}.
However, the interacting phase diagram of moir\'e TMD homobilayers that we found here is quite different from MATBG in the strong coupling limit \cite{bultinckGroundStateHidden2020a,bernevigTwistedBilayerGraphene2021a,lianTwistedBilayerGraphene2021}.
We attribute this to crucial symmetry differences between the two systems.
MATBG has enhanced symmetry that prevents the construction of localized Wannier orbitals for the bands in each valley, but admits a Chern basis where the interactions are diagonal in the chiral limit \cite{songAllMagicAngles2019a,zouBandStructureTwisted2018,poFaithfulTightbindingModels2019,poFragileTopologyWannier2018}. 
In contrast, the strong spin-orbit coupling in moir\'e TMDs reduces the symmetry, which allows for the possibility to construct localized Wannier orbitals to describe the flat bands in each valley, but makes the interaction strongly off-diagonal in the band (Chern) basis.  
	  
\textit{Electrical control of magnetism}.
Having established the prospects of multiferroicity in twisted TMDs homobilayers, we now explore a route to manipulate the magnetization $M$ with applied electric fields.
Within Landau theory, the lowest order symmetry-allowed coupling between the magnetic and the electric polarization ($P$) takes the form $H_{M-P} = \alpha M^2 P^2$ \cite{Sym_note}. 
Depending on the sign of the coupling $\alpha$, the two orders will attract ($\alpha <0$) or repel ($\alpha >0$) each other. 
In particular, if $\alpha > 0$, then an electric field $E$ can switch the magnetic order.

To see this, we consider the Landau free energy density $F = (-r_1 M^2 + u_1 M^4) + (-r_2 P^2 + u_2 P^4 - P E) + \alpha M^2 P^2$, where $\{r_1,r_2,u_1,u_2\}$ are the positive constants. 
Minimizing $F$ with respect to $M$ leads to $M^2 = (r_1 - \alpha P^2)/2$, implying the existence of non-zero magnetization for $r_1 > \alpha P^2$. 
Since $M^2 \geq 0$, increasing $P$ by tuning up the electric field $E$ will turn off the magnetic order once $P$ reaches the critical value $P_c = \sqrt{r_1/\alpha}$. 
We conclude that presence of ferromagnetism can be simply tuned by the external electric field. 

	\textit{Discussion.} 
	In this paper, we have presented a comprehensive study of the phase diagram of twisted TMD homobilayers in the presence of long range Coulomb interactions. Our main finding is the abundance of multiferroicity in the phase diagram, driven by spontaneous symmetry breaking and strong interactions. 
 Using this mutiferroic order, we proposed a pathway towards electrical control of magnetism.
 While such control been proposed and demonstrated experimentally in graphene-based platforms \cite{luSuperconductorsOrbitalMagnets2019,polshynElectricalSwitchingMagnetic2020,Taige}, the presence of strong spin-orbit coupling in TMDs offers additional routes to achieve a higher degree of tunability. 


To identify the multiferroic phase in experiments, one may use spin-polarized STM \cite{bode2003spin} to detect spin-polarization, while simultaneously inferring the absence of orbital magnetization (characteristic of Chern insulators) via a scanning Squid probe \cite{jose2017nanosquids}.
The electrical polarization can be probed via nanoscale electrometry \cite{dolde2011electric,Block2021,bian2021nanoscale,Sahay2021} using optically active color centers, such as boron vacancy (V$_B^-$) centers in hBN substrates \cite{Vaidya2023} that are sensitive to local electric fields.
Such color centers can also be used as nanoscale probes of ferromagnetism \cite{casola2018probing,hong2013nanoscale,CRD2019}, present in both multiferroic and Chern phases. 

    Finally, while we have focused on interacting physics at odd-integer fillings 
    where multiferroicity is more likely, 
    we anticipate interesting correlated physics 
    at half-filling ($\nu_h = 2$) too. 
    Specifically, we expect different magnetically ordered phases and possible quantum spin liquids competing with the quantum spin-Hall phase obtained in the non-interacting limit. 
    In addition, exotic phases such as fractional Chern insulators 
\cite{abouelkomsan2020particle,ledwith2020fractional,repellinChernBandsTwisted2020,liSpontaneousFractionalChern2021}, or skyrmion-pairing superconductivity \cite{GroverSenthil,Chatterjee19,Eslam,Assaad,CIZ2022,Kwan} may show up at fractional fillings of the flat bands.
    We leave the study of the full interacting phase diagram as a function of doping to future work.

 \begin{acknowledgements}
	\textit{Acknowledgements}. We are grateful to  Vladimir Calvera, Johan Carlstr\"om, Valentin Crépel, Liang Fu, Sid Morampudi, Dan Parker and Taige Wang for helpful discussions.
	A.A and E.J.B are supported by the Swedish Research Council (grant 2018-00313), and the Knut and Alice Wallenberg Foundation (KAW) via the Wallenberg Academy Fellows program (2018.0460).
	S.C. acknowledges support from the U.S. DOE, Office of Science, Office of Advanced Scientific Computing Research, under the Accelerated Research in Quantum Computing (ARQC) program and the ARO through the MURI program (grant number W911NF-17-1-0323).
 \end{acknowledgements}
	\bibliography{paper_v2}

\begin{thebibliography}{77}%
\makeatletter
\providecommand \@ifxundefined [1]{%
 \@ifx{#1\undefined}
}%
\providecommand \@ifnum [1]{%
 \ifnum #1\expandafter \@firstoftwo
 \else \expandafter \@secondoftwo
 \fi
}%
\providecommand \@ifx [1]{%
 \ifx #1\expandafter \@firstoftwo
 \else \expandafter \@secondoftwo
 \fi
}%
\providecommand \natexlab [1]{#1}%
\providecommand \enquote  [1]{``#1''}%
\providecommand \bibnamefont  [1]{#1}%
\providecommand \bibfnamefont [1]{#1}%
\providecommand \citenamefont [1]{#1}%
\providecommand \href@noop [0]{\@secondoftwo}%
\providecommand \href [0]{\begingroup \@sanitize@url \@href}%
\providecommand \@href[1]{\@@startlink{#1}\@@href}%
\providecommand \@@href[1]{\endgroup#1\@@endlink}%
\providecommand \@sanitize@url [0]{\catcode `\\12\catcode `\$12\catcode
  `\&12\catcode `\#12\catcode `\^12\catcode `\_12\catcode `\%12\relax}%
\providecommand \@@startlink[1]{}%
\providecommand \@@endlink[0]{}%
\providecommand \url  [0]{\begingroup\@sanitize@url \@url }%
\providecommand \@url [1]{\endgroup\@href {#1}{\urlprefix }}%
\providecommand \urlprefix  [0]{URL }%
\providecommand \Eprint [0]{\href }%
\providecommand \doibase [0]{http://dx.doi.org/}%
\providecommand \selectlanguage [0]{\@gobble}%
\providecommand \bibinfo  [0]{\@secondoftwo}%
\providecommand \bibfield  [0]{\@secondoftwo}%
\providecommand \translation [1]{[#1]}%
\providecommand \BibitemOpen [0]{}%
\providecommand \bibitemStop [0]{}%
\providecommand \bibitemNoStop [0]{.\EOS\space}%
\providecommand \EOS [0]{\spacefactor3000\relax}%
\providecommand \BibitemShut  [1]{\csname bibitem#1\endcsname}%
\let\auto@bib@innerbib\@empty
\bibitem [{\citenamefont {Geim}\ and\ \citenamefont
  {Grigorieva}(2013)}]{geimVanWaalsHeterostructures2013}%
  \BibitemOpen
  \bibfield  {author} {\bibinfo {author} {\bibfnamefont {A.~K.}\ \bibnamefont
  {Geim}}\ and\ \bibinfo {author} {\bibfnamefont {I.~V.}\ \bibnamefont
  {Grigorieva}},\ }\bibfield  {title} {\enquote {\bibinfo {title} {Van der
  {{Waals}} heterostructures},}\ }\href {\doibase 10.1038/nature12385}
  {\bibfield  {journal} {\bibinfo  {journal} {Nature}\ }\textbf {\bibinfo
  {volume} {499}},\ \bibinfo {pages} {419--425} (\bibinfo {year}
  {2013})}\BibitemShut {NoStop}%
\bibitem [{\citenamefont {Lado}\ and\ \citenamefont
  {Liljeroth}(2021)}]{ladoDesignerQuantumMatter2021}%
  \BibitemOpen
  \bibfield  {author} {\bibinfo {author} {\bibfnamefont {J.~L.}\ \bibnamefont
  {Lado}}\ and\ \bibinfo {author} {\bibfnamefont {P.}~\bibnamefont
  {Liljeroth}},\ }\href {\doibase 10.48550/arXiv.2102.11779} {\enquote
  {\bibinfo {title} {Designer quantum matter in van der {{Waals}}
  heterostructures},}\ } (\bibinfo {year} {2021}),\ \Eprint
  {http://arxiv.org/abs/2102.11779} {arXiv:2102.11779 [cond-mat]} \BibitemShut
  {NoStop}%
\bibitem [{\citenamefont {Balents}\ \emph {et~al.}(2020)\citenamefont
  {Balents}, \citenamefont {Dean}, \citenamefont {Efetov},\ and\ \citenamefont
  {Young}}]{balentsSuperconductivityStrongCorrelations2020}%
  \BibitemOpen
  \bibfield  {author} {\bibinfo {author} {\bibfnamefont {L.}~\bibnamefont
  {Balents}}, \bibinfo {author} {\bibfnamefont {C.~R.}\ \bibnamefont {Dean}},
  \bibinfo {author} {\bibfnamefont {D.~K.}\ \bibnamefont {Efetov}}, \ and\
  \bibinfo {author} {\bibfnamefont {A.~F.}\ \bibnamefont {Young}},\ }\bibfield
  {title} {\enquote {\bibinfo {title} {Superconductivity and strong
  correlations in moir\'e flat bands},}\ }\href {\doibase
  10.1038/s41567-020-0906-9} {\bibfield  {journal} {\bibinfo  {journal} {Nat.
  Phys.}\ }\textbf {\bibinfo {volume} {16}},\ \bibinfo {pages} {725--733}
  (\bibinfo {year} {2020})}\BibitemShut {NoStop}%
\bibitem [{\citenamefont {Cao}\ \emph {et~al.}(2018{\natexlab{a}})\citenamefont
  {Cao}, \citenamefont {Fatemi}, \citenamefont {Demir}, \citenamefont {Fang},
  \citenamefont {Tomarken}, \citenamefont {Luo}, \citenamefont
  {{Sanchez-Yamagishi}}, \citenamefont {Watanabe}, \citenamefont {Taniguchi},
  \citenamefont {Kaxiras}, \citenamefont {Ashoori},\ and\ \citenamefont
  {{Jarillo-Herrero}}}]{caoCorrelatedInsulatorBehaviour2018}%
  \BibitemOpen
  \bibfield  {author} {\bibinfo {author} {\bibfnamefont {Y.}~\bibnamefont
  {Cao}}, \bibinfo {author} {\bibfnamefont {V.}~\bibnamefont {Fatemi}},
  \bibinfo {author} {\bibfnamefont {A.}~\bibnamefont {Demir}}, \bibinfo
  {author} {\bibfnamefont {S.}~\bibnamefont {Fang}}, \bibinfo {author}
  {\bibfnamefont {S.~L.}\ \bibnamefont {Tomarken}}, \bibinfo {author}
  {\bibfnamefont {J.~Y.}\ \bibnamefont {Luo}}, \bibinfo {author} {\bibfnamefont
  {J.~D.}\ \bibnamefont {{Sanchez-Yamagishi}}}, \bibinfo {author}
  {\bibfnamefont {K.}~\bibnamefont {Watanabe}}, \bibinfo {author}
  {\bibfnamefont {T.}~\bibnamefont {Taniguchi}}, \bibinfo {author}
  {\bibfnamefont {E.}~\bibnamefont {Kaxiras}}, \bibinfo {author} {\bibfnamefont
  {R.~C.}\ \bibnamefont {Ashoori}}, \ and\ \bibinfo {author} {\bibfnamefont
  {P.}~\bibnamefont {{Jarillo-Herrero}}},\ }\bibfield  {title} {\enquote
  {\bibinfo {title} {Correlated insulator behaviour at half-filling in
  magic-angle graphene superlattices},}\ }\href {\doibase 10.1038/nature26154}
  {\bibfield  {journal} {\bibinfo  {journal} {Nature}\ }\textbf {\bibinfo
  {volume} {556}},\ \bibinfo {pages} {80--84} (\bibinfo {year}
  {2018}{\natexlab{a}})}\BibitemShut {NoStop}%
\bibitem [{\citenamefont {Cao}\ \emph {et~al.}(2018{\natexlab{b}})\citenamefont
  {Cao}, \citenamefont {Fatemi}, \citenamefont {Fang}, \citenamefont
  {Watanabe}, \citenamefont {Taniguchi}, \citenamefont {Kaxiras},\ and\
  \citenamefont
  {{Jarillo-Herrero}}}]{caoUnconventionalSuperconductivityMagicangle2018}%
  \BibitemOpen
  \bibfield  {author} {\bibinfo {author} {\bibfnamefont {Y.}~\bibnamefont
  {Cao}}, \bibinfo {author} {\bibfnamefont {V.}~\bibnamefont {Fatemi}},
  \bibinfo {author} {\bibfnamefont {S.}~\bibnamefont {Fang}}, \bibinfo {author}
  {\bibfnamefont {K.}~\bibnamefont {Watanabe}}, \bibinfo {author}
  {\bibfnamefont {T.}~\bibnamefont {Taniguchi}}, \bibinfo {author}
  {\bibfnamefont {E.}~\bibnamefont {Kaxiras}}, \ and\ \bibinfo {author}
  {\bibfnamefont {P.}~\bibnamefont {{Jarillo-Herrero}}},\ }\bibfield  {title}
  {\enquote {\bibinfo {title} {Unconventional superconductivity in magic-angle
  graphene superlattices},}\ }\href {\doibase 10.1038/nature26160} {\bibfield
  {journal} {\bibinfo  {journal} {Nature}\ }\textbf {\bibinfo {volume} {556}},\
  \bibinfo {pages} {43--50} (\bibinfo {year} {2018}{\natexlab{b}})}\BibitemShut
  {NoStop}%
\bibitem [{\citenamefont {Chen}\ \emph {et~al.}(2019)\citenamefont {Chen},
  \citenamefont {Sharpe}, \citenamefont {Gallagher}, \citenamefont {Rosen},
  \citenamefont {Fox}, \citenamefont {Jiang}, \citenamefont {Lyu},
  \citenamefont {Li}, \citenamefont {Watanabe}, \citenamefont {Taniguchi},
  \citenamefont {Jung}, \citenamefont {Shi}, \citenamefont
  {{Goldhaber-Gordon}}, \citenamefont {Zhang},\ and\ \citenamefont
  {Wang}}]{chenSignaturesTunableSuperconductivity2019}%
  \BibitemOpen
  \bibfield  {author} {\bibinfo {author} {\bibfnamefont {G.}~\bibnamefont
  {Chen}}, \bibinfo {author} {\bibfnamefont {A.~L.}\ \bibnamefont {Sharpe}},
  \bibinfo {author} {\bibfnamefont {P.}~\bibnamefont {Gallagher}}, \bibinfo
  {author} {\bibfnamefont {I.~T.}\ \bibnamefont {Rosen}}, \bibinfo {author}
  {\bibfnamefont {E.~J.}\ \bibnamefont {Fox}}, \bibinfo {author} {\bibfnamefont
  {L.}~\bibnamefont {Jiang}}, \bibinfo {author} {\bibfnamefont
  {B.}~\bibnamefont {Lyu}}, \bibinfo {author} {\bibfnamefont {H.}~\bibnamefont
  {Li}}, \bibinfo {author} {\bibfnamefont {K.}~\bibnamefont {Watanabe}},
  \bibinfo {author} {\bibfnamefont {T.}~\bibnamefont {Taniguchi}}, \bibinfo
  {author} {\bibfnamefont {J.}~\bibnamefont {Jung}}, \bibinfo {author}
  {\bibfnamefont {Z.}~\bibnamefont {Shi}}, \bibinfo {author} {\bibfnamefont
  {D.}~\bibnamefont {{Goldhaber-Gordon}}}, \bibinfo {author} {\bibfnamefont
  {Y.}~\bibnamefont {Zhang}}, \ and\ \bibinfo {author} {\bibfnamefont
  {F.}~\bibnamefont {Wang}},\ }\bibfield  {title} {\enquote {\bibinfo {title}
  {Signatures of tunable superconductivity in a trilayer graphene moir\'e
  superlattice},}\ }\href {\doibase 10.1038/s41586-019-1393-y} {\bibfield
  {journal} {\bibinfo  {journal} {Nature}\ }\textbf {\bibinfo {volume} {572}},\
  \bibinfo {pages} {215--219} (\bibinfo {year} {2019})}\BibitemShut {NoStop}%
\bibitem [{\citenamefont {Lu}\ \emph {et~al.}(2019)\citenamefont {Lu},
  \citenamefont {Stepanov}, \citenamefont {Yang}, \citenamefont {Xie},
  \citenamefont {Aamir}, \citenamefont {Das}, \citenamefont {Urgell},
  \citenamefont {Watanabe}, \citenamefont {Taniguchi}, \citenamefont {Zhang},
  \citenamefont {Bachtold}, \citenamefont {MacDonald},\ and\ \citenamefont
  {Efetov}}]{luSuperconductorsOrbitalMagnets2019}%
  \BibitemOpen
  \bibfield  {author} {\bibinfo {author} {\bibfnamefont {X.}~\bibnamefont
  {Lu}}, \bibinfo {author} {\bibfnamefont {P.}~\bibnamefont {Stepanov}},
  \bibinfo {author} {\bibfnamefont {W.}~\bibnamefont {Yang}}, \bibinfo {author}
  {\bibfnamefont {M.}~\bibnamefont {Xie}}, \bibinfo {author} {\bibfnamefont
  {M.~A.}\ \bibnamefont {Aamir}}, \bibinfo {author} {\bibfnamefont
  {I.}~\bibnamefont {Das}}, \bibinfo {author} {\bibfnamefont {C.}~\bibnamefont
  {Urgell}}, \bibinfo {author} {\bibfnamefont {K.}~\bibnamefont {Watanabe}},
  \bibinfo {author} {\bibfnamefont {T.}~\bibnamefont {Taniguchi}}, \bibinfo
  {author} {\bibfnamefont {G.}~\bibnamefont {Zhang}}, \bibinfo {author}
  {\bibfnamefont {A.}~\bibnamefont {Bachtold}}, \bibinfo {author}
  {\bibfnamefont {A.~H.}\ \bibnamefont {MacDonald}}, \ and\ \bibinfo {author}
  {\bibfnamefont {D.~K.}\ \bibnamefont {Efetov}},\ }\bibfield  {title}
  {\enquote {\bibinfo {title} {Superconductors, orbital magnets and correlated
  states in magic-angle bilayer graphene},}\ }\href {\doibase
  10.1038/s41586-019-1695-0} {\bibfield  {journal} {\bibinfo  {journal}
  {Nature}\ }\textbf {\bibinfo {volume} {574}},\ \bibinfo {pages} {653--657}
  (\bibinfo {year} {2019})}\BibitemShut {NoStop}%
\bibitem [{\citenamefont {Yankowitz}\ \emph {et~al.}(2019)\citenamefont
  {Yankowitz}, \citenamefont {Chen}, \citenamefont {Polshyn}, \citenamefont
  {Zhang}, \citenamefont {Watanabe}, \citenamefont {Taniguchi}, \citenamefont
  {Graf}, \citenamefont {Young},\ and\ \citenamefont
  {Dean}}]{yankowitzTuningSuperconductivityTwisted2019}%
  \BibitemOpen
  \bibfield  {author} {\bibinfo {author} {\bibfnamefont {M.}~\bibnamefont
  {Yankowitz}}, \bibinfo {author} {\bibfnamefont {S.}~\bibnamefont {Chen}},
  \bibinfo {author} {\bibfnamefont {H.}~\bibnamefont {Polshyn}}, \bibinfo
  {author} {\bibfnamefont {Y.}~\bibnamefont {Zhang}}, \bibinfo {author}
  {\bibfnamefont {K.}~\bibnamefont {Watanabe}}, \bibinfo {author}
  {\bibfnamefont {T.}~\bibnamefont {Taniguchi}}, \bibinfo {author}
  {\bibfnamefont {D.}~\bibnamefont {Graf}}, \bibinfo {author} {\bibfnamefont
  {A.~F.}\ \bibnamefont {Young}}, \ and\ \bibinfo {author} {\bibfnamefont
  {C.~R.}\ \bibnamefont {Dean}},\ }\bibfield  {title} {\enquote {\bibinfo
  {title} {Tuning superconductivity in twisted bilayer graphene},}\ }\href
  {\doibase 10.1126/science.aav1910} {\bibfield  {journal} {\bibinfo  {journal}
  {Science}\ }\textbf {\bibinfo {volume} {363}},\ \bibinfo {pages} {1059--1064}
  (\bibinfo {year} {2019})}\BibitemShut {NoStop}%
\bibitem [{\citenamefont {Tang}\ \emph {et~al.}(2020)\citenamefont {Tang},
  \citenamefont {Li}, \citenamefont {Li}, \citenamefont {Xu}, \citenamefont
  {Liu}, \citenamefont {Barmak}, \citenamefont {Watanabe}, \citenamefont
  {Taniguchi}, \citenamefont {MacDonald}, \citenamefont {Shan},\ and\
  \citenamefont {Mak}}]{tangSimulationHubbardModel2020a}%
  \BibitemOpen
  \bibfield  {author} {\bibinfo {author} {\bibfnamefont {Y.}~\bibnamefont
  {Tang}}, \bibinfo {author} {\bibfnamefont {L.}~\bibnamefont {Li}}, \bibinfo
  {author} {\bibfnamefont {T.}~\bibnamefont {Li}}, \bibinfo {author}
  {\bibfnamefont {Y.}~\bibnamefont {Xu}}, \bibinfo {author} {\bibfnamefont
  {S.}~\bibnamefont {Liu}}, \bibinfo {author} {\bibfnamefont {K.}~\bibnamefont
  {Barmak}}, \bibinfo {author} {\bibfnamefont {K.}~\bibnamefont {Watanabe}},
  \bibinfo {author} {\bibfnamefont {T.}~\bibnamefont {Taniguchi}}, \bibinfo
  {author} {\bibfnamefont {A.~H.}\ \bibnamefont {MacDonald}}, \bibinfo {author}
  {\bibfnamefont {J.}~\bibnamefont {Shan}}, \ and\ \bibinfo {author}
  {\bibfnamefont {K.~F.}\ \bibnamefont {Mak}},\ }\bibfield  {title} {\enquote
  {\bibinfo {title} {Simulation of {{Hubbard}} model physics in
  {{WSe2}}/{{WS2}} moir\'e superlattices},}\ }\href {\doibase
  10.1038/s41586-020-2085-3} {\bibfield  {journal} {\bibinfo  {journal}
  {Nature}\ }\textbf {\bibinfo {volume} {579}},\ \bibinfo {pages} {353--358}
  (\bibinfo {year} {2020})}\BibitemShut {NoStop}%
\bibitem [{\citenamefont {Xu}\ \emph {et~al.}(2022)\citenamefont {Xu},
  \citenamefont {Kang}, \citenamefont {Watanabe}, \citenamefont {Taniguchi},
  \citenamefont {Mak},\ and\ \citenamefont
  {Shan}}]{xuTunableBilayerHubbard2022}%
  \BibitemOpen
  \bibfield  {author} {\bibinfo {author} {\bibfnamefont {Y.}~\bibnamefont
  {Xu}}, \bibinfo {author} {\bibfnamefont {K.}~\bibnamefont {Kang}}, \bibinfo
  {author} {\bibfnamefont {K.}~\bibnamefont {Watanabe}}, \bibinfo {author}
  {\bibfnamefont {T.}~\bibnamefont {Taniguchi}}, \bibinfo {author}
  {\bibfnamefont {K.~F.}\ \bibnamefont {Mak}}, \ and\ \bibinfo {author}
  {\bibfnamefont {J.}~\bibnamefont {Shan}},\ }\bibfield  {title} {\enquote
  {\bibinfo {title} {A tunable bilayer {{Hubbard}} model in twisted
  {{WSe2}}},}\ }\href {\doibase 10.1038/s41565-022-01180-7} {\bibfield
  {journal} {\bibinfo  {journal} {Nat. Nanotechnol.}\ }\textbf {\bibinfo
  {volume} {17}},\ \bibinfo {pages} {934--939} (\bibinfo {year}
  {2022})}\BibitemShut {NoStop}%
\bibitem [{\citenamefont {Wang}\ \emph {et~al.}(2020)\citenamefont {Wang},
  \citenamefont {Shih}, \citenamefont {Ghiotto}, \citenamefont {Xian},
  \citenamefont {Rhodes}, \citenamefont {Tan}, \citenamefont {Claassen},
  \citenamefont {Kennes}, \citenamefont {Bai}, \citenamefont {Kim},
  \citenamefont {Watanabe}, \citenamefont {Taniguchi}, \citenamefont {Zhu},
  \citenamefont {Hone}, \citenamefont {Rubio}, \citenamefont {Pasupathy},\ and\
  \citenamefont {Dean}}]{wangCorrelatedElectronicPhases2020}%
  \BibitemOpen
  \bibfield  {author} {\bibinfo {author} {\bibfnamefont {L.}~\bibnamefont
  {Wang}}, \bibinfo {author} {\bibfnamefont {E.-M.}\ \bibnamefont {Shih}},
  \bibinfo {author} {\bibfnamefont {A.}~\bibnamefont {Ghiotto}}, \bibinfo
  {author} {\bibfnamefont {L.}~\bibnamefont {Xian}}, \bibinfo {author}
  {\bibfnamefont {D.~A.}\ \bibnamefont {Rhodes}}, \bibinfo {author}
  {\bibfnamefont {C.}~\bibnamefont {Tan}}, \bibinfo {author} {\bibfnamefont
  {M.}~\bibnamefont {Claassen}}, \bibinfo {author} {\bibfnamefont {D.~M.}\
  \bibnamefont {Kennes}}, \bibinfo {author} {\bibfnamefont {Y.}~\bibnamefont
  {Bai}}, \bibinfo {author} {\bibfnamefont {B.}~\bibnamefont {Kim}}, \bibinfo
  {author} {\bibfnamefont {K.}~\bibnamefont {Watanabe}}, \bibinfo {author}
  {\bibfnamefont {T.}~\bibnamefont {Taniguchi}}, \bibinfo {author}
  {\bibfnamefont {X.}~\bibnamefont {Zhu}}, \bibinfo {author} {\bibfnamefont
  {J.}~\bibnamefont {Hone}}, \bibinfo {author} {\bibfnamefont {A.}~\bibnamefont
  {Rubio}}, \bibinfo {author} {\bibfnamefont {A.~N.}\ \bibnamefont
  {Pasupathy}}, \ and\ \bibinfo {author} {\bibfnamefont {C.~R.}\ \bibnamefont
  {Dean}},\ }\bibfield  {title} {\enquote {\bibinfo {title} {Correlated
  electronic phases in twisted bilayer transition metal dichalcogenides},}\
  }\href {\doibase 10.1038/s41563-020-0708-6} {\bibfield  {journal} {\bibinfo
  {journal} {Nat. Mater.}\ }\textbf {\bibinfo {volume} {19}},\ \bibinfo {pages}
  {861--866} (\bibinfo {year} {2020})}\BibitemShut {NoStop}%
\bibitem [{\citenamefont {Xu}\ \emph {et~al.}(2020)\citenamefont {Xu},
  \citenamefont {Liu}, \citenamefont {Rhodes}, \citenamefont {Watanabe},
  \citenamefont {Taniguchi}, \citenamefont {Hone}, \citenamefont {Elser},
  \citenamefont {Mak},\ and\ \citenamefont
  {Shan}}]{xuCorrelatedInsulatingStates2020a}%
  \BibitemOpen
  \bibfield  {author} {\bibinfo {author} {\bibfnamefont {Y.}~\bibnamefont
  {Xu}}, \bibinfo {author} {\bibfnamefont {S.}~\bibnamefont {Liu}}, \bibinfo
  {author} {\bibfnamefont {D.~A.}\ \bibnamefont {Rhodes}}, \bibinfo {author}
  {\bibfnamefont {K.}~\bibnamefont {Watanabe}}, \bibinfo {author}
  {\bibfnamefont {T.}~\bibnamefont {Taniguchi}}, \bibinfo {author}
  {\bibfnamefont {J.}~\bibnamefont {Hone}}, \bibinfo {author} {\bibfnamefont
  {V.}~\bibnamefont {Elser}}, \bibinfo {author} {\bibfnamefont {K.~F.}\
  \bibnamefont {Mak}}, \ and\ \bibinfo {author} {\bibfnamefont
  {J.}~\bibnamefont {Shan}},\ }\bibfield  {title} {\enquote {\bibinfo {title}
  {Correlated insulating states at fractional fillings of moir\'e
  superlattices},}\ }\href {\doibase 10.1038/s41586-020-2868-6} {\bibfield
  {journal} {\bibinfo  {journal} {Nature}\ }\textbf {\bibinfo {volume} {587}},\
  \bibinfo {pages} {214--218} (\bibinfo {year} {2020})}\BibitemShut {NoStop}%
\bibitem [{\citenamefont {Huang}\ \emph {et~al.}(2021)\citenamefont {Huang},
  \citenamefont {Wang}, \citenamefont {Miao}, \citenamefont {Wang},
  \citenamefont {Li}, \citenamefont {Lian}, \citenamefont {Taniguchi},
  \citenamefont {Watanabe}, \citenamefont {Okamoto}, \citenamefont {Xiao},
  \citenamefont {Shi},\ and\ \citenamefont
  {Cui}}]{huangCorrelatedInsulatingStates2021}%
  \BibitemOpen
  \bibfield  {author} {\bibinfo {author} {\bibfnamefont {X.}~\bibnamefont
  {Huang}}, \bibinfo {author} {\bibfnamefont {T.}~\bibnamefont {Wang}},
  \bibinfo {author} {\bibfnamefont {S.}~\bibnamefont {Miao}}, \bibinfo {author}
  {\bibfnamefont {C.}~\bibnamefont {Wang}}, \bibinfo {author} {\bibfnamefont
  {Z.}~\bibnamefont {Li}}, \bibinfo {author} {\bibfnamefont {Z.}~\bibnamefont
  {Lian}}, \bibinfo {author} {\bibfnamefont {T.}~\bibnamefont {Taniguchi}},
  \bibinfo {author} {\bibfnamefont {K.}~\bibnamefont {Watanabe}}, \bibinfo
  {author} {\bibfnamefont {S.}~\bibnamefont {Okamoto}}, \bibinfo {author}
  {\bibfnamefont {D.}~\bibnamefont {Xiao}}, \bibinfo {author} {\bibfnamefont
  {S.-F.}\ \bibnamefont {Shi}}, \ and\ \bibinfo {author} {\bibfnamefont
  {Y.-T.}\ \bibnamefont {Cui}},\ }\bibfield  {title} {\enquote {\bibinfo
  {title} {Correlated insulating states at fractional fillings of the
  {{WS2}}/{{WSe2}} moir\'e lattice},}\ }\href {\doibase
  10.1038/s41567-021-01171-w} {\bibfield  {journal} {\bibinfo  {journal} {Nat.
  Phys.}\ }\textbf {\bibinfo {volume} {17}},\ \bibinfo {pages} {715--719}
  (\bibinfo {year} {2021})}\BibitemShut {NoStop}%
\bibitem [{\citenamefont {Regan}\ \emph {et~al.}(2020)\citenamefont {Regan},
  \citenamefont {Wang}, \citenamefont {Jin}, \citenamefont {Bakti~Utama},
  \citenamefont {Gao}, \citenamefont {Wei}, \citenamefont {Zhao}, \citenamefont
  {Zhao}, \citenamefont {Zhang}, \citenamefont {Yumigeta}, \citenamefont
  {Blei}, \citenamefont {Carlstr{\"o}m}, \citenamefont {Watanabe},
  \citenamefont {Taniguchi}, \citenamefont {Tongay}, \citenamefont {Crommie},
  \citenamefont {Zettl},\ and\ \citenamefont
  {Wang}}]{reganMottGeneralizedWigner2020a}%
  \BibitemOpen
  \bibfield  {author} {\bibinfo {author} {\bibfnamefont {E.~C.}\ \bibnamefont
  {Regan}}, \bibinfo {author} {\bibfnamefont {D.}~\bibnamefont {Wang}},
  \bibinfo {author} {\bibfnamefont {C.}~\bibnamefont {Jin}}, \bibinfo {author}
  {\bibfnamefont {M.~I.}\ \bibnamefont {Bakti~Utama}}, \bibinfo {author}
  {\bibfnamefont {B.}~\bibnamefont {Gao}}, \bibinfo {author} {\bibfnamefont
  {X.}~\bibnamefont {Wei}}, \bibinfo {author} {\bibfnamefont {S.}~\bibnamefont
  {Zhao}}, \bibinfo {author} {\bibfnamefont {W.}~\bibnamefont {Zhao}}, \bibinfo
  {author} {\bibfnamefont {Z.}~\bibnamefont {Zhang}}, \bibinfo {author}
  {\bibfnamefont {K.}~\bibnamefont {Yumigeta}}, \bibinfo {author}
  {\bibfnamefont {M.}~\bibnamefont {Blei}}, \bibinfo {author} {\bibfnamefont
  {J.~D.}\ \bibnamefont {Carlstr{\"o}m}}, \bibinfo {author} {\bibfnamefont
  {K.}~\bibnamefont {Watanabe}}, \bibinfo {author} {\bibfnamefont
  {T.}~\bibnamefont {Taniguchi}}, \bibinfo {author} {\bibfnamefont
  {S.}~\bibnamefont {Tongay}}, \bibinfo {author} {\bibfnamefont
  {M.}~\bibnamefont {Crommie}}, \bibinfo {author} {\bibfnamefont
  {A.}~\bibnamefont {Zettl}}, \ and\ \bibinfo {author} {\bibfnamefont
  {F.}~\bibnamefont {Wang}},\ }\bibfield  {title} {\enquote {\bibinfo {title}
  {Mott and generalized {{Wigner}} crystal states in {{WSe2}}/{{WS2}} moir\'e
  superlattices},}\ }\href {\doibase 10.1038/s41586-020-2092-4} {\bibfield
  {journal} {\bibinfo  {journal} {Nature}\ }\textbf {\bibinfo {volume} {579}},\
  \bibinfo {pages} {359--363} (\bibinfo {year} {2020})}\BibitemShut {NoStop}%
\bibitem [{\citenamefont {Li}\ \emph {et~al.}(2021{\natexlab{a}})\citenamefont
  {Li}, \citenamefont {Jiang}, \citenamefont {Li}, \citenamefont {Zhang},
  \citenamefont {Kang}, \citenamefont {Zhu}, \citenamefont {Watanabe},
  \citenamefont {Taniguchi}, \citenamefont {Chowdhury}, \citenamefont {Fu},
  \citenamefont {Shan},\ and\ \citenamefont
  {Mak}}]{liContinuousMottTransition2021}%
  \BibitemOpen
  \bibfield  {author} {\bibinfo {author} {\bibfnamefont {T.}~\bibnamefont
  {Li}}, \bibinfo {author} {\bibfnamefont {S.}~\bibnamefont {Jiang}}, \bibinfo
  {author} {\bibfnamefont {L.}~\bibnamefont {Li}}, \bibinfo {author}
  {\bibfnamefont {Y.}~\bibnamefont {Zhang}}, \bibinfo {author} {\bibfnamefont
  {K.}~\bibnamefont {Kang}}, \bibinfo {author} {\bibfnamefont {J.}~\bibnamefont
  {Zhu}}, \bibinfo {author} {\bibfnamefont {K.}~\bibnamefont {Watanabe}},
  \bibinfo {author} {\bibfnamefont {T.}~\bibnamefont {Taniguchi}}, \bibinfo
  {author} {\bibfnamefont {D.}~\bibnamefont {Chowdhury}}, \bibinfo {author}
  {\bibfnamefont {L.}~\bibnamefont {Fu}}, \bibinfo {author} {\bibfnamefont
  {J.}~\bibnamefont {Shan}}, \ and\ \bibinfo {author} {\bibfnamefont {K.~F.}\
  \bibnamefont {Mak}},\ }\bibfield  {title} {\enquote {\bibinfo {title}
  {Continuous {{Mott}} transition in semiconductor moir\'e superlattices},}\
  }\href {\doibase 10.1038/s41586-021-03853-0} {\bibfield  {journal} {\bibinfo
  {journal} {Nature}\ }\textbf {\bibinfo {volume} {597}},\ \bibinfo {pages}
  {350--354} (\bibinfo {year} {2021}{\natexlab{a}})}\BibitemShut {NoStop}%
\bibitem [{\citenamefont {Ghiotto}\ \emph {et~al.}(2021)\citenamefont
  {Ghiotto}, \citenamefont {Shih}, \citenamefont {Pereira}, \citenamefont
  {Rhodes}, \citenamefont {Kim}, \citenamefont {Zang}, \citenamefont {Millis},
  \citenamefont {Watanabe}, \citenamefont {Taniguchi}, \citenamefont {Hone},
  \citenamefont {Wang}, \citenamefont {Dean},\ and\ \citenamefont
  {Pasupathy}}]{ghiottoQuantumCriticalityTwisted2021}%
  \BibitemOpen
  \bibfield  {author} {\bibinfo {author} {\bibfnamefont {A.}~\bibnamefont
  {Ghiotto}}, \bibinfo {author} {\bibfnamefont {E.-M.}\ \bibnamefont {Shih}},
  \bibinfo {author} {\bibfnamefont {G.~S. S.~G.}\ \bibnamefont {Pereira}},
  \bibinfo {author} {\bibfnamefont {D.~A.}\ \bibnamefont {Rhodes}}, \bibinfo
  {author} {\bibfnamefont {B.}~\bibnamefont {Kim}}, \bibinfo {author}
  {\bibfnamefont {J.}~\bibnamefont {Zang}}, \bibinfo {author} {\bibfnamefont
  {A.~J.}\ \bibnamefont {Millis}}, \bibinfo {author} {\bibfnamefont
  {K.}~\bibnamefont {Watanabe}}, \bibinfo {author} {\bibfnamefont
  {T.}~\bibnamefont {Taniguchi}}, \bibinfo {author} {\bibfnamefont {J.~C.}\
  \bibnamefont {Hone}}, \bibinfo {author} {\bibfnamefont {L.}~\bibnamefont
  {Wang}}, \bibinfo {author} {\bibfnamefont {C.~R.}\ \bibnamefont {Dean}}, \
  and\ \bibinfo {author} {\bibfnamefont {A.~N.}\ \bibnamefont {Pasupathy}},\
  }\bibfield  {title} {\enquote {\bibinfo {title} {Quantum criticality in
  twisted transition metal dichalcogenides},}\ }\href {\doibase
  10.1038/s41586-021-03815-6} {\bibfield  {journal} {\bibinfo  {journal}
  {Nature}\ }\textbf {\bibinfo {volume} {597}},\ \bibinfo {pages} {345--349}
  (\bibinfo {year} {2021})}\BibitemShut {NoStop}%
\bibitem [{\citenamefont {Li}\ \emph {et~al.}(2021{\natexlab{b}})\citenamefont
  {Li}, \citenamefont {Jiang}, \citenamefont {Shen}, \citenamefont {Zhang},
  \citenamefont {Li}, \citenamefont {Tao}, \citenamefont {Devakul},
  \citenamefont {Watanabe}, \citenamefont {Taniguchi}, \citenamefont {Fu},
  \citenamefont {Shan},\ and\ \citenamefont
  {Mak}}]{liQuantumAnomalousHall2021}%
  \BibitemOpen
  \bibfield  {author} {\bibinfo {author} {\bibfnamefont {T.}~\bibnamefont
  {Li}}, \bibinfo {author} {\bibfnamefont {S.}~\bibnamefont {Jiang}}, \bibinfo
  {author} {\bibfnamefont {B.}~\bibnamefont {Shen}}, \bibinfo {author}
  {\bibfnamefont {Y.}~\bibnamefont {Zhang}}, \bibinfo {author} {\bibfnamefont
  {L.}~\bibnamefont {Li}}, \bibinfo {author} {\bibfnamefont {Z.}~\bibnamefont
  {Tao}}, \bibinfo {author} {\bibfnamefont {T.}~\bibnamefont {Devakul}},
  \bibinfo {author} {\bibfnamefont {K.}~\bibnamefont {Watanabe}}, \bibinfo
  {author} {\bibfnamefont {T.}~\bibnamefont {Taniguchi}}, \bibinfo {author}
  {\bibfnamefont {L.}~\bibnamefont {Fu}}, \bibinfo {author} {\bibfnamefont
  {J.}~\bibnamefont {Shan}}, \ and\ \bibinfo {author} {\bibfnamefont {K.~F.}\
  \bibnamefont {Mak}},\ }\bibfield  {title} {\enquote {\bibinfo {title}
  {Quantum anomalous {{Hall}} effect from intertwined moir\'e bands},}\ }\href
  {\doibase 10.1038/s41586-021-04171-1} {\bibfield  {journal} {\bibinfo
  {journal} {Nature}\ }\textbf {\bibinfo {volume} {600}},\ \bibinfo {pages}
  {641--646} (\bibinfo {year} {2021}{\natexlab{b}})}\BibitemShut {NoStop}%
\bibitem [{\citenamefont {Zhao}\ \emph {et~al.}(2022)\citenamefont {Zhao},
  \citenamefont {Kang}, \citenamefont {Li}, \citenamefont {Tschirhart},
  \citenamefont {Redekop}, \citenamefont {Watanabe}, \citenamefont {Taniguchi},
  \citenamefont {Young}, \citenamefont {Shan},\ and\ \citenamefont
  {Mak}}]{zhaoRealizationHaldaneChern2022}%
  \BibitemOpen
  \bibfield  {author} {\bibinfo {author} {\bibfnamefont {W.}~\bibnamefont
  {Zhao}}, \bibinfo {author} {\bibfnamefont {K.}~\bibnamefont {Kang}}, \bibinfo
  {author} {\bibfnamefont {L.}~\bibnamefont {Li}}, \bibinfo {author}
  {\bibfnamefont {C.}~\bibnamefont {Tschirhart}}, \bibinfo {author}
  {\bibfnamefont {E.}~\bibnamefont {Redekop}}, \bibinfo {author} {\bibfnamefont
  {K.}~\bibnamefont {Watanabe}}, \bibinfo {author} {\bibfnamefont
  {T.}~\bibnamefont {Taniguchi}}, \bibinfo {author} {\bibfnamefont
  {A.}~\bibnamefont {Young}}, \bibinfo {author} {\bibfnamefont
  {J.}~\bibnamefont {Shan}}, \ and\ \bibinfo {author} {\bibfnamefont {K.~F.}\
  \bibnamefont {Mak}},\ }\href {\doibase 10.48550/arXiv.2207.02312} {\enquote
  {\bibinfo {title} {Realization of the {{Haldane Chern}} insulator in a
  moir\textbackslash 'e lattice},}\ } (\bibinfo {year} {2022}),\ \Eprint
  {http://arxiv.org/abs/2207.02312} {arXiv:2207.02312 [cond-mat]} \BibitemShut
  {NoStop}%
\bibitem [{\citenamefont {Tao}\ \emph {et~al.}(2022)\citenamefont {Tao},
  \citenamefont {Shen}, \citenamefont {Jiang}, \citenamefont {Li},
  \citenamefont {Li}, \citenamefont {Ma}, \citenamefont {Zhao}, \citenamefont
  {Hu}, \citenamefont {Pistunova}, \citenamefont {Watanabe}, \citenamefont
  {Taniguchi}, \citenamefont {Heinz}, \citenamefont {Mak},\ and\ \citenamefont
  {Shan}}]{taoValleycoherentQuantumAnomalous2022}%
  \BibitemOpen
  \bibfield  {author} {\bibinfo {author} {\bibfnamefont {Z.}~\bibnamefont
  {Tao}}, \bibinfo {author} {\bibfnamefont {B.}~\bibnamefont {Shen}}, \bibinfo
  {author} {\bibfnamefont {S.}~\bibnamefont {Jiang}}, \bibinfo {author}
  {\bibfnamefont {T.}~\bibnamefont {Li}}, \bibinfo {author} {\bibfnamefont
  {L.}~\bibnamefont {Li}}, \bibinfo {author} {\bibfnamefont {L.}~\bibnamefont
  {Ma}}, \bibinfo {author} {\bibfnamefont {W.}~\bibnamefont {Zhao}}, \bibinfo
  {author} {\bibfnamefont {J.}~\bibnamefont {Hu}}, \bibinfo {author}
  {\bibfnamefont {K.}~\bibnamefont {Pistunova}}, \bibinfo {author}
  {\bibfnamefont {K.}~\bibnamefont {Watanabe}}, \bibinfo {author}
  {\bibfnamefont {T.}~\bibnamefont {Taniguchi}}, \bibinfo {author}
  {\bibfnamefont {T.~F.}\ \bibnamefont {Heinz}}, \bibinfo {author}
  {\bibfnamefont {K.~F.}\ \bibnamefont {Mak}}, \ and\ \bibinfo {author}
  {\bibfnamefont {J.}~\bibnamefont {Shan}},\ }\href {\doibase
  10.48550/arXiv.2208.07452} {\enquote {\bibinfo {title} {Valley-coherent
  quantum anomalous {{Hall}} state in {{AB-stacked MoTe2}}/{{WSe2}}
  bilayers},}\ } (\bibinfo {year} {2022}),\ \Eprint
  {http://arxiv.org/abs/2208.07452} {arXiv:2208.07452 [cond-mat]} \BibitemShut
  {NoStop}%
\bibitem [{\citenamefont {Polshyn}\ \emph {et~al.}(2020)\citenamefont
  {Polshyn}, \citenamefont {Zhu}, \citenamefont {Kumar}, \citenamefont {Zhang},
  \citenamefont {Yang}, \citenamefont {Tschirhart}, \citenamefont {Serlin},
  \citenamefont {Watanabe}, \citenamefont {Taniguchi}, \citenamefont
  {MacDonald},\ and\ \citenamefont
  {Young}}]{polshynElectricalSwitchingMagnetic2020}%
  \BibitemOpen
  \bibfield  {author} {\bibinfo {author} {\bibfnamefont {H.}~\bibnamefont
  {Polshyn}}, \bibinfo {author} {\bibfnamefont {J.}~\bibnamefont {Zhu}},
  \bibinfo {author} {\bibfnamefont {M.~A.}\ \bibnamefont {Kumar}}, \bibinfo
  {author} {\bibfnamefont {Y.}~\bibnamefont {Zhang}}, \bibinfo {author}
  {\bibfnamefont {F.}~\bibnamefont {Yang}}, \bibinfo {author} {\bibfnamefont
  {C.~L.}\ \bibnamefont {Tschirhart}}, \bibinfo {author} {\bibfnamefont
  {M.}~\bibnamefont {Serlin}}, \bibinfo {author} {\bibfnamefont
  {K.}~\bibnamefont {Watanabe}}, \bibinfo {author} {\bibfnamefont
  {T.}~\bibnamefont {Taniguchi}}, \bibinfo {author} {\bibfnamefont {A.~H.}\
  \bibnamefont {MacDonald}}, \ and\ \bibinfo {author} {\bibfnamefont {A.~F.}\
  \bibnamefont {Young}},\ }\bibfield  {title} {\enquote {\bibinfo {title}
  {Electrical switching of magnetic order in an orbital {{Chern}} insulator},}\
  }\href {\doibase 10.1038/s41586-020-2963-8} {\bibfield  {journal} {\bibinfo
  {journal} {Nature}\ }\textbf {\bibinfo {volume} {588}},\ \bibinfo {pages}
  {66--70} (\bibinfo {year} {2020})}\BibitemShut {NoStop}%
\bibitem [{\citenamefont {Chen}\ \emph {et~al.}(2020)\citenamefont {Chen},
  \citenamefont {Sharpe}, \citenamefont {Fox}, \citenamefont {Zhang},
  \citenamefont {Wang}, \citenamefont {Jiang}, \citenamefont {Lyu},
  \citenamefont {Li}, \citenamefont {Watanabe}, \citenamefont {Taniguchi},
  \citenamefont {Shi}, \citenamefont {Senthil}, \citenamefont
  {{Goldhaber-Gordon}}, \citenamefont {Zhang},\ and\ \citenamefont
  {Wang}}]{chenTunableCorrelatedChern2020}%
  \BibitemOpen
  \bibfield  {author} {\bibinfo {author} {\bibfnamefont {G.}~\bibnamefont
  {Chen}}, \bibinfo {author} {\bibfnamefont {A.~L.}\ \bibnamefont {Sharpe}},
  \bibinfo {author} {\bibfnamefont {E.~J.}\ \bibnamefont {Fox}}, \bibinfo
  {author} {\bibfnamefont {Y.-H.}\ \bibnamefont {Zhang}}, \bibinfo {author}
  {\bibfnamefont {S.}~\bibnamefont {Wang}}, \bibinfo {author} {\bibfnamefont
  {L.}~\bibnamefont {Jiang}}, \bibinfo {author} {\bibfnamefont
  {B.}~\bibnamefont {Lyu}}, \bibinfo {author} {\bibfnamefont {H.}~\bibnamefont
  {Li}}, \bibinfo {author} {\bibfnamefont {K.}~\bibnamefont {Watanabe}},
  \bibinfo {author} {\bibfnamefont {T.}~\bibnamefont {Taniguchi}}, \bibinfo
  {author} {\bibfnamefont {Z.}~\bibnamefont {Shi}}, \bibinfo {author}
  {\bibfnamefont {T.}~\bibnamefont {Senthil}}, \bibinfo {author} {\bibfnamefont
  {D.}~\bibnamefont {{Goldhaber-Gordon}}}, \bibinfo {author} {\bibfnamefont
  {Y.}~\bibnamefont {Zhang}}, \ and\ \bibinfo {author} {\bibfnamefont
  {F.}~\bibnamefont {Wang}},\ }\bibfield  {title} {\enquote {\bibinfo {title}
  {Tunable correlated {{Chern}} insulator and ferromagnetism in a moir\'e
  superlattice},}\ }\href {\doibase 10.1038/s41586-020-2049-7} {\bibfield
  {journal} {\bibinfo  {journal} {Nature}\ }\textbf {\bibinfo {volume} {579}},\
  \bibinfo {pages} {56--61} (\bibinfo {year} {2020})}\BibitemShut {NoStop}%
\bibitem [{\citenamefont {Chen}\ \emph {et~al.}(2022)\citenamefont {Chen},
  \citenamefont {Sharpe}, \citenamefont {Fox}, \citenamefont {Wang},
  \citenamefont {Lyu}, \citenamefont {Jiang}, \citenamefont {Li}, \citenamefont
  {Watanabe}, \citenamefont {Taniguchi}, \citenamefont {Crommie}, \citenamefont
  {Kastner}, \citenamefont {Shi}, \citenamefont {{Goldhaber-Gordon}},
  \citenamefont {Zhang},\ and\ \citenamefont
  {Wang}}]{chenTunableOrbitalFerromagnetism2022}%
  \BibitemOpen
  \bibfield  {author} {\bibinfo {author} {\bibfnamefont {G.}~\bibnamefont
  {Chen}}, \bibinfo {author} {\bibfnamefont {A.~L.}\ \bibnamefont {Sharpe}},
  \bibinfo {author} {\bibfnamefont {E.~J.}\ \bibnamefont {Fox}}, \bibinfo
  {author} {\bibfnamefont {S.}~\bibnamefont {Wang}}, \bibinfo {author}
  {\bibfnamefont {B.}~\bibnamefont {Lyu}}, \bibinfo {author} {\bibfnamefont
  {L.}~\bibnamefont {Jiang}}, \bibinfo {author} {\bibfnamefont
  {H.}~\bibnamefont {Li}}, \bibinfo {author} {\bibfnamefont {K.}~\bibnamefont
  {Watanabe}}, \bibinfo {author} {\bibfnamefont {T.}~\bibnamefont {Taniguchi}},
  \bibinfo {author} {\bibfnamefont {M.~F.}\ \bibnamefont {Crommie}}, \bibinfo
  {author} {\bibfnamefont {M.~A.}\ \bibnamefont {Kastner}}, \bibinfo {author}
  {\bibfnamefont {Z.}~\bibnamefont {Shi}}, \bibinfo {author} {\bibfnamefont
  {D.}~\bibnamefont {{Goldhaber-Gordon}}}, \bibinfo {author} {\bibfnamefont
  {Y.}~\bibnamefont {Zhang}}, \ and\ \bibinfo {author} {\bibfnamefont
  {F.}~\bibnamefont {Wang}},\ }\bibfield  {title} {\enquote {\bibinfo {title}
  {Tunable {{Orbital Ferromagnetism}} at {{Noninteger Filling}} of a {{Moir\'e
  Superlattice}}},}\ }\href {\doibase 10.1021/acs.nanolett.1c03699} {\bibfield
  {journal} {\bibinfo  {journal} {Nano Lett.}\ }\textbf {\bibinfo {volume}
  {22}},\ \bibinfo {pages} {238--245} (\bibinfo {year} {2022})}\BibitemShut
  {NoStop}%
\bibitem [{\citenamefont {Wang}\ \emph {et~al.}(2022)\citenamefont {Wang},
  \citenamefont {Xiao}, \citenamefont {Park}, \citenamefont {Zhu},
  \citenamefont {Wang}, \citenamefont {Taniguchi}, \citenamefont {Watanabe},
  \citenamefont {Yan}, \citenamefont {Xiao}, \citenamefont {Gamelin},
  \citenamefont {Yao},\ and\ \citenamefont
  {Xu}}]{wangLightinducedFerromagnetismMoire2022}%
  \BibitemOpen
  \bibfield  {author} {\bibinfo {author} {\bibfnamefont {X.}~\bibnamefont
  {Wang}}, \bibinfo {author} {\bibfnamefont {C.}~\bibnamefont {Xiao}}, \bibinfo
  {author} {\bibfnamefont {H.}~\bibnamefont {Park}}, \bibinfo {author}
  {\bibfnamefont {J.}~\bibnamefont {Zhu}}, \bibinfo {author} {\bibfnamefont
  {C.}~\bibnamefont {Wang}}, \bibinfo {author} {\bibfnamefont {T.}~\bibnamefont
  {Taniguchi}}, \bibinfo {author} {\bibfnamefont {K.}~\bibnamefont {Watanabe}},
  \bibinfo {author} {\bibfnamefont {J.}~\bibnamefont {Yan}}, \bibinfo {author}
  {\bibfnamefont {D.}~\bibnamefont {Xiao}}, \bibinfo {author} {\bibfnamefont
  {D.~R.}\ \bibnamefont {Gamelin}}, \bibinfo {author} {\bibfnamefont
  {W.}~\bibnamefont {Yao}}, \ and\ \bibinfo {author} {\bibfnamefont
  {X.}~\bibnamefont {Xu}},\ }\bibfield  {title} {\enquote {\bibinfo {title}
  {Light-induced ferromagnetism in moir\'e superlattices},}\ }\href {\doibase
  10.1038/s41586-022-04472-z} {\bibfield  {journal} {\bibinfo  {journal}
  {Nature}\ }\textbf {\bibinfo {volume} {604}},\ \bibinfo {pages} {468--473}
  (\bibinfo {year} {2022})}\BibitemShut {NoStop}%
\bibitem [{\citenamefont {Lin}\ \emph {et~al.}(2022)\citenamefont {Lin},
  \citenamefont {Zhang}, \citenamefont {Morissette}, \citenamefont {Wang},
  \citenamefont {Liu}, \citenamefont {Rhodes}, \citenamefont {Watanabe},
  \citenamefont {Taniguchi}, \citenamefont {Hone},\ and\ \citenamefont
  {Li}}]{linSpinorbitDrivenFerromagnetism2022a}%
  \BibitemOpen
  \bibfield  {author} {\bibinfo {author} {\bibfnamefont {J.-X.}\ \bibnamefont
  {Lin}}, \bibinfo {author} {\bibfnamefont {Y.-H.}\ \bibnamefont {Zhang}},
  \bibinfo {author} {\bibfnamefont {E.}~\bibnamefont {Morissette}}, \bibinfo
  {author} {\bibfnamefont {Z.}~\bibnamefont {Wang}}, \bibinfo {author}
  {\bibfnamefont {S.}~\bibnamefont {Liu}}, \bibinfo {author} {\bibfnamefont
  {D.}~\bibnamefont {Rhodes}}, \bibinfo {author} {\bibfnamefont
  {K.}~\bibnamefont {Watanabe}}, \bibinfo {author} {\bibfnamefont
  {T.}~\bibnamefont {Taniguchi}}, \bibinfo {author} {\bibfnamefont
  {J.}~\bibnamefont {Hone}}, \ and\ \bibinfo {author} {\bibfnamefont
  {J.~I.~A.}\ \bibnamefont {Li}},\ }\bibfield  {title} {\enquote {\bibinfo
  {title} {Spin-orbit\textendash driven ferromagnetism at half moir\'e filling
  in magic-angle twisted bilayer graphene},}\ }\href {\doibase
  10.1126/science.abh2889} {\bibfield  {journal} {\bibinfo  {journal}
  {Science}\ }\textbf {\bibinfo {volume} {375}},\ \bibinfo {pages} {437--441}
  (\bibinfo {year} {2022})}\BibitemShut {NoStop}%
\bibitem [{\citenamefont {Sharpe}\ \emph {et~al.}(2019)\citenamefont {Sharpe},
  \citenamefont {Fox}, \citenamefont {Barnard}, \citenamefont {Finney},
  \citenamefont {Watanabe}, \citenamefont {Taniguchi}, \citenamefont
  {Kastner},\ and\ \citenamefont
  {{Goldhaber-Gordon}}}]{sharpeEmergentFerromagnetismThreequarters2019}%
  \BibitemOpen
  \bibfield  {author} {\bibinfo {author} {\bibfnamefont {A.~L.}\ \bibnamefont
  {Sharpe}}, \bibinfo {author} {\bibfnamefont {E.~J.}\ \bibnamefont {Fox}},
  \bibinfo {author} {\bibfnamefont {A.~W.}\ \bibnamefont {Barnard}}, \bibinfo
  {author} {\bibfnamefont {J.}~\bibnamefont {Finney}}, \bibinfo {author}
  {\bibfnamefont {K.}~\bibnamefont {Watanabe}}, \bibinfo {author}
  {\bibfnamefont {T.}~\bibnamefont {Taniguchi}}, \bibinfo {author}
  {\bibfnamefont {M.~A.}\ \bibnamefont {Kastner}}, \ and\ \bibinfo {author}
  {\bibfnamefont {D.}~\bibnamefont {{Goldhaber-Gordon}}},\ }\bibfield  {title}
  {\enquote {\bibinfo {title} {Emergent ferromagnetism near three-quarters
  filling in twisted bilayer graphene},}\ }\href {\doibase
  10.1126/science.aaw3780} {\bibfield  {journal} {\bibinfo  {journal}
  {Science}\ }\textbf {\bibinfo {volume} {365}},\ \bibinfo {pages} {605--608}
  (\bibinfo {year} {2019})}\BibitemShut {NoStop}%
\bibitem [{\citenamefont {Liu}\ and\ \citenamefont
  {Dai}(2021)}]{liuOrbitalMagneticStates2021}%
  \BibitemOpen
  \bibfield  {author} {\bibinfo {author} {\bibfnamefont {J.}~\bibnamefont
  {Liu}}\ and\ \bibinfo {author} {\bibfnamefont {X.}~\bibnamefont {Dai}},\
  }\bibfield  {title} {\enquote {\bibinfo {title} {Orbital magnetic states in
  moir\'e graphene systems},}\ }\href {\doibase 10.1038/s42254-021-00297-3}
  {\bibfield  {journal} {\bibinfo  {journal} {Nat Rev Phys}\ }\textbf {\bibinfo
  {volume} {3}},\ \bibinfo {pages} {367--382} (\bibinfo {year}
  {2021})}\BibitemShut {NoStop}%
\bibitem [{\citenamefont {Serlin}\ \emph {et~al.}(2020)\citenamefont {Serlin},
  \citenamefont {Tschirhart}, \citenamefont {Polshyn}, \citenamefont {Zhang},
  \citenamefont {Zhu}, \citenamefont {Watanabe}, \citenamefont {Taniguchi},
  \citenamefont {Balents},\ and\ \citenamefont
  {Young}}]{serlinIntrinsicQuantizedAnomalous2020}%
  \BibitemOpen
  \bibfield  {author} {\bibinfo {author} {\bibfnamefont {M.}~\bibnamefont
  {Serlin}}, \bibinfo {author} {\bibfnamefont {C.~L.}\ \bibnamefont
  {Tschirhart}}, \bibinfo {author} {\bibfnamefont {H.}~\bibnamefont {Polshyn}},
  \bibinfo {author} {\bibfnamefont {Y.}~\bibnamefont {Zhang}}, \bibinfo
  {author} {\bibfnamefont {J.}~\bibnamefont {Zhu}}, \bibinfo {author}
  {\bibfnamefont {K.}~\bibnamefont {Watanabe}}, \bibinfo {author}
  {\bibfnamefont {T.}~\bibnamefont {Taniguchi}}, \bibinfo {author}
  {\bibfnamefont {L.}~\bibnamefont {Balents}}, \ and\ \bibinfo {author}
  {\bibfnamefont {A.~F.}\ \bibnamefont {Young}},\ }\bibfield  {title} {\enquote
  {\bibinfo {title} {Intrinsic quantized anomalous {{Hall}} effect in a moir\'e
  heterostructure},}\ }\href {\doibase 10.1126/science.aay5533} {\bibfield
  {journal} {\bibinfo  {journal} {Science}\ }\textbf {\bibinfo {volume}
  {367}},\ \bibinfo {pages} {900--903} (\bibinfo {year} {2020})}\BibitemShut
  {NoStop}%
\bibitem [{\citenamefont {Polshyn}\ \emph {et~al.}(2022)\citenamefont
  {Polshyn}, \citenamefont {Zhang}, \citenamefont {Kumar}, \citenamefont
  {Soejima}, \citenamefont {Ledwith}, \citenamefont {Watanabe}, \citenamefont
  {Taniguchi}, \citenamefont {Vishwanath}, \citenamefont {Zaletel},\ and\
  \citenamefont {Young}}]{polshynTopologicalChargeDensity2022}%
  \BibitemOpen
  \bibfield  {author} {\bibinfo {author} {\bibfnamefont {H.}~\bibnamefont
  {Polshyn}}, \bibinfo {author} {\bibfnamefont {Y.}~\bibnamefont {Zhang}},
  \bibinfo {author} {\bibfnamefont {M.~A.}\ \bibnamefont {Kumar}}, \bibinfo
  {author} {\bibfnamefont {T.}~\bibnamefont {Soejima}}, \bibinfo {author}
  {\bibfnamefont {P.}~\bibnamefont {Ledwith}}, \bibinfo {author} {\bibfnamefont
  {K.}~\bibnamefont {Watanabe}}, \bibinfo {author} {\bibfnamefont
  {T.}~\bibnamefont {Taniguchi}}, \bibinfo {author} {\bibfnamefont
  {A.}~\bibnamefont {Vishwanath}}, \bibinfo {author} {\bibfnamefont {M.~P.}\
  \bibnamefont {Zaletel}}, \ and\ \bibinfo {author} {\bibfnamefont {A.~F.}\
  \bibnamefont {Young}},\ }\bibfield  {title} {\enquote {\bibinfo {title}
  {Topological charge density waves at half-integer filling of a moir\'e
  superlattice},}\ }\href {\doibase 10.1038/s41567-021-01418-6} {\bibfield
  {journal} {\bibinfo  {journal} {Nat. Phys.}\ }\textbf {\bibinfo {volume}
  {18}},\ \bibinfo {pages} {42--47} (\bibinfo {year} {2022})}\BibitemShut
  {NoStop}%
\bibitem [{\citenamefont {Eerenstein}\ \emph {et~al.}(2006)\citenamefont
  {Eerenstein}, \citenamefont {Mathur},\ and\ \citenamefont
  {Scott}}]{eerensteinMultiferroicMagnetoelectricMaterials2006}%
  \BibitemOpen
  \bibfield  {author} {\bibinfo {author} {\bibfnamefont {W.}~\bibnamefont
  {Eerenstein}}, \bibinfo {author} {\bibfnamefont {N.~D.}\ \bibnamefont
  {Mathur}}, \ and\ \bibinfo {author} {\bibfnamefont {J.~F.}\ \bibnamefont
  {Scott}},\ }\bibfield  {title} {\enquote {\bibinfo {title} {Multiferroic and
  magnetoelectric materials},}\ }\href {\doibase 10.1038/nature05023}
  {\bibfield  {journal} {\bibinfo  {journal} {Nature}\ }\textbf {\bibinfo
  {volume} {442}},\ \bibinfo {pages} {759--765} (\bibinfo {year}
  {2006})}\BibitemShut {NoStop}%
\bibitem [{\citenamefont {Parsonnet}\ \emph {et~al.}(2022)\citenamefont
  {Parsonnet}, \citenamefont {Caretta}, \citenamefont {Nagarajan},
  \citenamefont {Zhang}, \citenamefont {Taghinejad}, \citenamefont {Behera},
  \citenamefont {Huang}, \citenamefont {Kavle}, \citenamefont {Fernandez},
  \citenamefont {Nikonov}, \citenamefont {Li}, \citenamefont {Young},
  \citenamefont {Analytis},\ and\ \citenamefont
  {Ramesh}}]{parsonnetNonvolatileElectricField2022}%
  \BibitemOpen
  \bibfield  {author} {\bibinfo {author} {\bibfnamefont {E.}~\bibnamefont
  {Parsonnet}}, \bibinfo {author} {\bibfnamefont {L.}~\bibnamefont {Caretta}},
  \bibinfo {author} {\bibfnamefont {V.}~\bibnamefont {Nagarajan}}, \bibinfo
  {author} {\bibfnamefont {H.}~\bibnamefont {Zhang}}, \bibinfo {author}
  {\bibfnamefont {H.}~\bibnamefont {Taghinejad}}, \bibinfo {author}
  {\bibfnamefont {P.}~\bibnamefont {Behera}}, \bibinfo {author} {\bibfnamefont
  {X.}~\bibnamefont {Huang}}, \bibinfo {author} {\bibfnamefont
  {P.}~\bibnamefont {Kavle}}, \bibinfo {author} {\bibfnamefont
  {A.}~\bibnamefont {Fernandez}}, \bibinfo {author} {\bibfnamefont
  {D.}~\bibnamefont {Nikonov}}, \bibinfo {author} {\bibfnamefont
  {H.}~\bibnamefont {Li}}, \bibinfo {author} {\bibfnamefont {I.}~\bibnamefont
  {Young}}, \bibinfo {author} {\bibfnamefont {J.}~\bibnamefont {Analytis}}, \
  and\ \bibinfo {author} {\bibfnamefont {R.}~\bibnamefont {Ramesh}},\
  }\bibfield  {title} {\enquote {\bibinfo {title} {Nonvolatile {{Electric Field
  Control}} of {{Thermal Magnons}} in the {{Absence}} of an {{Applied Magnetic
  Field}}},}\ }\href {\doibase 10.1103/PhysRevLett.129.087601} {\bibfield
  {journal} {\bibinfo  {journal} {Phys. Rev. Lett.}\ }\textbf {\bibinfo
  {volume} {129}},\ \bibinfo {pages} {087601} (\bibinfo {year}
  {2022})}\BibitemShut {NoStop}%
\bibitem [{\citenamefont {Manipatruni}\ \emph {et~al.}(2019)\citenamefont
  {Manipatruni}, \citenamefont {Nikonov}, \citenamefont {Lin}, \citenamefont
  {Gosavi}, \citenamefont {Liu}, \citenamefont {Prasad}, \citenamefont {Huang},
  \citenamefont {Bonturim}, \citenamefont {Ramesh},\ and\ \citenamefont
  {Young}}]{manipatruniScalableEnergyefficientMagnetoelectric2019}%
  \BibitemOpen
  \bibfield  {author} {\bibinfo {author} {\bibfnamefont {S.}~\bibnamefont
  {Manipatruni}}, \bibinfo {author} {\bibfnamefont {D.~E.}\ \bibnamefont
  {Nikonov}}, \bibinfo {author} {\bibfnamefont {C.-C.}\ \bibnamefont {Lin}},
  \bibinfo {author} {\bibfnamefont {T.~A.}\ \bibnamefont {Gosavi}}, \bibinfo
  {author} {\bibfnamefont {H.}~\bibnamefont {Liu}}, \bibinfo {author}
  {\bibfnamefont {B.}~\bibnamefont {Prasad}}, \bibinfo {author} {\bibfnamefont
  {Y.-L.}\ \bibnamefont {Huang}}, \bibinfo {author} {\bibfnamefont
  {E.}~\bibnamefont {Bonturim}}, \bibinfo {author} {\bibfnamefont
  {R.}~\bibnamefont {Ramesh}}, \ and\ \bibinfo {author} {\bibfnamefont {I.~A.}\
  \bibnamefont {Young}},\ }\bibfield  {title} {\enquote {\bibinfo {title}
  {Scalable energy-efficient magnetoelectric spin\textendash orbit logic},}\
  }\href {\doibase 10.1038/s41586-018-0770-2} {\bibfield  {journal} {\bibinfo
  {journal} {Nature}\ }\textbf {\bibinfo {volume} {565}},\ \bibinfo {pages}
  {35--42} (\bibinfo {year} {2019})}\BibitemShut {NoStop}%
\bibitem [{\citenamefont {Chen}\ and\ \citenamefont
  {Sigrist}(2015)}]{chenDissipationlessMultiferroicMagnonics2015}%
  \BibitemOpen
  \bibfield  {author} {\bibinfo {author} {\bibfnamefont {W.}~\bibnamefont
  {Chen}}\ and\ \bibinfo {author} {\bibfnamefont {M.}~\bibnamefont {Sigrist}},\
  }\bibfield  {title} {\enquote {\bibinfo {title} {Dissipationless
  {{Multiferroic Magnonics}}},}\ }\href {\doibase
  10.1103/PhysRevLett.114.157203} {\bibfield  {journal} {\bibinfo  {journal}
  {Phys. Rev. Lett.}\ }\textbf {\bibinfo {volume} {114}},\ \bibinfo {pages}
  {157203} (\bibinfo {year} {2015})}\BibitemShut {NoStop}%
\bibitem [{\citenamefont {Wu}\ \emph {et~al.}(2019)\citenamefont {Wu},
  \citenamefont {Lovorn}, \citenamefont {Tutuc}, \citenamefont {Martin},\ and\
  \citenamefont {MacDonald}}]{wuTopologicalInsulatorsTwisted2019a}%
  \BibitemOpen
  \bibfield  {author} {\bibinfo {author} {\bibfnamefont {F.}~\bibnamefont
  {Wu}}, \bibinfo {author} {\bibfnamefont {T.}~\bibnamefont {Lovorn}}, \bibinfo
  {author} {\bibfnamefont {E.}~\bibnamefont {Tutuc}}, \bibinfo {author}
  {\bibfnamefont {I.}~\bibnamefont {Martin}}, \ and\ \bibinfo {author}
  {\bibfnamefont {A.~H.}\ \bibnamefont {MacDonald}},\ }\bibfield  {title}
  {\enquote {\bibinfo {title} {Topological {{Insulators}} in {{Twisted
  Transition Metal Dichalcogenide Homobilayers}}},}\ }\href {\doibase
  10.1103/PhysRevLett.122.086402} {\bibfield  {journal} {\bibinfo  {journal}
  {Phys. Rev. Lett.}\ }\textbf {\bibinfo {volume} {122}},\ \bibinfo {pages}
  {086402} (\bibinfo {year} {2019})}\BibitemShut {NoStop}%
\bibitem [{\citenamefont {Pan}\ \emph {et~al.}(2020)\citenamefont {Pan},
  \citenamefont {Wu},\ and\ \citenamefont
  {Das~Sarma}}]{panBandTopologyHubbard2020a}%
  \BibitemOpen
  \bibfield  {author} {\bibinfo {author} {\bibfnamefont {H.}~\bibnamefont
  {Pan}}, \bibinfo {author} {\bibfnamefont {F.}~\bibnamefont {Wu}}, \ and\
  \bibinfo {author} {\bibfnamefont {S.}~\bibnamefont {Das~Sarma}},\ }\bibfield
  {title} {\enquote {\bibinfo {title} {Band topology, {{Hubbard}} model,
  {{Heisenberg}} model, and {{Dzyaloshinskii-Moriya}} interaction in twisted
  bilayer \$\{\textbackslash
  mathrm\{\vphantom{\}\}}{{WSe}}\vphantom\{\}\vphantom\{\}\_\{2\}\$},}\ }\href
  {\doibase 10.1103/PhysRevResearch.2.033087} {\bibfield  {journal} {\bibinfo
  {journal} {Phys. Rev. Research}\ }\textbf {\bibinfo {volume} {2}},\ \bibinfo
  {pages} {033087} (\bibinfo {year} {2020})}\BibitemShut {NoStop}%
\bibitem [{\citenamefont {Zhu}\ \emph {et~al.}(2011)\citenamefont {Zhu},
  \citenamefont {Cheng},\ and\ \citenamefont
  {Schwingenschl{\"o}gl}}]{zhuGiantSpinorbitinducedSpin2011}%
  \BibitemOpen
  \bibfield  {author} {\bibinfo {author} {\bibfnamefont {Z.~Y.}\ \bibnamefont
  {Zhu}}, \bibinfo {author} {\bibfnamefont {Y.~C.}\ \bibnamefont {Cheng}}, \
  and\ \bibinfo {author} {\bibfnamefont {U.}~\bibnamefont
  {Schwingenschl{\"o}gl}},\ }\bibfield  {title} {\enquote {\bibinfo {title}
  {Giant spin-orbit-induced spin splitting in two-dimensional transition-metal
  dichalcogenide semiconductors},}\ }\href {\doibase
  10.1103/PhysRevB.84.153402} {\bibfield  {journal} {\bibinfo  {journal} {Phys.
  Rev. B}\ }\textbf {\bibinfo {volume} {84}},\ \bibinfo {pages} {153402}
  (\bibinfo {year} {2011})}\BibitemShut {NoStop}%
\bibitem [{sup()}]{supplement}%
  \BibitemOpen
  \href@noop {} {}\bibinfo {note} {See Supplemental Material which contains
  Refs
  \onlinecite{varneyInteractionEffectsQuantum2010,shaoInterplayLocalOrder2021}
  for model details and symmetries, additional results and comments on the
  exact diagonalization}\BibitemShut {NoStop}%
\bibitem [{\citenamefont {Devakul}\ \emph {et~al.}(2021)\citenamefont
  {Devakul}, \citenamefont {Cr{\'e}pel}, \citenamefont {Zhang},\ and\
  \citenamefont {Fu}}]{devakulMagicTwistedTransition2021}%
  \BibitemOpen
  \bibfield  {author} {\bibinfo {author} {\bibfnamefont {T.}~\bibnamefont
  {Devakul}}, \bibinfo {author} {\bibfnamefont {V.}~\bibnamefont {Cr{\'e}pel}},
  \bibinfo {author} {\bibfnamefont {Y.}~\bibnamefont {Zhang}}, \ and\ \bibinfo
  {author} {\bibfnamefont {L.}~\bibnamefont {Fu}},\ }\bibfield  {title}
  {\enquote {\bibinfo {title} {Magic in twisted transition metal dichalcogenide
  bilayers},}\ }\href {\doibase 10.1038/s41467-021-27042-9} {\bibfield
  {journal} {\bibinfo  {journal} {Nat Commun}\ }\textbf {\bibinfo {volume}
  {12}},\ \bibinfo {pages} {6730} (\bibinfo {year} {2021})},\ \Eprint
  {http://arxiv.org/abs/2106.11954} {arXiv:2106.11954} \BibitemShut {NoStop}%
\bibitem [{Vzn()}]{Vznote}%
  \BibitemOpen
  \href@noop {} {}\bibinfo {note} {The calculations for $P_L$ are performed
  with a tiny displacement field $V_z = 10 \> \mu\textrm{eV}$ to enforce the
  layer symmetry breaking since the Hamiltonian \eqref{eq:intHam} is $C_{2y}$
  symmetric.}\BibitemShut {Stop}%
\bibitem [{eps()}]{eps_note}%
  \BibitemOpen
  \href@noop {} {}\bibinfo {note} {Such $\epsilon$ is typical for dielectric
  hBN substrates used in experiments.}\BibitemShut {Stop}%
\bibitem [{\citenamefont {Kane}\ and\ \citenamefont
  {Mele}(2005)}]{kaneQuantumSpinHall2005}%
  \BibitemOpen
  \bibfield  {author} {\bibinfo {author} {\bibfnamefont {C.~L.}\ \bibnamefont
  {Kane}}\ and\ \bibinfo {author} {\bibfnamefont {E.~J.}\ \bibnamefont
  {Mele}},\ }\bibfield  {title} {\enquote {\bibinfo {title} {Quantum {{Spin
  Hall Effect}} in {{Graphene}}},}\ }\href {\doibase
  10.1103/PhysRevLett.95.226801} {\bibfield  {journal} {\bibinfo  {journal}
  {Phys. Rev. Lett.}\ }\textbf {\bibinfo {volume} {95}},\ \bibinfo {pages}
  {226801} (\bibinfo {year} {2005})}\BibitemShut {NoStop}%
\bibitem [{\citenamefont {Devakul}\ and\ \citenamefont
  {Fu}(2022)}]{devakulQuantumAnomalousHall2022a}%
  \BibitemOpen
  \bibfield  {author} {\bibinfo {author} {\bibfnamefont {T.}~\bibnamefont
  {Devakul}}\ and\ \bibinfo {author} {\bibfnamefont {L.}~\bibnamefont {Fu}},\
  }\bibfield  {title} {\enquote {\bibinfo {title} {Quantum {{Anomalous Hall
  Effect}} from {{Inverted Charge Transfer Gap}}},}\ }\href {\doibase
  10.1103/PhysRevX.12.021031} {\bibfield  {journal} {\bibinfo  {journal} {Phys.
  Rev. X}\ }\textbf {\bibinfo {volume} {12}},\ \bibinfo {pages} {021031}
  (\bibinfo {year} {2022})}\BibitemShut {NoStop}%
\bibitem [{\citenamefont {Zhang}\ \emph {et~al.}(2020)\citenamefont {Zhang},
  \citenamefont {Yuan},\ and\ \citenamefont
  {Fu}}]{zhangMoirQuantumChemistry2020}%
  \BibitemOpen
  \bibfield  {author} {\bibinfo {author} {\bibfnamefont {Y.}~\bibnamefont
  {Zhang}}, \bibinfo {author} {\bibfnamefont {N.~F.~Q.}\ \bibnamefont {Yuan}},
  \ and\ \bibinfo {author} {\bibfnamefont {L.}~\bibnamefont {Fu}},\ }\bibfield
  {title} {\enquote {\bibinfo {title} {Moir\textbackslash 'e quantum chemistry:
  {{Charge}} transfer in transition metal dichalcogenide superlattices},}\
  }\href {\doibase 10.1103/PhysRevB.102.201115} {\bibfield  {journal} {\bibinfo
   {journal} {Phys. Rev. B}\ }\textbf {\bibinfo {volume} {102}},\ \bibinfo
  {pages} {201115} (\bibinfo {year} {2020})}\BibitemShut {NoStop}%
\bibitem [{\citenamefont {Zhang}\ \emph {et~al.}(2021)\citenamefont {Zhang},
  \citenamefont {Liu},\ and\ \citenamefont
  {Fu}}]{zhangElectronicStructuresCharge2021a}%
  \BibitemOpen
  \bibfield  {author} {\bibinfo {author} {\bibfnamefont {Y.}~\bibnamefont
  {Zhang}}, \bibinfo {author} {\bibfnamefont {T.}~\bibnamefont {Liu}}, \ and\
  \bibinfo {author} {\bibfnamefont {L.}~\bibnamefont {Fu}},\ }\bibfield
  {title} {\enquote {\bibinfo {title} {Electronic structures, charge transfer,
  and charge order in twisted transition metal dichalcogenide bilayers},}\
  }\href {\doibase 10.1103/PhysRevB.103.155142} {\bibfield  {journal} {\bibinfo
   {journal} {Phys. Rev. B}\ }\textbf {\bibinfo {volume} {103}},\ \bibinfo
  {pages} {155142} (\bibinfo {year} {2021})}\BibitemShut {NoStop}%
\bibitem [{\citenamefont {Slagle}\ and\ \citenamefont
  {Fu}(2020)}]{slagleChargeTransferExcitations2020}%
  \BibitemOpen
  \bibfield  {author} {\bibinfo {author} {\bibfnamefont {K.}~\bibnamefont
  {Slagle}}\ and\ \bibinfo {author} {\bibfnamefont {L.}~\bibnamefont {Fu}},\
  }\bibfield  {title} {\enquote {\bibinfo {title} {Charge transfer excitations,
  pair density waves, and superconductivity in moir\textbackslash 'e
  materials},}\ }\href {\doibase 10.1103/PhysRevB.102.235423} {\bibfield
  {journal} {\bibinfo  {journal} {Phys. Rev. B}\ }\textbf {\bibinfo {volume}
  {102}},\ \bibinfo {pages} {235423} (\bibinfo {year} {2020})}\BibitemShut
  {NoStop}%
\bibitem [{\citenamefont {Haavisto}\ \emph {et~al.}(2022)\citenamefont
  {Haavisto}, \citenamefont {Lado},\ and\ \citenamefont
  {Fumega}}]{haavistoTopologicalMultiferroicOrder2022a}%
  \BibitemOpen
  \bibfield  {author} {\bibinfo {author} {\bibfnamefont {M.}~\bibnamefont
  {Haavisto}}, \bibinfo {author} {\bibfnamefont {J.~L.}\ \bibnamefont {Lado}},
  \ and\ \bibinfo {author} {\bibfnamefont {A.~O.}\ \bibnamefont {Fumega}},\
  }\bibfield  {title} {\enquote {\bibinfo {title} {Topological multiferroic
  order in twisted transition metal dichalcogenide bilayers},}\ }\href
  {\doibase 10.21468/SciPostPhys.13.3.052} {\bibfield  {journal} {\bibinfo
  {journal} {SciPost Phys.}\ }\textbf {\bibinfo {volume} {13}},\ \bibinfo
  {pages} {052} (\bibinfo {year} {2022})},\ \Eprint
  {http://arxiv.org/abs/2204.03360} {arXiv:2204.03360 [cond-mat]} \BibitemShut
  {NoStop}%
\bibitem [{\citenamefont {Bultinck}\ \emph {et~al.}(2020)\citenamefont
  {Bultinck}, \citenamefont {Khalaf}, \citenamefont {Liu}, \citenamefont
  {Chatterjee}, \citenamefont {Vishwanath},\ and\ \citenamefont
  {Zaletel}}]{bultinckGroundStateHidden2020a}%
  \BibitemOpen
  \bibfield  {author} {\bibinfo {author} {\bibfnamefont {N.}~\bibnamefont
  {Bultinck}}, \bibinfo {author} {\bibfnamefont {E.}~\bibnamefont {Khalaf}},
  \bibinfo {author} {\bibfnamefont {S.}~\bibnamefont {Liu}}, \bibinfo {author}
  {\bibfnamefont {S.}~\bibnamefont {Chatterjee}}, \bibinfo {author}
  {\bibfnamefont {A.}~\bibnamefont {Vishwanath}}, \ and\ \bibinfo {author}
  {\bibfnamefont {M.~P.}\ \bibnamefont {Zaletel}},\ }\bibfield  {title}
  {\enquote {\bibinfo {title} {Ground {{State}} and {{Hidden Symmetry}} of
  {{Magic-Angle Graphene}} at {{Even Integer Filling}}},}\ }\href {\doibase
  10.1103/PhysRevX.10.031034} {\bibfield  {journal} {\bibinfo  {journal} {Phys.
  Rev. X}\ }\textbf {\bibinfo {volume} {10}},\ \bibinfo {pages} {031034}
  (\bibinfo {year} {2020})}\BibitemShut {NoStop}%
\bibitem [{\citenamefont {Bernevig}\ \emph {et~al.}(2021)\citenamefont
  {Bernevig}, \citenamefont {Song}, \citenamefont {Regnault},\ and\
  \citenamefont {Lian}}]{bernevigTwistedBilayerGraphene2021a}%
  \BibitemOpen
  \bibfield  {author} {\bibinfo {author} {\bibfnamefont {B.~A.}\ \bibnamefont
  {Bernevig}}, \bibinfo {author} {\bibfnamefont {Z.-D.}\ \bibnamefont {Song}},
  \bibinfo {author} {\bibfnamefont {N.}~\bibnamefont {Regnault}}, \ and\
  \bibinfo {author} {\bibfnamefont {B.}~\bibnamefont {Lian}},\ }\bibfield
  {title} {\enquote {\bibinfo {title} {Twisted bilayer graphene. {{III}}.
  {{Interacting Hamiltonian}} and exact symmetries},}\ }\href {\doibase
  10.1103/PhysRevB.103.205413} {\bibfield  {journal} {\bibinfo  {journal}
  {Phys. Rev. B}\ }\textbf {\bibinfo {volume} {103}},\ \bibinfo {pages}
  {205413} (\bibinfo {year} {2021})}\BibitemShut {NoStop}%
\bibitem [{\citenamefont {Lian}\ \emph {et~al.}(2021)\citenamefont {Lian},
  \citenamefont {Song}, \citenamefont {Regnault}, \citenamefont {Efetov},
  \citenamefont {Yazdani},\ and\ \citenamefont
  {Bernevig}}]{lianTwistedBilayerGraphene2021}%
  \BibitemOpen
  \bibfield  {author} {\bibinfo {author} {\bibfnamefont {B.}~\bibnamefont
  {Lian}}, \bibinfo {author} {\bibfnamefont {Z.-D.}\ \bibnamefont {Song}},
  \bibinfo {author} {\bibfnamefont {N.}~\bibnamefont {Regnault}}, \bibinfo
  {author} {\bibfnamefont {D.~K.}\ \bibnamefont {Efetov}}, \bibinfo {author}
  {\bibfnamefont {A.}~\bibnamefont {Yazdani}}, \ and\ \bibinfo {author}
  {\bibfnamefont {B.~A.}\ \bibnamefont {Bernevig}},\ }\bibfield  {title}
  {\enquote {\bibinfo {title} {Twisted bilayer graphene. {{IV}}. {{Exact}}
  insulator ground states and phase diagram},}\ }\href {\doibase
  10.1103/PhysRevB.103.205414} {\bibfield  {journal} {\bibinfo  {journal}
  {Phys. Rev. B}\ }\textbf {\bibinfo {volume} {103}},\ \bibinfo {pages}
  {205414} (\bibinfo {year} {2021})}\BibitemShut {NoStop}%
\bibitem [{\citenamefont {Tarnopolsky}\ \emph {et~al.}(2019)\citenamefont
  {Tarnopolsky}, \citenamefont {Kruchkov},\ and\ \citenamefont
  {Vishwanath}}]{tarnopolskyOriginMagicAngles2019}%
  \BibitemOpen
  \bibfield  {author} {\bibinfo {author} {\bibfnamefont {G.}~\bibnamefont
  {Tarnopolsky}}, \bibinfo {author} {\bibfnamefont {A.~J.}\ \bibnamefont
  {Kruchkov}}, \ and\ \bibinfo {author} {\bibfnamefont {A.}~\bibnamefont
  {Vishwanath}},\ }\bibfield  {title} {\enquote {\bibinfo {title} {Origin of
  {{Magic Angles}} in {{Twisted Bilayer Graphene}}},}\ }\href {\doibase
  10.1103/PhysRevLett.122.106405} {\bibfield  {journal} {\bibinfo  {journal}
  {Phys. Rev. Lett.}\ }\textbf {\bibinfo {volume} {122}},\ \bibinfo {pages}
  {106405} (\bibinfo {year} {2019})}\BibitemShut {NoStop}%
\bibitem [{\citenamefont {Song}\ \emph {et~al.}(2019)\citenamefont {Song},
  \citenamefont {Wang}, \citenamefont {Shi}, \citenamefont {Li}, \citenamefont
  {Fang},\ and\ \citenamefont {Bernevig}}]{songAllMagicAngles2019a}%
  \BibitemOpen
  \bibfield  {author} {\bibinfo {author} {\bibfnamefont {Z.}~\bibnamefont
  {Song}}, \bibinfo {author} {\bibfnamefont {Z.}~\bibnamefont {Wang}}, \bibinfo
  {author} {\bibfnamefont {W.}~\bibnamefont {Shi}}, \bibinfo {author}
  {\bibfnamefont {G.}~\bibnamefont {Li}}, \bibinfo {author} {\bibfnamefont
  {C.}~\bibnamefont {Fang}}, \ and\ \bibinfo {author} {\bibfnamefont {B.~A.}\
  \bibnamefont {Bernevig}},\ }\bibfield  {title} {\enquote {\bibinfo {title}
  {All {{Magic Angles}} in {{Twisted Bilayer Graphene}} are {{Topological}}},}\
  }\href {\doibase 10.1103/PhysRevLett.123.036401} {\bibfield  {journal}
  {\bibinfo  {journal} {Phys. Rev. Lett.}\ }\textbf {\bibinfo {volume} {123}},\
  \bibinfo {pages} {036401} (\bibinfo {year} {2019})}\BibitemShut {NoStop}%
\bibitem [{\citenamefont {Zou}\ \emph {et~al.}(2018)\citenamefont {Zou},
  \citenamefont {Po}, \citenamefont {Vishwanath},\ and\ \citenamefont
  {Senthil}}]{zouBandStructureTwisted2018}%
  \BibitemOpen
  \bibfield  {author} {\bibinfo {author} {\bibfnamefont {L.}~\bibnamefont
  {Zou}}, \bibinfo {author} {\bibfnamefont {H.~C.}\ \bibnamefont {Po}},
  \bibinfo {author} {\bibfnamefont {A.}~\bibnamefont {Vishwanath}}, \ and\
  \bibinfo {author} {\bibfnamefont {T.}~\bibnamefont {Senthil}},\ }\bibfield
  {title} {\enquote {\bibinfo {title} {Band structure of twisted bilayer
  graphene: {{Emergent}} symmetries, commensurate approximants, and {{Wannier}}
  obstructions},}\ }\href {\doibase 10.1103/PhysRevB.98.085435} {\bibfield
  {journal} {\bibinfo  {journal} {Phys. Rev. B}\ }\textbf {\bibinfo {volume}
  {98}},\ \bibinfo {pages} {085435} (\bibinfo {year} {2018})}\BibitemShut
  {NoStop}%
\bibitem [{\citenamefont {Po}\ \emph {et~al.}(2019)\citenamefont {Po},
  \citenamefont {Zou}, \citenamefont {Senthil},\ and\ \citenamefont
  {Vishwanath}}]{poFaithfulTightbindingModels2019}%
  \BibitemOpen
  \bibfield  {author} {\bibinfo {author} {\bibfnamefont {H.~C.}\ \bibnamefont
  {Po}}, \bibinfo {author} {\bibfnamefont {L.}~\bibnamefont {Zou}}, \bibinfo
  {author} {\bibfnamefont {T.}~\bibnamefont {Senthil}}, \ and\ \bibinfo
  {author} {\bibfnamefont {A.}~\bibnamefont {Vishwanath}},\ }\bibfield  {title}
  {\enquote {\bibinfo {title} {Faithful tight-binding models and fragile
  topology of magic-angle bilayer graphene},}\ }\href {\doibase
  10.1103/PhysRevB.99.195455} {\bibfield  {journal} {\bibinfo  {journal} {Phys.
  Rev. B}\ }\textbf {\bibinfo {volume} {99}},\ \bibinfo {pages} {195455}
  (\bibinfo {year} {2019})}\BibitemShut {NoStop}%
\bibitem [{\citenamefont {Po}\ \emph {et~al.}(2018)\citenamefont {Po},
  \citenamefont {Watanabe},\ and\ \citenamefont
  {Vishwanath}}]{poFragileTopologyWannier2018}%
  \BibitemOpen
  \bibfield  {author} {\bibinfo {author} {\bibfnamefont {H.~C.}\ \bibnamefont
  {Po}}, \bibinfo {author} {\bibfnamefont {H.}~\bibnamefont {Watanabe}}, \ and\
  \bibinfo {author} {\bibfnamefont {A.}~\bibnamefont {Vishwanath}},\ }\bibfield
   {title} {\enquote {\bibinfo {title} {Fragile {{Topology}} and {{Wannier
  Obstructions}}},}\ }\href {\doibase 10.1103/PhysRevLett.121.126402}
  {\bibfield  {journal} {\bibinfo  {journal} {Phys. Rev. Lett.}\ }\textbf
  {\bibinfo {volume} {121}},\ \bibinfo {pages} {126402} (\bibinfo {year}
  {2018})}\BibitemShut {NoStop}%
\bibitem [{Sym()}]{Sym_note}%
  \BibitemOpen
  \href@noop {} {}\bibinfo {note} {Linear terms in $M$ and $P$ are not allowed
  by symmetries, as time-reversal $\mathcal{T}: (M \to - M, P \to P)$ while
  $C_{2y} \mathcal{T}: (M \to M, P \to -P)$. The lowest order symmetry-allowed
  coupling is $M^2 P^2$.}\BibitemShut {Stop}%
\bibitem [{\citenamefont {Wang}\ \emph {et~al.}(2024)\citenamefont {Wang},
  \citenamefont {Vila}, \citenamefont {Zaletel},\ and\ \citenamefont
  {Chatterjee}}]{Taige}%
  \BibitemOpen
  \bibfield  {author} {\bibinfo {author} {\bibfnamefont {T.}~\bibnamefont
  {Wang}}, \bibinfo {author} {\bibfnamefont {M.}~\bibnamefont {Vila}}, \bibinfo
  {author} {\bibfnamefont {M.~P.}\ \bibnamefont {Zaletel}}, \ and\ \bibinfo
  {author} {\bibfnamefont {S.}~\bibnamefont {Chatterjee}},\ }\bibfield  {title}
  {\enquote {\bibinfo {title} {Electrical control of spin and valley in
  spin-orbit coupled graphene multilayers},}\ }\href {\doibase
  10.1103/PhysRevLett.132.116504} {\bibfield  {journal} {\bibinfo  {journal}
  {Phys. Rev. Lett.}\ }\textbf {\bibinfo {volume} {132}},\ \bibinfo {pages}
  {116504} (\bibinfo {year} {2024})}\BibitemShut {NoStop}%
\bibitem [{\citenamefont {Bode}(2003)}]{bode2003spin}%
  \BibitemOpen
  \bibfield  {author} {\bibinfo {author} {\bibfnamefont {M.}~\bibnamefont
  {Bode}},\ }\bibfield  {title} {\enquote {\bibinfo {title} {Spin-polarized
  scanning tunnelling microscopy},}\ }\href@noop {} {\bibfield  {journal}
  {\bibinfo  {journal} {Reports on Progress in Physics}\ }\textbf {\bibinfo
  {volume} {66}},\ \bibinfo {pages} {523} (\bibinfo {year} {2003})}\BibitemShut
  {NoStop}%
\bibitem [{\citenamefont {Jos{\'e} Mart{\'\i}nez-P{\'e}rez}\ and\ \citenamefont
  {Koelle}(2017)}]{jose2017nanosquids}%
  \BibitemOpen
  \bibfield  {author} {\bibinfo {author} {\bibfnamefont {M.}~\bibnamefont
  {Jos{\'e} Mart{\'\i}nez-P{\'e}rez}}\ and\ \bibinfo {author} {\bibfnamefont
  {D.}~\bibnamefont {Koelle}},\ }\bibfield  {title} {\enquote {\bibinfo {title}
  {Nanosquids: Basics \& recent advances},}\ }\href@noop {} {\bibfield
  {journal} {\bibinfo  {journal} {Physical Sciences Reviews}\ }\textbf
  {\bibinfo {volume} {2}},\ \bibinfo {pages} {20175001} (\bibinfo {year}
  {2017})}\BibitemShut {NoStop}%
\bibitem [{\citenamefont {Dolde}\ \emph {et~al.}(2011)\citenamefont {Dolde},
  \citenamefont {Fedder}, \citenamefont {Doherty}, \citenamefont {N{\"o}bauer},
  \citenamefont {Rempp}, \citenamefont {Balasubramanian}, \citenamefont {Wolf},
  \citenamefont {Reinhard}, \citenamefont {Hollenberg}, \citenamefont {Jelezko}
  \emph {et~al.}}]{dolde2011electric}%
  \BibitemOpen
  \bibfield  {author} {\bibinfo {author} {\bibfnamefont {F.}~\bibnamefont
  {Dolde}}, \bibinfo {author} {\bibfnamefont {H.}~\bibnamefont {Fedder}},
  \bibinfo {author} {\bibfnamefont {M.~W.}\ \bibnamefont {Doherty}}, \bibinfo
  {author} {\bibfnamefont {T.}~\bibnamefont {N{\"o}bauer}}, \bibinfo {author}
  {\bibfnamefont {F.}~\bibnamefont {Rempp}}, \bibinfo {author} {\bibfnamefont
  {G.}~\bibnamefont {Balasubramanian}}, \bibinfo {author} {\bibfnamefont
  {T.}~\bibnamefont {Wolf}}, \bibinfo {author} {\bibfnamefont {F.}~\bibnamefont
  {Reinhard}}, \bibinfo {author} {\bibfnamefont {L.~C.}\ \bibnamefont
  {Hollenberg}}, \bibinfo {author} {\bibfnamefont {F.}~\bibnamefont {Jelezko}},
   \emph {et~al.},\ }\bibfield  {title} {\enquote {\bibinfo {title}
  {Electric-field sensing using single diamond spins},}\ }\href@noop {}
  {\bibfield  {journal} {\bibinfo  {journal} {Nature Physics}\ }\textbf
  {\bibinfo {volume} {7}},\ \bibinfo {pages} {459--463} (\bibinfo {year}
  {2011})}\BibitemShut {NoStop}%
\bibitem [{\citenamefont {Block}\ \emph {et~al.}(2021)\citenamefont {Block},
  \citenamefont {Kobrin}, \citenamefont {Jarmola}, \citenamefont {Hsieh},
  \citenamefont {Zu}, \citenamefont {Figueroa}, \citenamefont {Acosta},
  \citenamefont {Minguzzi}, \citenamefont {Maze}, \citenamefont {Budker},\ and\
  \citenamefont {Yao}}]{Block2021}%
  \BibitemOpen
  \bibfield  {author} {\bibinfo {author} {\bibfnamefont {M.}~\bibnamefont
  {Block}}, \bibinfo {author} {\bibfnamefont {B.}~\bibnamefont {Kobrin}},
  \bibinfo {author} {\bibfnamefont {A.}~\bibnamefont {Jarmola}}, \bibinfo
  {author} {\bibfnamefont {S.}~\bibnamefont {Hsieh}}, \bibinfo {author}
  {\bibfnamefont {C.}~\bibnamefont {Zu}}, \bibinfo {author} {\bibfnamefont
  {N.}~\bibnamefont {Figueroa}}, \bibinfo {author} {\bibfnamefont
  {V.}~\bibnamefont {Acosta}}, \bibinfo {author} {\bibfnamefont
  {J.}~\bibnamefont {Minguzzi}}, \bibinfo {author} {\bibfnamefont
  {J.}~\bibnamefont {Maze}}, \bibinfo {author} {\bibfnamefont {D.}~\bibnamefont
  {Budker}}, \ and\ \bibinfo {author} {\bibfnamefont {N.}~\bibnamefont {Yao}},\
  }\bibfield  {title} {\enquote {\bibinfo {title} {Optically enhanced electric
  field sensing using nitrogen-vacancy ensembles},}\ }\href {\doibase
  10.1103/PhysRevApplied.16.024024} {\bibfield  {journal} {\bibinfo  {journal}
  {Phys. Rev. Appl.}\ }\textbf {\bibinfo {volume} {16}},\ \bibinfo {pages}
  {024024} (\bibinfo {year} {2021})}\BibitemShut {NoStop}%
\bibitem [{\citenamefont {Bian}\ \emph {et~al.}(2021)\citenamefont {Bian},
  \citenamefont {Zheng}, \citenamefont {Zeng}, \citenamefont {Chen},
  \citenamefont {St{\"o}hr}, \citenamefont {Denisenko}, \citenamefont {Yang},
  \citenamefont {Wrachtrup},\ and\ \citenamefont {Jiang}}]{bian2021nanoscale}%
  \BibitemOpen
  \bibfield  {author} {\bibinfo {author} {\bibfnamefont {K.}~\bibnamefont
  {Bian}}, \bibinfo {author} {\bibfnamefont {W.}~\bibnamefont {Zheng}},
  \bibinfo {author} {\bibfnamefont {X.}~\bibnamefont {Zeng}}, \bibinfo {author}
  {\bibfnamefont {X.}~\bibnamefont {Chen}}, \bibinfo {author} {\bibfnamefont
  {R.}~\bibnamefont {St{\"o}hr}}, \bibinfo {author} {\bibfnamefont
  {A.}~\bibnamefont {Denisenko}}, \bibinfo {author} {\bibfnamefont
  {S.}~\bibnamefont {Yang}}, \bibinfo {author} {\bibfnamefont {J.}~\bibnamefont
  {Wrachtrup}}, \ and\ \bibinfo {author} {\bibfnamefont {Y.}~\bibnamefont
  {Jiang}},\ }\bibfield  {title} {\enquote {\bibinfo {title} {Nanoscale
  electric-field imaging based on a quantum sensor and its charge-state control
  under ambient condition},}\ }\href@noop {} {\bibfield  {journal} {\bibinfo
  {journal} {Nature Communications}\ }\textbf {\bibinfo {volume} {12}},\
  \bibinfo {pages} {2457} (\bibinfo {year} {2021})}\BibitemShut {NoStop}%
\bibitem [{\citenamefont {{Sahay}}\ \emph {et~al.}(2021)\citenamefont
  {{Sahay}}, \citenamefont {{Hsieh}}, \citenamefont {{Parsonnet}},
  \citenamefont {{Martin}}, \citenamefont {{Ramesh}}, \citenamefont {{Yao}},\
  and\ \citenamefont {{Chatterjee}}}]{Sahay2021}%
  \BibitemOpen
  \bibfield  {author} {\bibinfo {author} {\bibfnamefont {R.}~\bibnamefont
  {{Sahay}}}, \bibinfo {author} {\bibfnamefont {S.}~\bibnamefont {{Hsieh}}},
  \bibinfo {author} {\bibfnamefont {E.}~\bibnamefont {{Parsonnet}}}, \bibinfo
  {author} {\bibfnamefont {L.~W.}\ \bibnamefont {{Martin}}}, \bibinfo {author}
  {\bibfnamefont {R.}~\bibnamefont {{Ramesh}}}, \bibinfo {author}
  {\bibfnamefont {N.~Y.}\ \bibnamefont {{Yao}}}, \ and\ \bibinfo {author}
  {\bibfnamefont {S.}~\bibnamefont {{Chatterjee}}},\ }\bibfield  {title}
  {\enquote {\bibinfo {title} {{Noise Electrometry of Polar and Dielectric
  Materials}},}\ }\href {\doibase 10.48550/arXiv.2111.09315} {\bibfield
  {journal} {\bibinfo  {journal} {arXiv e-prints}\ ,\ \bibinfo {eid}
  {arXiv:2111.09315}} (\bibinfo {year} {2021})},\ \Eprint
  {http://arxiv.org/abs/2111.09315} {arXiv:2111.09315 [cond-mat.mtrl-sci]}
  \BibitemShut {NoStop}%
\bibitem [{\citenamefont {{Vaidya}}\ \emph {et~al.}(2023)\citenamefont
  {{Vaidya}}, \citenamefont {{Gao}}, \citenamefont {{Dikshit}}, \citenamefont
  {{Aharonovich}},\ and\ \citenamefont {{Li}}}]{Vaidya2023}%
  \BibitemOpen
  \bibfield  {author} {\bibinfo {author} {\bibfnamefont {S.}~\bibnamefont
  {{Vaidya}}}, \bibinfo {author} {\bibfnamefont {X.}~\bibnamefont {{Gao}}},
  \bibinfo {author} {\bibfnamefont {S.}~\bibnamefont {{Dikshit}}}, \bibinfo
  {author} {\bibfnamefont {I.}~\bibnamefont {{Aharonovich}}}, \ and\ \bibinfo
  {author} {\bibfnamefont {T.}~\bibnamefont {{Li}}},\ }\bibfield  {title}
  {\enquote {\bibinfo {title} {{Quantum sensing and imaging with spin defects
  in hexagonal boron nitride}},}\ }\href {\doibase
  10.1080/23746149.2023.2206049} {\bibfield  {journal} {\bibinfo  {journal}
  {Advances in Physics X}\ }\textbf {\bibinfo {volume} {8}},\ \bibinfo {eid}
  {2206049} (\bibinfo {year} {2023})},\ \Eprint
  {http://arxiv.org/abs/2302.11169} {arXiv:2302.11169 [quant-ph]} \BibitemShut
  {NoStop}%
\bibitem [{\citenamefont {Casola}\ \emph {et~al.}(2018)\citenamefont {Casola},
  \citenamefont {Van Der~Sar},\ and\ \citenamefont
  {Yacoby}}]{casola2018probing}%
  \BibitemOpen
  \bibfield  {author} {\bibinfo {author} {\bibfnamefont {F.}~\bibnamefont
  {Casola}}, \bibinfo {author} {\bibfnamefont {T.}~\bibnamefont {Van Der~Sar}},
  \ and\ \bibinfo {author} {\bibfnamefont {A.}~\bibnamefont {Yacoby}},\
  }\bibfield  {title} {\enquote {\bibinfo {title} {Probing condensed matter
  physics with magnetometry based on nitrogen-vacancy centres in diamond},}\
  }\href@noop {} {\bibfield  {journal} {\bibinfo  {journal} {Nature Reviews
  Materials}\ }\textbf {\bibinfo {volume} {3}},\ \bibinfo {pages} {1--13}
  (\bibinfo {year} {2018})}\BibitemShut {NoStop}%
\bibitem [{\citenamefont {Hong}\ \emph {et~al.}(2013)\citenamefont {Hong},
  \citenamefont {Grinolds}, \citenamefont {Pham}, \citenamefont {Le~Sage},
  \citenamefont {Luan}, \citenamefont {Walsworth},\ and\ \citenamefont
  {Yacoby}}]{hong2013nanoscale}%
  \BibitemOpen
  \bibfield  {author} {\bibinfo {author} {\bibfnamefont {S.}~\bibnamefont
  {Hong}}, \bibinfo {author} {\bibfnamefont {M.~S.}\ \bibnamefont {Grinolds}},
  \bibinfo {author} {\bibfnamefont {L.~M.}\ \bibnamefont {Pham}}, \bibinfo
  {author} {\bibfnamefont {D.}~\bibnamefont {Le~Sage}}, \bibinfo {author}
  {\bibfnamefont {L.}~\bibnamefont {Luan}}, \bibinfo {author} {\bibfnamefont
  {R.~L.}\ \bibnamefont {Walsworth}}, \ and\ \bibinfo {author} {\bibfnamefont
  {A.}~\bibnamefont {Yacoby}},\ }\bibfield  {title} {\enquote {\bibinfo {title}
  {Nanoscale magnetometry with nv centers in diamond},}\ }\href@noop {}
  {\bibfield  {journal} {\bibinfo  {journal} {MRS bulletin}\ }\textbf {\bibinfo
  {volume} {38}},\ \bibinfo {pages} {155--161} (\bibinfo {year}
  {2013})}\BibitemShut {NoStop}%
\bibitem [{\citenamefont {Chatterjee}\ \emph {et~al.}(2019)\citenamefont
  {Chatterjee}, \citenamefont {Rodriguez-Nieva},\ and\ \citenamefont
  {Demler}}]{CRD2019}%
  \BibitemOpen
  \bibfield  {author} {\bibinfo {author} {\bibfnamefont {S.}~\bibnamefont
  {Chatterjee}}, \bibinfo {author} {\bibfnamefont {J.~F.}\ \bibnamefont
  {Rodriguez-Nieva}}, \ and\ \bibinfo {author} {\bibfnamefont {E.}~\bibnamefont
  {Demler}},\ }\bibfield  {title} {\enquote {\bibinfo {title} {Diagnosing
  phases of magnetic insulators via noise magnetometry with spin qubits},}\
  }\href {\doibase 10.1103/PhysRevB.99.104425} {\bibfield  {journal} {\bibinfo
  {journal} {Phys. Rev. B}\ }\textbf {\bibinfo {volume} {99}},\ \bibinfo
  {pages} {104425} (\bibinfo {year} {2019})}\BibitemShut {NoStop}%
\bibitem [{\citenamefont {Abouelkomsan}\ \emph {et~al.}(2020)\citenamefont
  {Abouelkomsan}, \citenamefont {Liu},\ and\ \citenamefont
  {Bergholtz}}]{abouelkomsan2020particle}%
  \BibitemOpen
  \bibfield  {author} {\bibinfo {author} {\bibfnamefont {A.}~\bibnamefont
  {Abouelkomsan}}, \bibinfo {author} {\bibfnamefont {Z.}~\bibnamefont {Liu}}, \
  and\ \bibinfo {author} {\bibfnamefont {E.~J.}\ \bibnamefont {Bergholtz}},\
  }\bibfield  {title} {\enquote {\bibinfo {title} {Particle-hole duality,
  emergent fermi liquids, and fractional chern insulators in moir{\'e}
  flatbands},}\ }\href {\doibase 10.1103/PhysRevLett.124.106803} {\bibfield
  {journal} {\bibinfo  {journal} {Physical review letters}\ }\textbf {\bibinfo
  {volume} {124}},\ \bibinfo {pages} {106803} (\bibinfo {year}
  {2020})}\BibitemShut {NoStop}%
\bibitem [{\citenamefont {Ledwith}\ \emph {et~al.}(2020)\citenamefont
  {Ledwith}, \citenamefont {Tarnopolsky}, \citenamefont {Khalaf},\ and\
  \citenamefont {Vishwanath}}]{ledwith2020fractional}%
  \BibitemOpen
  \bibfield  {author} {\bibinfo {author} {\bibfnamefont {P.~J.}\ \bibnamefont
  {Ledwith}}, \bibinfo {author} {\bibfnamefont {G.}~\bibnamefont
  {Tarnopolsky}}, \bibinfo {author} {\bibfnamefont {E.}~\bibnamefont {Khalaf}},
  \ and\ \bibinfo {author} {\bibfnamefont {A.}~\bibnamefont {Vishwanath}},\
  }\bibfield  {title} {\enquote {\bibinfo {title} {Fractional chern insulator
  states in twisted bilayer graphene: An analytical approach},}\ }\href
  {https://journals.aps.org/prresearch/abstract/10.1103/PhysRevResearch.2.023237}
  {\bibfield  {journal} {\bibinfo  {journal} {Physical Review Research}\
  }\textbf {\bibinfo {volume} {2}},\ \bibinfo {pages} {023237} (\bibinfo {year}
  {2020})}\BibitemShut {NoStop}%
\bibitem [{\citenamefont {Repellin}\ and\ \citenamefont
  {Senthil}(2020)}]{repellinChernBandsTwisted2020}%
  \BibitemOpen
  \bibfield  {author} {\bibinfo {author} {\bibfnamefont {C.}~\bibnamefont
  {Repellin}}\ and\ \bibinfo {author} {\bibfnamefont {T.}~\bibnamefont
  {Senthil}},\ }\bibfield  {title} {\enquote {\bibinfo {title} {Chern bands of
  twisted bilayer graphene: {{Fractional Chern}} insulators and spin phase
  transition},}\ }\href {\doibase 10.1103/PhysRevResearch.2.023238} {\bibfield
  {journal} {\bibinfo  {journal} {Physical Review Research}\ }\textbf {\bibinfo
  {volume} {2}},\ \bibinfo {pages} {023238} (\bibinfo {year}
  {2020})}\BibitemShut {NoStop}%
\bibitem [{\citenamefont {Li}\ \emph {et~al.}(2021{\natexlab{c}})\citenamefont
  {Li}, \citenamefont {Kumar}, \citenamefont {Sun},\ and\ \citenamefont
  {Lin}}]{liSpontaneousFractionalChern2021}%
  \BibitemOpen
  \bibfield  {author} {\bibinfo {author} {\bibfnamefont {H.}~\bibnamefont
  {Li}}, \bibinfo {author} {\bibfnamefont {U.}~\bibnamefont {Kumar}}, \bibinfo
  {author} {\bibfnamefont {K.}~\bibnamefont {Sun}}, \ and\ \bibinfo {author}
  {\bibfnamefont {S.-Z.}\ \bibnamefont {Lin}},\ }\bibfield  {title} {\enquote
  {\bibinfo {title} {Spontaneous fractional {{Chern}} insulators in transition
  metal dichalcogenide moir\'e superlattices},}\ }\href {\doibase
  10.1103/PhysRevResearch.3.L032070} {\bibfield  {journal} {\bibinfo  {journal}
  {Phys. Rev. Research}\ }\textbf {\bibinfo {volume} {3}},\ \bibinfo {pages}
  {L032070} (\bibinfo {year} {2021}{\natexlab{c}})}\BibitemShut {NoStop}%
\bibitem [{\citenamefont {Grover}\ and\ \citenamefont
  {Senthil}(2008)}]{GroverSenthil}%
  \BibitemOpen
  \bibfield  {author} {\bibinfo {author} {\bibfnamefont {T.}~\bibnamefont
  {Grover}}\ and\ \bibinfo {author} {\bibfnamefont {T.}~\bibnamefont
  {Senthil}},\ }\bibfield  {title} {\enquote {\bibinfo {title} {Topological
  spin hall states, charged skyrmions, and superconductivity in two
  dimensions},}\ }\href {\doibase 10.1103/PhysRevLett.100.156804} {\bibfield
  {journal} {\bibinfo  {journal} {Phys. Rev. Lett.}\ }\textbf {\bibinfo
  {volume} {100}},\ \bibinfo {pages} {156804} (\bibinfo {year}
  {2008})}\BibitemShut {NoStop}%
\bibitem [{\citenamefont {Chatterjee}\ \emph {et~al.}(2020)\citenamefont
  {Chatterjee}, \citenamefont {Bultinck},\ and\ \citenamefont
  {Zaletel}}]{Chatterjee19}%
  \BibitemOpen
  \bibfield  {author} {\bibinfo {author} {\bibfnamefont {S.}~\bibnamefont
  {Chatterjee}}, \bibinfo {author} {\bibfnamefont {N.}~\bibnamefont
  {Bultinck}}, \ and\ \bibinfo {author} {\bibfnamefont {M.~P.}\ \bibnamefont
  {Zaletel}},\ }\bibfield  {title} {\enquote {\bibinfo {title} {Symmetry
  breaking and skyrmionic transport in twisted bilayer graphene},}\ }\href
  {\doibase 10.1103/PhysRevB.101.165141} {\bibfield  {journal} {\bibinfo
  {journal} {Phys. Rev. B}\ }\textbf {\bibinfo {volume} {101}},\ \bibinfo
  {pages} {165141} (\bibinfo {year} {2020})}\BibitemShut {NoStop}%
\bibitem [{\citenamefont {{Khalaf}}\ \emph {et~al.}(2021)\citenamefont
  {{Khalaf}}, \citenamefont {{Chatterjee}}, \citenamefont {{Bultinck}},
  \citenamefont {{Zaletel}},\ and\ \citenamefont {{Vishwanath}}}]{Eslam}%
  \BibitemOpen
  \bibfield  {author} {\bibinfo {author} {\bibfnamefont {E.}~\bibnamefont
  {{Khalaf}}}, \bibinfo {author} {\bibfnamefont {S.}~\bibnamefont
  {{Chatterjee}}}, \bibinfo {author} {\bibfnamefont {N.}~\bibnamefont
  {{Bultinck}}}, \bibinfo {author} {\bibfnamefont {M.~P.}\ \bibnamefont
  {{Zaletel}}}, \ and\ \bibinfo {author} {\bibfnamefont {A.}~\bibnamefont
  {{Vishwanath}}},\ }\bibfield  {title} {\enquote {\bibinfo {title} {{Charged
  skyrmions and topological origin of superconductivity in magic-angle
  graphene}},}\ }\href {\doibase 10.1126/sciadv.abf5299} {\bibfield  {journal}
  {\bibinfo  {journal} {Science Advances}\ }\textbf {\bibinfo {volume} {7}},\
  \bibinfo {pages} {eabf5299} (\bibinfo {year} {2021})},\ \Eprint
  {http://arxiv.org/abs/2004.00638} {arXiv:2004.00638 [cond-mat.str-el]}
  \BibitemShut {NoStop}%
\bibitem [{\citenamefont {Wang}\ \emph {et~al.}(2021)\citenamefont {Wang},
  \citenamefont {Liu}, \citenamefont {Sato}, \citenamefont {Hohenadler},
  \citenamefont {Wang}, \citenamefont {Guo},\ and\ \citenamefont
  {Assaad}}]{Assaad}%
  \BibitemOpen
  \bibfield  {author} {\bibinfo {author} {\bibfnamefont {Z.}~\bibnamefont
  {Wang}}, \bibinfo {author} {\bibfnamefont {Y.}~\bibnamefont {Liu}}, \bibinfo
  {author} {\bibfnamefont {T.}~\bibnamefont {Sato}}, \bibinfo {author}
  {\bibfnamefont {M.}~\bibnamefont {Hohenadler}}, \bibinfo {author}
  {\bibfnamefont {C.}~\bibnamefont {Wang}}, \bibinfo {author} {\bibfnamefont
  {W.}~\bibnamefont {Guo}}, \ and\ \bibinfo {author} {\bibfnamefont {F.~F.}\
  \bibnamefont {Assaad}},\ }\bibfield  {title} {\enquote {\bibinfo {title}
  {Doping-induced quantum spin hall insulator to superconductor transition},}\
  }\href {\doibase 10.1103/PhysRevLett.126.205701} {\bibfield  {journal}
  {\bibinfo  {journal} {Phys. Rev. Lett.}\ }\textbf {\bibinfo {volume} {126}},\
  \bibinfo {pages} {205701} (\bibinfo {year} {2021})}\BibitemShut {NoStop}%
\bibitem [{\citenamefont {Chatterjee}\ \emph {et~al.}(2022)\citenamefont
  {Chatterjee}, \citenamefont {Ippoliti},\ and\ \citenamefont
  {Zaletel}}]{CIZ2022}%
  \BibitemOpen
  \bibfield  {author} {\bibinfo {author} {\bibfnamefont {S.}~\bibnamefont
  {Chatterjee}}, \bibinfo {author} {\bibfnamefont {M.}~\bibnamefont
  {Ippoliti}}, \ and\ \bibinfo {author} {\bibfnamefont {M.~P.}\ \bibnamefont
  {Zaletel}},\ }\bibfield  {title} {\enquote {\bibinfo {title} {Skyrmion
  superconductivity: Dmrg evidence for a topological route to
  superconductivity},}\ }\href {\doibase 10.1103/PhysRevB.106.035421}
  {\bibfield  {journal} {\bibinfo  {journal} {Phys. Rev. B}\ }\textbf {\bibinfo
  {volume} {106}},\ \bibinfo {pages} {035421} (\bibinfo {year}
  {2022})}\BibitemShut {NoStop}%
\bibitem [{\citenamefont {Kwan}\ \emph {et~al.}(2022)\citenamefont {Kwan},
  \citenamefont {Wagner}, \citenamefont {Bultinck}, \citenamefont {Simon},\
  and\ \citenamefont {Parameswaran}}]{Kwan}%
  \BibitemOpen
  \bibfield  {author} {\bibinfo {author} {\bibfnamefont {Y.~H.}\ \bibnamefont
  {Kwan}}, \bibinfo {author} {\bibfnamefont {G.}~\bibnamefont {Wagner}},
  \bibinfo {author} {\bibfnamefont {N.}~\bibnamefont {Bultinck}}, \bibinfo
  {author} {\bibfnamefont {S.~H.}\ \bibnamefont {Simon}}, \ and\ \bibinfo
  {author} {\bibfnamefont {S.~A.}\ \bibnamefont {Parameswaran}},\ }\bibfield
  {title} {\enquote {\bibinfo {title} {Skyrmions in twisted bilayer graphene:
  Stability, pairing, and crystallization},}\ }\href {\doibase
  10.1103/PhysRevX.12.031020} {\bibfield  {journal} {\bibinfo  {journal} {Phys.
  Rev. X}\ }\textbf {\bibinfo {volume} {12}},\ \bibinfo {pages} {031020}
  (\bibinfo {year} {2022})}\BibitemShut {NoStop}%
\bibitem [{\citenamefont {Varney}\ \emph {et~al.}(2010)\citenamefont {Varney},
  \citenamefont {Sun}, \citenamefont {Rigol},\ and\ \citenamefont
  {Galitski}}]{varneyInteractionEffectsQuantum2010}%
  \BibitemOpen
  \bibfield  {author} {\bibinfo {author} {\bibfnamefont {C.~N.}\ \bibnamefont
  {Varney}}, \bibinfo {author} {\bibfnamefont {K.}~\bibnamefont {Sun}},
  \bibinfo {author} {\bibfnamefont {M.}~\bibnamefont {Rigol}}, \ and\ \bibinfo
  {author} {\bibfnamefont {V.}~\bibnamefont {Galitski}},\ }\bibfield  {title}
  {\enquote {\bibinfo {title} {Interaction effects and quantum phase
  transitions in topological insulators},}\ }\href {\doibase
  10.1103/PhysRevB.82.115125} {\bibfield  {journal} {\bibinfo  {journal} {Phys.
  Rev. B}\ }\textbf {\bibinfo {volume} {82}},\ \bibinfo {pages} {115125}
  (\bibinfo {year} {2010})}\BibitemShut {NoStop}%
\bibitem [{\citenamefont {Shao}\ \emph {et~al.}(2021)\citenamefont {Shao},
  \citenamefont {Castro}, \citenamefont {Hu},\ and\ \citenamefont
  {Mondaini}}]{shaoInterplayLocalOrder2021}%
  \BibitemOpen
  \bibfield  {author} {\bibinfo {author} {\bibfnamefont {C.}~\bibnamefont
  {Shao}}, \bibinfo {author} {\bibfnamefont {E.~V.}\ \bibnamefont {Castro}},
  \bibinfo {author} {\bibfnamefont {S.}~\bibnamefont {Hu}}, \ and\ \bibinfo
  {author} {\bibfnamefont {R.}~\bibnamefont {Mondaini}},\ }\bibfield  {title}
  {\enquote {\bibinfo {title} {Interplay of local order and topology in the
  extended {{Haldane-Hubbard}} model},}\ }\href {\doibase
  10.1103/PhysRevB.103.035125} {\bibfield  {journal} {\bibinfo  {journal}
  {Phys. Rev. B}\ }\textbf {\bibinfo {volume} {103}},\ \bibinfo {pages}
  {035125} (\bibinfo {year} {2021})}\BibitemShut {NoStop}%
\end{thebibliography}%
 \appendix

\begin{widetext}
     		\renewcommand{\theequation}{S\arabic{equation}}
		\setcounter{equation}{0}
		\renewcommand{\thefigure}{S\arabic{figure}}
		\setcounter{figure}{0}
		\renewcommand{\thetable}{S\arabic{table}}
		\setcounter{table}{0}
  
		\section{Supplemental Material}
		In this supplemental material, we elaborate on the model presented in the main text, provide additional results and details regarding the methods used. 
		\renewcommand{\theequation}{S\arabic{equation}}
		\setcounter{equation}{0}
		\renewcommand{\thefigure}{S\arabic{figure}}
		\setcounter{figure}{0}
		\renewcommand{\thetable}{S\arabic{table}}
		\setcounter{table}{0}
		\section{Model}
		We are concerned with continuum models that describe twisted TMD homobilayers. The low energy physics in each layer comes from the spin-orbit split valence bands maxima around the corners of the Brillouin zone, the valley $\mathbf{K}_{\pm}$ points. The construction of the continuum model follows from the approach introduced in Ref. \cite{wuTopologicalInsulatorsTwisted2019a}. The Hamiltonian around a single valley (say valley $\mathbf{K}_+$) is given by \begin{multline}
			\label{eq:tmdhomobilayer}
			H^+_{\text{T-TMD}} = \sum_{\mathbf{k}} (\psi^{\dagger}_{t,+}(\mathbf{k}) h_{t,+}(\mathbf{k}) \psi_{t,+}(\mathbf{k}) + \psi^{\dagger}_{b,+}(\mathbf{k})h_{b,+} \psi_{b,+}(\mathbf{k})) \\ + \sum_{k} \Psi^{\dagger}_{+}(\mathbf{k}) V_z \sigma_z \Psi_{+}(\mathbf{k}) +\sum_{\mathbf{k},l_1,l_2,i = 1,2,\dots,6} \Psi^\dagger_{+}(\mathbf{k}+\mathbf{G}_i) T(\mathbf{G}_i) \Psi_{+}(\mathbf{k}) 
		\end{multline} where $\Psi^\dagger_{+}(\mathbf{k}) = (\psi_t^\dagger(\mathbf{k}),\psi_b^\dagger(\mathbf{k}))$ with $\psi^\dagger_{l,+}(\mathbf{k})$ is an operator that creates an electron with momentum $\mathbf{k}$ around valley $\mathbf{K}_+$ in layer $l = t,b$. $h_{l,+}(\mathbf{k})$ is the TMD monolayer low energy Hamiltonian around valley $\mathbf{K}_+$ in layer $l =t,b$. It's modelled as a free quadratic dispersion, $h_{t,+}(\mathbf{k}) = -\hbar^2(\mathbf{k}-\kappa_-)^2/2m^*$ and $h_{b,+}(\mathbf{k}) = -\hbar^2(\mathbf{k}-\kappa_+)^2/2m^*$ where $m^*$ is the effective mass and $\kappa_{\pm}$ are the corners of the moir\'e Brillouin zone around one valley as shown in Fig. \ref{fig_clusters}(a). $V_z$ is an applied displacement field between the top and bottom layers. The matrix $T_{l_1l_2}(\mathbf{G}_i)$ describes the moir\'e potential with $\mathbf{G}_i = R_{(i-1)\pi/3} \mathbf{G}_1$ with $\mathbf{G}_1 = (4\pi/\sqrt{3}a_M,0)$ and $a_M$ is the lattice constant of the moir\'e superlattice. \begin{equation}
		T_{l_1l_2}(\mathbf{G}_1) = \begin{pmatrix}
			V e^{i\psi}& w \\
			w & V e^{-i\psi}
		\end{pmatrix}
	\end{equation}
		The matrix $T_{l_1l_2}$ is constrained by the $C_3$ symmetry of the bilayer system to be $T_{l_1l_2}(R_{2\pi/3} \mathbf{G}_i) = T_{l_1l_2}(\mathbf{G}_i)$ and $T_{l_1l_2}(\mathbf{G}_i) = T_{l_1l_2}^*(-\mathbf{G}_i)$. The diagonal terms of $T_{l_1l_2}$ represent the moir\'e potential in the same layer (top or bottom) while the off-diagonal terms represent interlayer hopping. Hamiltonians of the form \eqref{eq:tmdhomobilayer} have the following symmetries. 
		\begin{itemize}
			\item Moir\'e translation symmetry $t_\mathbf{r}$:  $ t_\mathbf{\mathbf{\rho}} \> \Psi_{\tau}(\mathbf{k}) \> t^{-1}_\mathbf{\mathbb{\rho}} = e^{i \mathbf{k}\cdot \mathbf{\rho}} \Psi_{\tau}(\mathbf{k}) $
			\item Time reversal symmetry that flips the two valleys $\mathcal{T}$ : $\mathcal{T} \Psi_{\pm}(\mathbf{k}) \mathcal{T}^{-1} =  \Psi_{\mp}(\mathbf{k})$
			\item Three-fold rotation $C_{3z}$ : $ C_{3z} \Psi_{\tau}(\mathbf{k}) C_{3z}^{-1} = \Psi_{\tau}(C_{3z} \mathbf{k})  $
			\item In-plane reflection around the $y$ axis $C_{2y}$ that flips the layers. This operation is defined with respect to the origin of the twist of the bilayer, that is the $\Gamma$ point in Fig. \ref{fig_clusters}(a). In momentum space, this operation flips the valley also so $C_{2y}$ : $ C_{2y} \Psi_{\pm}(\mathbf{k}) C_{2y}^{-1} = \sigma_{x} \Psi_{\mp}(C_{2y} \mathbf{k}) $ with $C_{2y} \mathbf{k} = (-k_x,k_y)$ and $\sigma_{x}$ is Pauli $x$ in the layer space. Combining $C_{2y}$ with time reversal symmetry $\mathcal{T}$ flips back the valley so the combined action of $C_{2y} \mathcal{T}$ is a single valley symmetry. $C_{2y} \mathcal{T} $ : $ C_{2y} \mathcal{T} \Psi_{\tau}(\mathbf{k}) (C_{2y}\mathcal{T})^{-1} = \sigma_{x} \Psi_{\tau}(C_{2y}\mathcal{T} \> \mathbf{k}) $ with $C_{2y} \mathcal{T} \> \mathbf{k} = (k_x,-k_y)$			
		\end{itemize}
	In the main text, we used twisted $\mathrm{MoTe}_2$ for our numerical simulations. It is modelled by the following parameters $(m^*,V,w,\psi) = (0.62,8 \> \textrm{meV},-8.5 \> \textrm{meV},-89.6^\circ)$ \cite{wuTopologicalInsulatorsTwisted2019a}. For twist angle $\theta = 1.2^\circ$, the moir\'e lattice constant is $a_M \approx 14.4 \> \rm nm$. 
	\section{Interaction Hamiltonian}
	As discussed in the main text, we focus on the top two flat bands in each valley. We project the long-range Coulomb interaction onto these set of bands. We consider interacting \textit{holes} in the moir\'e valence bands. We define the hole operators $\Phi_{\tau}^{\dagger}(\mathbf{k}) = \Psi_{\tau}(-\mathbf{k})$. By diagonalizing the Hamiltonian \eqref{eq:tmdhomobilayer}, they  are given in terms of the band operators $\{c_{\alpha\tau}^\dagger(\mathbf{k}),c_{\alpha\tau}(\mathbf{k})\}$ by \begin{equation}
	\label{eq:electronoperator}
	   \Phi^\dagger_{\tau}(\mathbf{k} + m \mathbf{G}_1 + n \mathbf{G}_2)  = \sum_{\alpha \in \textrm{all bands}} u^{mn}_{\alpha \tau}(\mathbf{k})  c^\dagger_{\alpha \tau}(\mathbf{k})
	\end{equation}
	where $u^{mn}_{\alpha \tau}(\mathbf{k})$ is the $m,n$ component of the Bloch wavefunctions for band $\alpha$ in valley $\tau = \pm$ obtained in the plane-wave basis expansion $\mathbf{k} + m \mathbf{G}_1 + n\mathbf{G}_2$.  
 The density operator in one valley is given by \begin{equation}
	\label{eq_densityoperator}
	    \rho_{\tau}(\mathbf{q}) = \sum_{\mathbf{k}} \Phi^\dagger_{\tau}(\mathbf{k} + \mathbf{q}) \Phi_{\tau}(\mathbf{k}) =  \sum_{\mathbf{k}, \alpha\beta \in \textrm{all bands}} \lambda^{\alpha\beta}_{\tau}(\mathbf{k}+\mathbf{q},\mathbf{k}) c^{\dagger}_{\alpha\tau}(\mathbf{k} + \mathbf{q}) c_{\beta\tau}(\mathbf{k}) 
	\end{equation}
	where $\lambda^{\alpha\beta}_{\tau}(\mathbf{k}+\mathbf{q},\mathbf{k}) \equiv \langle u_{\alpha \tau}(\mathbf{k}+\mathbf{q})|u_{\beta \tau}(\mathbf{k}) \rangle$ are the form factors calculated from the Bloch eigenstates $u_{\alpha\tau}(\mathbf{k})$ of the bands. The \textit{projection} to the top two bands corresponds to restricting the sum in equation \eqref{eq_densityoperator} to the top two bands $\alpha = 1,2$ and discarding the rest. We then consider density-density interactions defined as \begin{equation}
	\label{eq:intHam_supp}
	    H_{\rm int} = \frac{1}{N_c}\sum_{\mathbf{q} \tau_1 \tau_2} :\rho_{\tau_1}(\mathbf{q})U(\mathbf{q})\rho_{\tau_2}(-\mathbf{q}): 
	\end{equation}
	where $: \> :$ denotes normal ordering with respect to charge neutrality (no holes in the system), $N_c$ is the number of moir\'e unit cells and $U(\mathbf{q})$ is taken to be the dual-gated Coulomb interactions  $U(\mathbf{q}) = 2 \pi U_0 \tanh(d_{g} |\mathbf{q}|)/(\sqrt{3}|\mathbf{q}|a_M)$ with $a_M$ is the moir\'e lattice constant. The screening length $d_g$ represents the distance between the gates and the moir\'e superlattice and $U_0$ is the strength of interaction. We have neglected intervalley interaction terms as they are much weaker, they scale roughly like $a_0/a_M$ where $a_0$ is the lattice constant of the monolayer system. The Hamiltonian \eqref{eq:intHam_supp} has a $U(1) \times U(1)$ symmetry corresponds to independent charge conservation within each valley. The way the dual-gated Coulomb potential is defined makes the constant $U_0$ corresponds to the bare Coulomb potential at a separation of one moir\'e lattice constant, $U_0 = \frac{e^2}{4 \pi \epsilon \epsilon_0 a_M}$ with $\epsilon$ is the dielectric constant.		\begin{figure}[t!]
		\centering
		\includegraphics[width=0.8\linewidth]{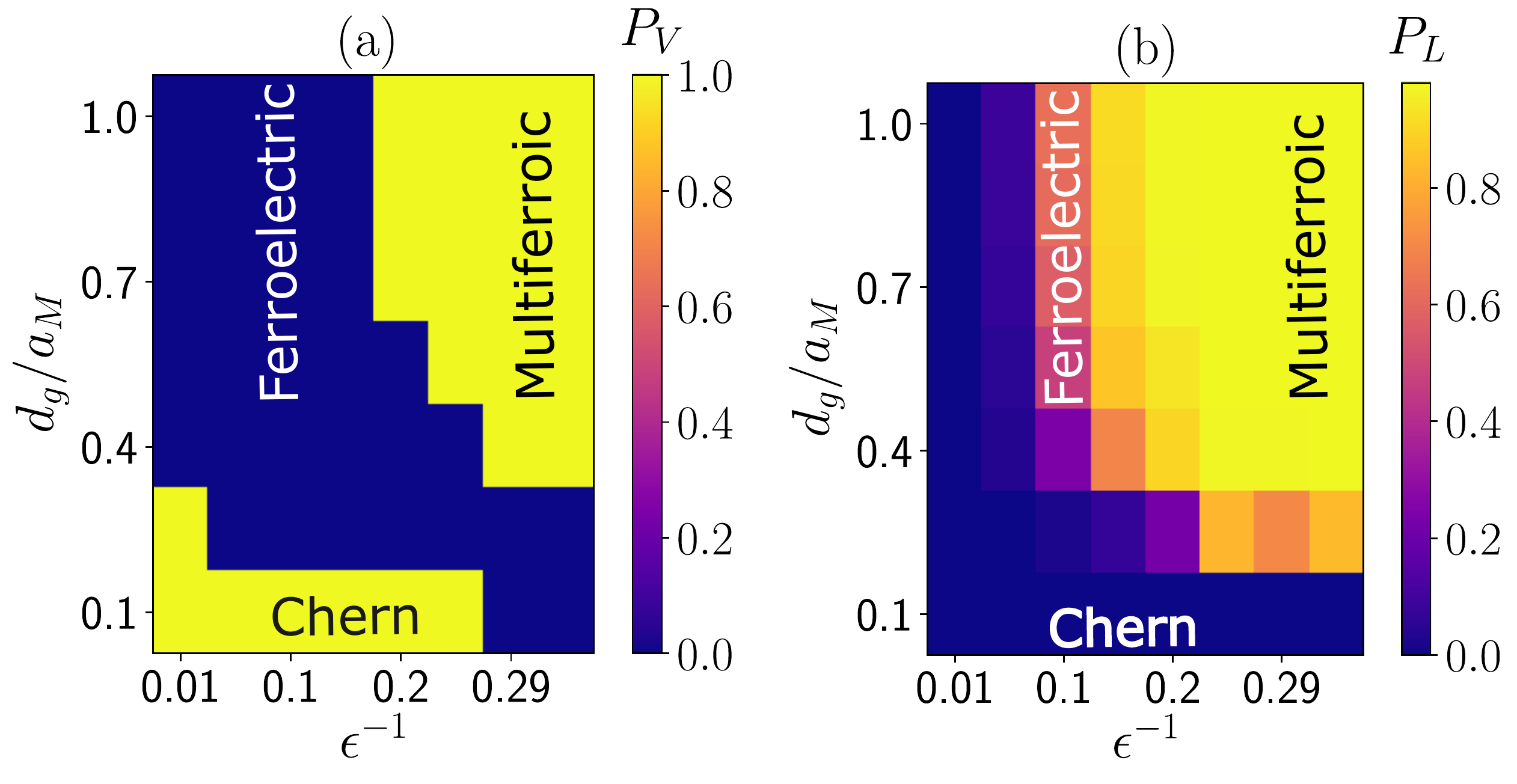}
		\caption{Phase diagram at $\nu_h = 1$ obtained from ED projected onto the first three valence bands (c.f Fig 2. in the main text) as a function of the interaction strength $\epsilon^{-1}$ and the interaction range $d_g/a_M$ where $a_M$ is the moir\'e lattice constant. $P_V$ and $P_L$ denote valley and layer polarization respectively as defined in the main text. The $P_L$ calculations are done with a tiny $V_z = 10^{-5}$ meV to enforce symmetry breaking. Calculations were done using cluster C6 (see Fig. \ref{fig_clusters}(b)).}
		\label{fig_3bandED}
	\end{figure}
    \section{Three band ED at $\nu_h = 1$}
    In this section, we investigate the effects of band mixing with the third valence band at $\nu_h = 1$. We include the third band (in each valley) in our ED calculation and find qualitatively similar results to the two band per valley case presented in the main text. As shown in Figs. \ref{fig_3bandED}(a) and \ref{fig_3bandED}(b), we find the existence of the multiferroic phase for strong interaction strength values of $\epsilon^{-1}$ as evident in the maximal valley polarization $P_V$ and layer polarization $P_L$. The multiferroic phase competes with a ferroelectric phase and a Chern insulator phase as a function of the interaction strength $\epsilon^{-1}$ and the interaction range $d_g$. 
			\begin{figure}[t!]
		\centering
		\includegraphics[width=0.8\linewidth]{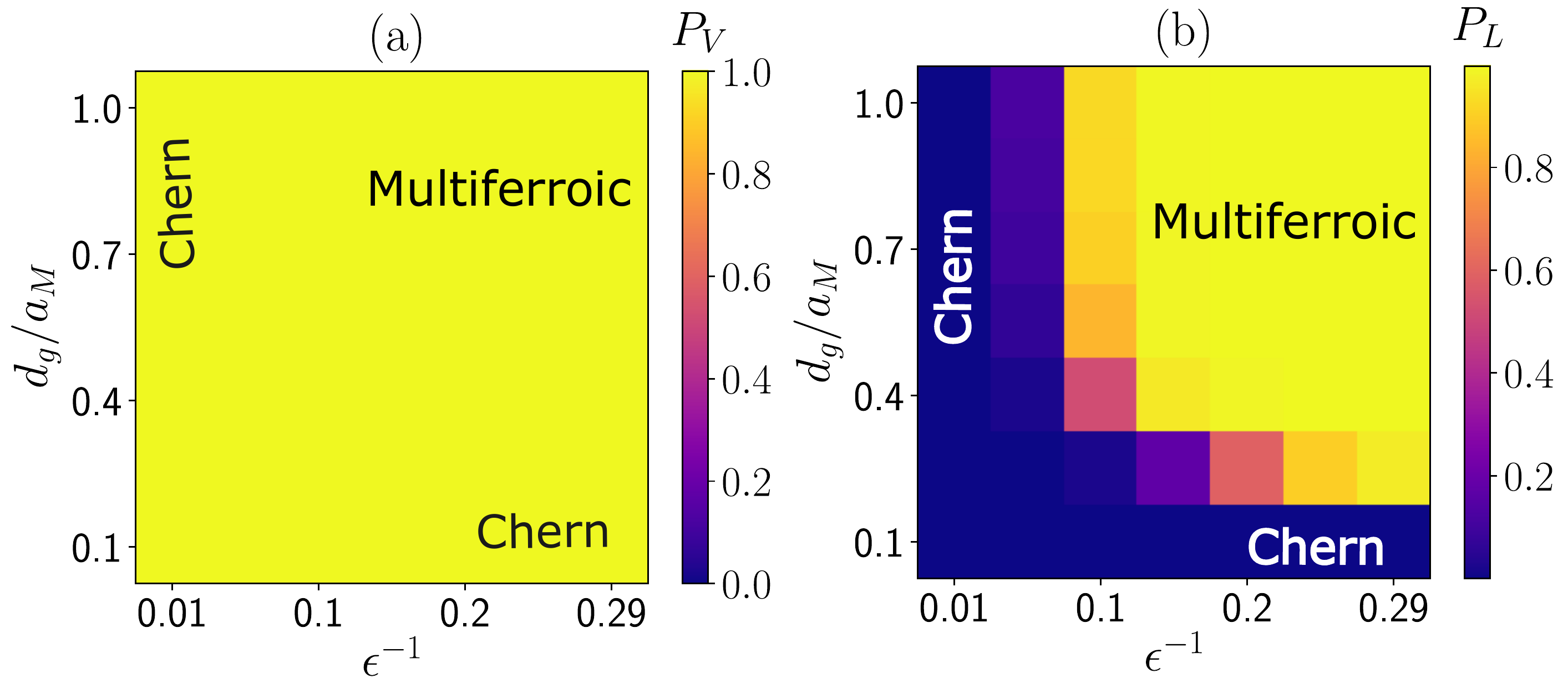}
		\caption{Phase diagram for fillings $\nu_h = 1$ obtained on cluster C8 (Fig. \ref{fig_clusters}(b)) . The quantities $P_V$, $P_L$ are defined in equation (2) in the main text. $d_g/a_M$ is ratio of the screening length from the gates to the moir\'e lattice constant while $\epsilon^{-1}$ is the inverse of the dielectric constant. The $P_L$ calculations are done with a tiny $V_z = 10^{-5}$ meV to enforce symmetry breaking. }
		\label{fig_C8plots}
	\end{figure}
    \section{ED results on the C8 cluster for $\nu_h = 1$}
    In this section, we present results obtained from exact diagonalization on the C8 cluster (c.f Fig. \ref{fig_clusters}(b)) at filling $\nu_h = 1$. On this cluster, we observe valley polarization in the whole parameter space studied as shown in Fig. \ref{fig_C8plots}(a). In the valley polarized space, we observe both a Chern insulator and a multiferroic phase as highlighted in Fig. \ref{fig_C8plots}(b). We notice a sharp transition betweeh the Chern insulator and the multiferroic phase on this cluster with no signs of intermediate ferroelectric phases. 
    			\begin{figure}[t!]
		\centering
		\includegraphics[width=\linewidth]{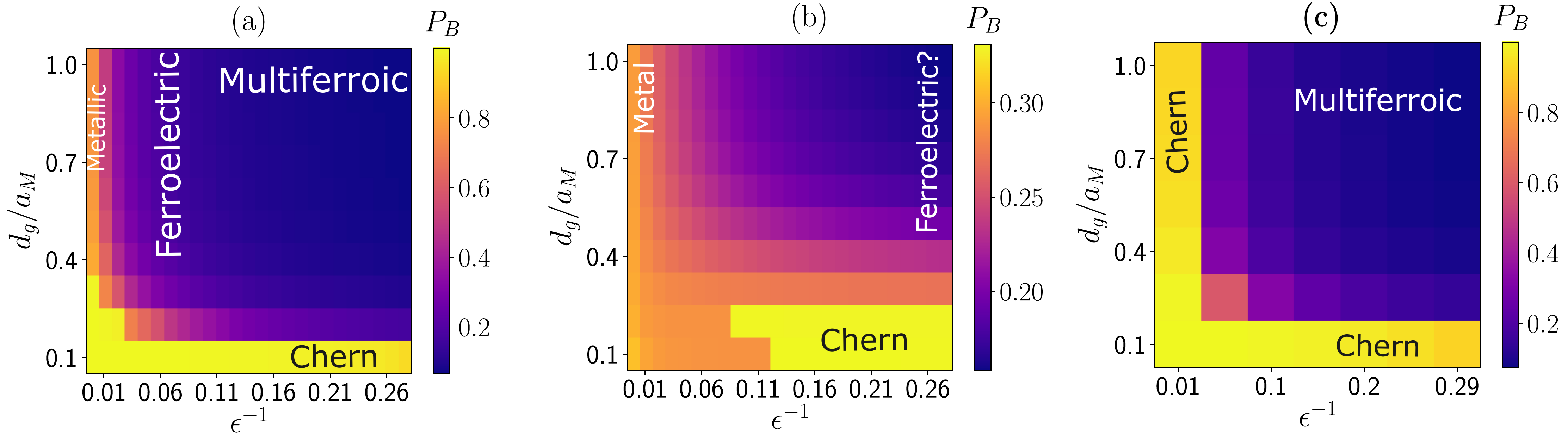}
		\caption{Band Polarization $P_B$ defined in equation \eqref{eq:PB_supp} for (a) filling $\nu_h = 1$ on cluster C6. (b) filling $\nu_h = 3$ on cluster C8. (c) filling $\nu_h = 1$ on cluster C8.}
		\label{fig_P_Bplots}
	\end{figure}
\section{Band Polarization}
In the main text, we have defined the observables $P_V$ and $P_L$ in equation (2) in the main text to characterize the phase diagram. Here, we present results for another quantity, $P_B$ that measures the band polarization, i.e - the difference in occupations between the top and bottom bands. It contains information about the degree of band mixing. It's defined as follow \begin{equation}
\label{eq:PB_supp}
    P_B =  \dfrac{1}{N_e} \sum_{\mathbf{k} \tau }  \braket{n_{1 \tau}(\mathbf{k})} - \braket{n_{2 \tau}(\mathbf{k})}
\end{equation}
In Fig. \ref{fig_P_Bplots}, we show $P_B$ at fillings  $\nu_h = 1$ and $\nu_h = 3$. We notice that $P_B$ is very small in the multiferroic phase. This can be understood as a consequence of layer polarization which is off-diagonal in the band basis as discussed in the main text.

  \section{$\nu_h = 1$ vs $\nu_h = 3$}
  In this section, we elaborate on the difference between filling $\nu_h = 1$ and $\nu_h = 3$. As shown in the main text, there is an absence of spin-valley polarization (and multiferroicity) at filling $\nu_h = 3$ compared to filling $\nu_h = 1$. Within our band-projected exact diagonalization, keeping two bands per valley, the Hamiltonians describing the two fillings are related by a particle-hole transformation. Quite generally, upon particle-hole transformation $c^{\dagger}_{\alpha\tau}(\mathbf{k})\rightarrow d_{\alpha \tau}(-\mathbf{k})$, the Hamiltonian in equation (1) in the main text will transform as \begin{equation}
  \label{eq:PH-H}
  \begin{aligned}
      H \rightarrow H = \sum_{\mathbf{k}\alpha \tau} \epsilon_{\alpha\tau}(\mathbf{k}) d^{\dagger}_{\alpha\tau}(-\mathbf{k})d_{\alpha \tau}(-\mathbf{k}) + \sum_{\mathbf{k}\alpha\tau} E_{\alpha \beta \tau}(\mathbf{k})  d^{\dagger}_{\alpha\tau}(-\mathbf{k})d_{\alpha \tau}(-\mathbf{k}) \\ + \dfrac{1}{N_c} \sum_{\mathbf{k}_1\mathbf{k}_2\mathbf{q}\mathbf{\tau}_1\mathbf{\tau}_2} U(\mathbf{q}) \lambda^{\alpha \beta}_{\tau_1}(\mathbf{k}_1+\mathbf{q},\mathbf{k}_1) \lambda_{\tau_2}^{\gamma \delta}(\mathbf{k}_2 - \mathbf{q},\mathbf{k}_2) d_{\beta \tau_1}^{\dagger}(-\mathbf{k}_1) d^{\dagger}_{\delta \tau_2}(-\mathbf{k}_2)d_{\gamma\tau_2}(-\mathbf{k}_2 + \mathbf{q}) d_{\alpha \tau_1}(-\mathbf{k}_1 - \mathbf{q})
  \end{aligned}
  \end{equation}
  The first term represents the non-interacting disperion. The third term represents the interactions between the particles. The second term represents an interaction-induced single particle term $E_{\alpha\tau}(\mathbf{k})$, it's given explicitly by 
  \begin{equation}
  \begin{aligned}
      \label{eq:intdispersion}
      E_{\alpha \beta \tau}(\mathbf{k}) = \dfrac{-2}{N_c} \sum_{\mathbf{G} ,\mathbf{k}' ,\alpha ,\beta ,\gamma}  U(\mathbf{G}) \lambda^{\alpha\beta}_{\tau_1}(\mathbf{k} + \mathbf{G},\mathbf{k}) \lambda^{\gamma \gamma}_{\tau} (\mathbf{k}' - \mathbf{G},\mathbf{k}')  \\ + 
      \dfrac{2}{N_c} \sum_{\mathbf{G}\mathbf{k}'\alpha \beta \gamma} U(\mathbf{G}) \lambda^{\alpha \beta}_{\tau}(\mathbf{k}' + \mathbf{G},\mathbf{k}) \lambda^{\gamma \alpha}_{\tau_1}(\mathbf{k} - \mathbf{G},\mathbf{k}') 
      \end{aligned}
  \end{equation}
We are going to focus on the regime where interactions dominate over the non-interacting dispersion (multiferroicity persists in the strong interaction limit).  Let $\nu_{p}$ be the particle-hole dual filling of $\nu_h$. Filling $\nu_h = 1 $ corresponds to $\nu_p = 3$ for the two-band per valley projected model. For filling $\nu_p = 1$, the Hamiltonian \eqref{eq:PH-H} (without the non-interacting dispersion) has a similar form to the non particle-hole transformed Hamiltonian descriping $\nu_h = 1 $ except for the interaction-induced single particle terms \eqref{eq:intdispersion}. At filling $\nu_h = 1$, there is robust multiferroicity for strong and long-range interaction but at filling $\nu_h = 3$ (equivalently $\nu_p = 1$), the spin-valley polarization vanishes. We postulate that such a difference can be attributed to the interaction-induced single particle terms \eqref{eq:intdispersion}. The effective bandwidth that arises from these terms can destabilize the spin-valley ferromagnetism. In Fig. \ref{fig_renenergy}, we plot the band structure obtained from diagonalizing the interaction induced single particle terms \eqref{eq:intdispersion}. We find the bandwidth of the top and bottom bands to be roughly $\widetilde{W}_1 = 0.0587 U_0$ and $\widetilde{W}_2 = 0.055 U_0$ respectively where $U_0$ is the strength of the interaction. When $U_0  = 20$ meV (roughly $\epsilon = 4.3$), this gives $\widetilde{W}_1  = 1.174$ meV and $\widetilde{W}_2  = 1.1$ meV. To compare, the non-interacting bandwidth at $\theta = 1.2^\circ$ is given by $W_1  = 0.2593$ meV and $W_2 = 0.922$ meV. We see that the interaction-induced dispersion can be greater than the non-interacting bandwidth. Since such a bandwidth depends on the strength of the interaction, this means that it cannot be overcome for stronger interactions.  
		\begin{figure}[t]
			\centering
			\includegraphics[width=0.6\linewidth]{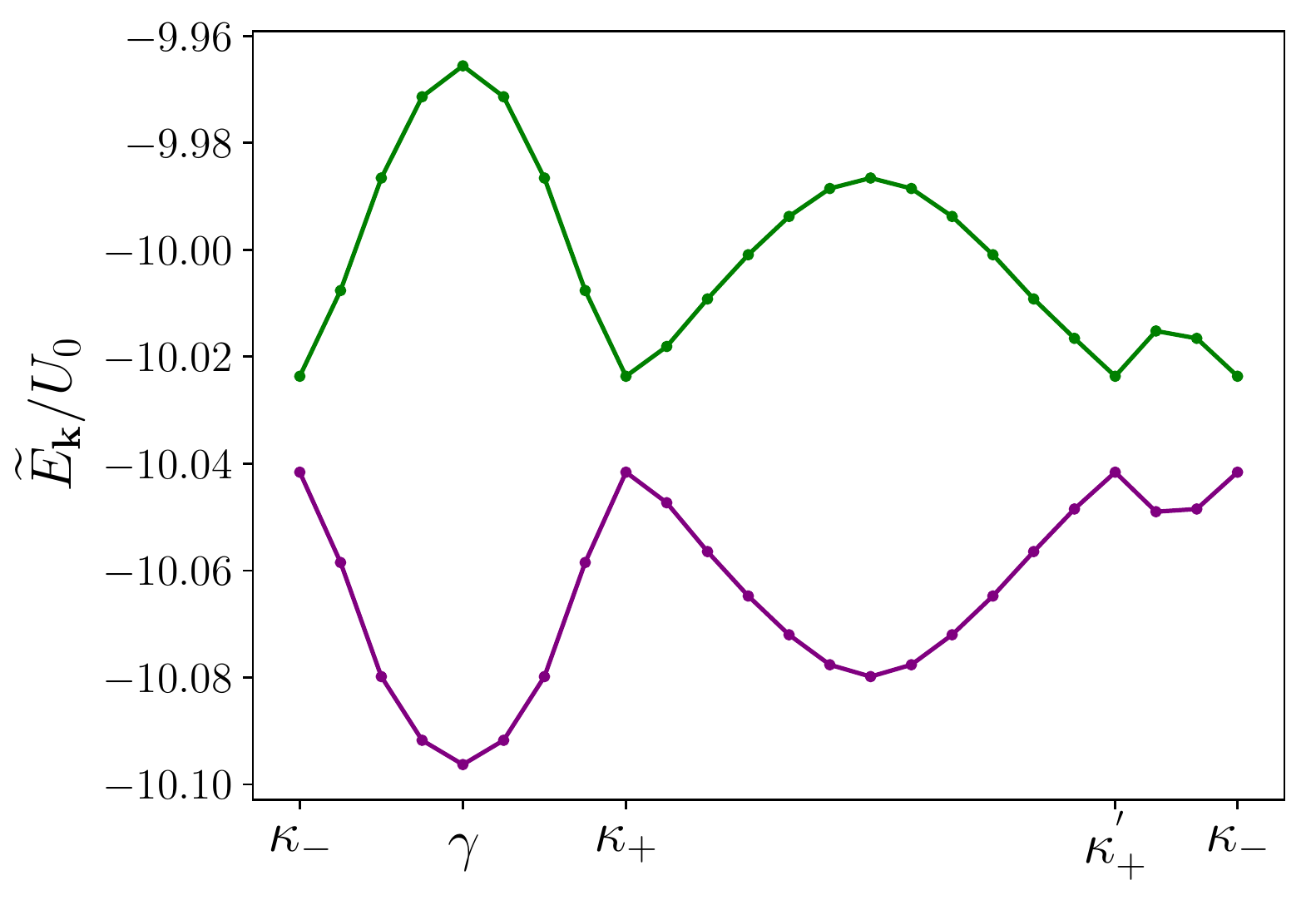}
			\caption{Interaction-induced particle energies obtained from diagonalizing \eqref{eq:intdispersion} in a single valley. $U_0$ is the strength of interaction defined in the main text.}
			\label{fig_renenergy}
		\end{figure}
  		\begin{figure}[t]
			\centering
			\includegraphics[width=\linewidth]{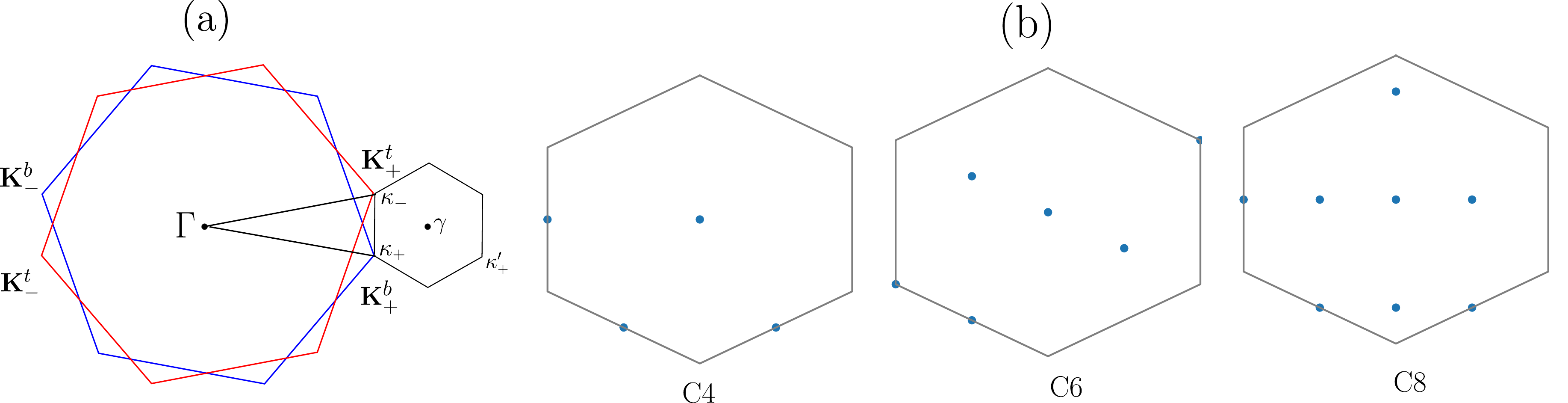}
			\caption{(a) Rotated Brillouin zones for the two layers in addition the moir\'e Brillouin zone around one valley.  (b) The finite clusters used in the exact diagonalization. Each point in the grid carry additional label $(\tau,\alpha)$ that label the valleys $\tau = \pm$ and the flat bands $\alpha =1,2$. So the number of momentum points in the clusters is four times what is shown. }
			\label{fig_clusters}
		\end{figure}
		\section{Exact Diagonalization}

		The main method we employ in this paper is exact diagonalization on finite clusters of the Hamiltonian (1) in the main text. We work directly in momentum space so the finite clusters used are grids in momentum space. In Fig. \ref{fig_clusters}(b), we show the clusters that were used in generating and interpreting the results in the main text. Only cluster C6 that features both the $\mathbf{K}_{\pm}$ points and the $\mathbf{M}$ point which is the main reason why we choose this cluster to present the results in the main text as it contains the highest number of different symmetry points. 
		
It is worth mentioning that we find general agreement among the results obtained from the different clusters, namely the robust existence of the phases, the metal, the multiferroic phase and the Chern insulator.
We also note that we observe on cluster C6 signatures of an intermediate ferroelectric phase between the metal and the multiferrioic and similarily between the Chern insulator and the mutliferroic as discussed in the main text. The intermediate phase has zero spin-valley polarization $P_V$ with signs of breaking the $C_{2y}$ symmetry and possible anti-ferromagnetic stripe correlations.
However, the finite size limitations prevents us from pin-pointing if this phase exists or it is a finite-size artifact. On clusters C4 and C8, we observe the absence of such a phase. However clusters C4 and C8 don't feature the $K$ point which could play an important role in choosing different phases \cite{varneyInteractionEffectsQuantum2010,shaoInterplayLocalOrder2021}.

\section{Neutral excitations above the ferroelectic phase}

In this section, we discuss how to estimate the energy cost to transfer one charge between the two layers in the ferroelectric phase. Recall the maximally layer-polarized basis defined in the main text as $\tilde{c}^\dagger_{\beta \tau}(\mathbf{k}) = \sum_{\alpha} U_{\beta \alpha} (\mathbf{k}) c^{\dagger}_{\alpha \tau}(\mathbf{k}) $, where $U_{\beta \alpha}(\mathbf{k})$ is a $2 \times 2$ unitary matrix that is optimized to maximize the layer polarization in each valley. As presented, the ferroelectric phase maximizes the layer polarization $P_L$ defined in equation (2) in the main text.

The elementary neutral excitation in this setting is an excitation corresponding to transferring one hole to the other layer in the ferroelectric phase.  The energy cost for such an excitation is indiciative of how robust the ferroelectric phase in our calculations. We estimate such a cost directly by doing the ED calculation in the layer-polarized basis. For simplicity, we consider the flat band limit, i.e., we neglect effects of the band-width which is quite small compared to the interaction scale. 

We start by rewriting the interacting Hamiltonian \eqref{eq:intHam_supp}:
\begin{equation}
     H_{\rm int} = \frac{1}{N_c}\sum_{\mathbf{q} \tau_1 \tau_2} :\rho_{\tau_1}(\mathbf{q})U(\mathbf{q})\rho_{\tau_2}(-\mathbf{q}): 
\end{equation}
The density operator $\rho_{\tau}(\mathbf{q})$ expressed in the layer basis is given by
\begin{equation}
\label{eq:densityoperator_layer}
    \rho_{\tau}(\mathbf{q}) =  \sum_{\mathbf{k}, \alpha\beta \gamma \delta} U^{\dagger}_{\alpha \gamma}(\mathbf{k} + \mathbf{q}) \lambda^{\alpha\beta}_{\tau} (\mathbf{k}+\mathbf{q},\mathbf{k}) U_{\beta \delta}(\mathbf{k})\tilde{c}^{\dagger}_{\gamma\tau}(\mathbf{k} + \mathbf{q}) \tilde{c}_{\delta\tau}(\mathbf{k}) 
\end{equation}
We further assume spin-valley polarization, consistent with our findings in the multi-ferroic phase. 
A complication arises from the fact that the density operator has off-diagonal components in the layer basis since there is no particle number conservation in each layer. To proceed, we impose an emergent U(1) symmetry in the layer space, such that the number of holes in each layer is conserved, and drop the off-diagonal terms in \eqref{eq:densityoperator_layer}. 
Since the orbitals are almost entirely layer polarized, we do not expect this approximation to significantly affect our estimate.
We then diagonalize the interacting Hamiltonian in different sectors denoted by $N = (N_{\rm top} - N_{\rm bottom})/2$  where $N_{\rm top}$ and $N_{\rm bottom}$ are the number of holes in the top and bottom layers respectively. We then compare the energy of the maximally layer polarized sector $N_{\rm top} = N$ and the energy in the sector with one charge transferred to the other layer $N_{\rm bottom} = 1$. In Fig. \ref{fig_ferroelectricgap}, we plot this gap as a function of the dielectric constant, and find that it is of the order of tens of meV.
In particular, at $\epsilon^{-1} \approx 0.23$, the energy cost to move a hole to the opposite layer is approximately 17 meV, as quoted in the main text.
\begin{figure}
\includegraphics[width=0.5\linewidth]{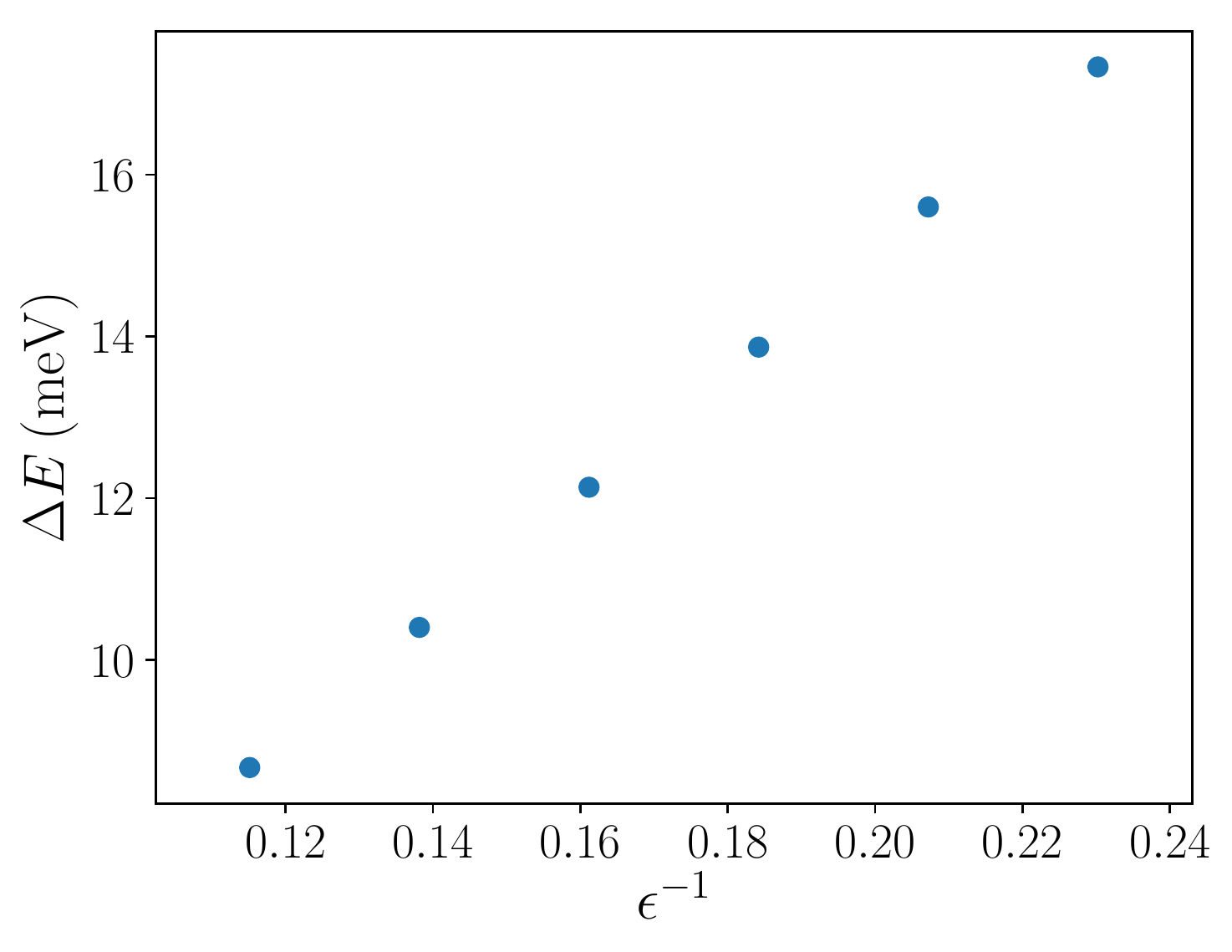}
		\caption{The energy cost to transfer a single charge across layers in the ferroelectric phase, defined as $\Delta E = E(N_{\rm top} = N-1, N_{\rm bottom} = 0) - E(N_{\rm top} = N, N_{\rm bottom} = 0)$, as a function of the dielectric constant $\epsilon$.}
		\label{fig_ferroelectricgap}
	\end{figure}
  \end{widetext}
\end{document}